# Attacking COVID-19 Progression using Multi-Drug Therapy for Synergetic Target Engagement

Mathew Coban[1], Juliet Morrison PhD[2], William D. Freeman MD[3], Evette Radisky PhD[1], Karine G. Le Roch PhD[4], Thomas R. Caulfield, PhD[1,5-8]

**Affiliations**

[1] *Department of Cancer Biology, Mayo Clinic, 4500 San Pablo Road South, Jacksonville, FL, 32224 USA*

[2] *Department* of Microbiology and Plant Pathology*, University of California, Riverside, 900 University, Riverside, CA, 92521 USA*

[3] *Department of Neurology, Mayo Clinic, 4500 San Pablo South, Jacksonville, FL, 32224 USA*

[4] *Department of Molecular, Cell and Systems Biology, University of California, Riverside, 900 University, Riverside, CA, 92521 USA*

[5] *Department of Neuroscience, Mayo Clinic, 4500 San Pablo South, Jacksonville, FL, 32224 USA*

[6] *Department of Neurosurgery, Mayo Clinic, 4500 San Pablo South, Jacksonville, FL, 32224 USA*

[7] *Department of Health Science Research (BSI), Mayo Clinic, 4500 San Pablo South, Jacksonville, FL, 32224 USA*

[8] *Department of Clinical Genomics (Enterprise), Mayo Clinic, Rochester, MN 55905, USA*

**Correspondence to:**
Thomas R. Caulfield, PhD,
Dept of Neuroscience, Cancer Biology, Neurosurgery, Health Science Research, & Clinical Genomics
Mayo Clinic, 4500 San Pablo Road South
Jacksonville, FL 32224
Telephone: +1 904-953-6072,
E-mail: caulfield.thomas@mayo.edu

## SUMMARY

COVID-19 is a devastating respiratory and inflammatory illness caused by a new coronavirus that is rapidly spreading throughout the human population. Over the past 6 months, severe acute respiratory syndrome coronavirus 2 (SARS-CoV-2), the virus responsible for COVID-19, has already infected over 11.6 million (25% located in United States) and killed more than 540K people around the world. As we face one of the most challenging times in our recent history, there is an urgent need to identify drug candidates that can attack SARS-CoV-2 on multiple fronts. We have therefore initiated a computational dynamics drug pipeline using molecular modeling, structure simulation, docking and machine learning models to predict the inhibitory activity of several million compounds against two essential SARS-CoV-2 viral proteins and their host protein interactors; S/Ace2, Tmprss2, Cathepsins L and K, and $M^{pro}$ to prevent binding, membrane fusion and replication of the virus, respectively. All together we generated an ensemble of structural conformations that increase high quality docking outcomes to screen over >6 million compounds including all FDA-approved drugs, drugs under clinical trial (>3000) and an additional >30 million selected chemotypes from fragment libraries. Our results yielded an initial set of 350 high value compounds from both new and FDA-approved compounds that can now be tested experimentally in appropriate biological model systems. We anticipate that our results will initiate screening campaigns and accelerate the discovery of COVID-19 treatments.

## INTRODUCTION

COVID-19 is a disease cause by severe acute respiratory syndrome coronavirus 2 (SARS-CoV-2). It was identified in Wuhan city, in the Hubei province of China in December 2019 (Chen et al., 2020; Huang et al., 2020; Zhu et al., 2020). The virus is spread between people via small droplets produce by talking, sneezing and coughing. The disease was declared a global pandemic by the World health organization (WHO) on March 11th, 2020. While a large proportion of the cases results in mild symptoms such as fever, cough, fatigues, loss of smell and taste, as well as shortness of breath, some cases progress into more acute respiratory symptoms such as pneumonia, multiple-organ failure, septic shock and blood clots. These more severe symptoms can lead to death and are likely to be precipitated by a cytokine storm after infection and multiplication of the virus in humans. Indeed, recent data indicate that the levels of IL-6 correlate with respiratory and organ failures (Gubernatorova et al., 2020). So far, the estimated death rate of SARS-CoV-2 is above 1.3%, which is more than 10 times higher than the death rate of seasonal influenza (Abdollahi et al., 2020). Older patients and patients who have serious underlying medical conditions such as hypertension, diabetes, and asthma are at higher risk for severe disease outcomes (Tian et al., 2020). A clear understanding of the genetics and molecular mechanisms controlling severe illness remains to be determined.

SARS-CoV-2 is a positive-sense, single-stranded RNA betacoronavirus, closely related to SARS-CoV-1, which caused severe acute respiratory syndrome (SARS) in 2003, and Middle East respiratory syndrome coronavirus (MERS-CoV), which caused MERS in 2012. Positive-strand RNA viruses are a large fraction of known viruses including common pathogens such as rhinoviruses that cause common colds, as well as dengue virus, hepatitis C virus (HCV), West Nile virus. The first genome sequence of SARS-CoV-2 was released in early January on the open access virological website (http://virological.org/) (Zhou et al., 2020). Its genome is ~29.8 kb and possesses 14 open reading frames (ORFs), encoding 27 proteins (Wu et al., 2020a). The genome contains four structural proteins: spike (S) glycoprotein, envelope (E) protein, membrane (M) protein, and nucleocapsid (N) protein. The E and M proteins form the viral envelope, while the N protein binds to the virus's RNA genome. The spike glycoprotein is a key surface protein that interacts with cell surface receptor, angiotensin-converting enzyme 2 (Ace2) mediating entrance of the virus into host cells (Zhu et al., 2018). In addition to its dependence on the binding of S to Ace2, cell entry also requires priming of S by the host serine protease, transmembrane serine protease 2 (Tmprss2). Tmprss2 proteolytically processes S, promoting membrane fusion, cell invasion and viral uptake (Heurich et al., 2014; Hoffmann et al., 2020). Blocking viral entry by targeting S/Ace2 interaction or Tmprss2-mediated priming may constitute an effective treatment strategy for COVID-19. The non-structural proteins, which include the main viral protease (nsp5 or $M^{pro}$) and RNA polymerase (nsp12), regulate virus replication and assembly. They are expressed as two long polypeptides, pp1a and pp1ab, which are proteolytically processed by $M^{pro}$. The key role of $M^{pro}$ in viral replication makes it a good therapeutic target as well. A third group of proteins are described as accessory proteins. This group is the least understood, but its members are thought to counteract host innate immunity (Kim et al., 2020, Cell 181, 914–921) (**Fig. 1A**).

There is currently no treatment or vaccine available to prevent or treat COVID-19 (Baden and Rubin, 2020; Lurie et al., 2020) (https://www.fda.gov/news-events/press-announcements/coronavirus-covid-19-update-daily-roundup-june-1-2020). While the FDA has granted emergency use authorization (EUA) for the 65-year-old antimalarial drug, hydroxychloroquine, COVID-19 treatment based on early results from clinical trial in China and France (Gao et al., 2020; Gautret et al., 2020a; Gautret et al., 2020b; Million et al., 2020), more recent results reported that hydroxychloroquine does not decrease viral replication, pneumonia or hospital mortality, and may in fact increase cardiac arrest in patients infected with COVID-19 (Mehra et al., 2020; Rosenberg et al., 2020). The accuracy of the statistical analyses in these studies raised serious concerns in the scientific community. More accurate data are needed to reach a conclusion about the effect of hydroxychloroquine in COVID-19 patients. In another recent study published in the New England Journal of Medicine, the antiviral remdesivir, an unapproved drug that was originally developed to fight Ebola, seemed to improve patients with severe breathing problems (Beigel et al., 2020) and has also recently been granted EUA by the FDA. Repurposing drugs that are designed to treat other diseases is one of the quickest ways to find therapeutics to control the current pandemic. Such drugs have already been tested for toxicity issues and can be granted EUA by the FDA to help doctors to treat COVID-19 patients.

Another efficient way to attack the virus is to use drug cocktails to target multiple enzymes/pathways used by the virus. Combination therapy has the advantage of being less likely to select for treatment-resistant viral mutants. Such a strategy has been successfully used to treat hepatitis C virus (HCV) and human-immunodeficiency virus (HIV) infections. In the case of HCV, the treatment, Enpclusa, combines sofosbuvir, which inhibits the viral RNA-dependent RNA polymerase (NS5B), and velpatasvir, a defective substrate that inhibits NS5A. Antiretroviral therapy (ART) against HIV combines drugs from different drug classes to target disparate aspects of the HIV replication cycle. These drug classes include nucleoside reverse transcriptase inhibitors, non-nucleoside reverse transcriptase inhibitors, protease inhibitors, fusion inhibitors, CCR5 antagonists, post-attachment inhibitors, and integrase inhibitors. One example from HIV-AIDS literature is the randomized comparison of 4 groups of patients comparing monotherapy to combination therapies: zidovudine (ZDV) monotherapy; ZDV zidovudine and didanosine; ZDV plus zalcitabine; or didanosine monotherapy. This randomized trial showed positive results when ZDT was combined with didanosine or zalcitabine, and for didanosine compared to ZDT monotherapy in raising CD4 counts greater than 50% (Hammer et al., 1996). Combination therapy has become standard of care initial treatment in other infectious diseases such as *Mycobacterium tuberculosis* and failure to cure with monotherapy and requires multidrug therapy (MDT) (Collaborative Group for the Meta-Analysis of Individual Patient Data in et al., 2018). Similar MDT is also found effective in hepatitis C virus infection using glecaprevir and pibrentasivr combination therapies which lead to sustained virological response rates as far out as 12 weeks' post-treatment (Wang et al., 2019).

We propose an effective combination therapy for COVID-19 could target the SARS-CoV-2 replication cycle at multiple levels to synergistically inhibit viral spread and dissemination. Using a computational pipeline that aimed to expeditiously identify lead compounds against COVID-19, we combined compound library preparation, molecular modeling, and structure simulations to generate an ensemble of conformations and increase high quality docking outcomes against two essential SARS-CoV-2 viral proteins and their host protein interactions; S/Ace2, Tmprss2, Cathepsin L and K, and $M^{pro}$ that are known to control both viral *binding*, *entry* and virus *replication* (**Fig. 1A**). Our in silico approach (**Fig. 1B**), which will most likely lead into experimental virus screening, structural characterization of binding interactions by X-ray crystallography, and compound safety profiling. Virtual screening (VS) is a rational driven controller for identification of new hits from compound libraries (Willett, 2006) using either ligand-based (LBvs) or structure-based (SBvs) virtual screening (Dror et al., 2004). LBvs tactics use structural and biological data of known active compounds to select favorable candidates with biological activity from experiments (Jahn et al., 2009; Maldonado et al., 2006). SBvs approaches, on the other hand, examine quantitative structure-activity relationships (QSAR), clustering, pharmacophore and 3D shape matching (Villoutreix et al., 2007). The utility of VS is evident in the growth of our knowledge base of new compounds and existing drugs as well as the expansion of our structural databases. SBvs is generally the preferred approach when access to the target 3D-information derived from NMR, X-ray crystallography or homology models (Jahn et al., 2009; Maldonado et al., 2006) is possible. Molecular docking (docking) is the most common SBvs approach used today (Bottegoni et al., 2009; Corbeil et al., 2012; Fernandez-Recio et al., 2005; Friesner et al., 2006; McGann, 2012; Morris et al., 2009) and searches for the ideal position and orientation (called "pose") of the small molecule within a target's binding site, which gives a score for the pose. When including knowledge of experimentally known compounds ("actives") from a 3D target, LBvs and SBvs can be combined to increase likelihood of obtaining new actives from searches (Kruger and Evers, 2010).

Hit identification in VS also requires careful selection of the methods used based on the goal of the project (e.g. compound databases and libraries can be either proprietary, commercial or public) (Bender, 2010). ZINC is one such large public database often used in VS (Irwin and Shoichet, 2005), which contains millions of compounds. By contrast, other libraries have structure-activity relationships (SAR) databases (Scior et al., 2007) that integrate information about compound interactions with their known targets. DrugBank, Chem-Space are other attractive sources of compounds for drug repurposing (or repositioning) (Ashburn and Thor, 2004; Duenas-Gonzalez et al., 2008; O'Connor and Roth, 2005) (Wishart et al., 2008), and maintain drug diversity that is useful for scaffold development (Gozalbes et al., 2008; Schreiber, 2000).

Advances in computing power have increased utility of in silico screening capabilities and balanced the need for accuracy with virtual high-throughput screening approximations and assumptions (Anthony, 2009; Lee et al., 2008; McGaughey et al., 2007; Plewczynski et al., 2009), while recent techniques have improved accuracy without sacrificing CPU time (Caulfield and Devkota, 2012; Caulfield et al., 2011; Jiang et al., 2014; MacKerell et al., 1998; Phillips et al., 2005) (**Fig. 1B**). Further innovations in docking methods have improved the exactness of empirical docking equations (Corbeil et al., 2012; Fernandez-Recio et al., 2005; Friesner et al., 2006; Kalid et al., 2012; Kruger and Evers, 2010; McGann, 2012). Accuracy is improved by incorporating molecular flexibility with simulations (Caulfield, 2012; Caulfield et al., 2019; Caulfield and Medina-Franco, 2011; Caulfield et al., 2011; Caulfield et al., 2014; Kayode et al., 2016), thus capturing conformational information on structural changes that directly impact compound docking results.

Here, we present in silico screening of both the approved FDA compound library and >30 million compounds representing new chemical entities (NCEs) (Clecildo Barreto Bezerra et al., 2018; Ekins et al., 2014; Janes et al., 2018; Pillaiyar et al., 2020). Other libraries consisting of approved drugs, natural products, and a subset of the ZINC data base were also included based on relationship with SARS-CoV-2 virus (Corsello et al., 2017; Lagarde et al., 2018; Riva et al., 2020)(Irwin and Shoichet, 2005). Our findings include >350 compounds, including both NCEs (310) and FDA repurposing compounds (40). Our approach combines VS and careful library selection with advanced docking techniques to efficiently search the behemoth chemical landscape of possible organic compounds (Bohacek et al., 1996) and identify high value hits toward key SARS-CoV-2 targets.

## RESULTS

To target the COVID19 problem on multiple fronts (e.g. Ace2:S protein, Tmprss2, $M^{pro}$, and Cathepsin L and K), as well as improve our screening accuracy using our selected repurposing libraries and new chemical entity libraries (ZINC database), we implemented a novel method that integrates protein flexibility/shape, adaptive biasing algorithms, machine learning from drug data, and final Z-score matrix weighting to our drug modeling. We matched all FDA compounds with our realistic (X-ray derived) protein structures over a dynamic range of protein conformations with accelerated dynamics using our algorithms, such as Maxwell's demon molecular dynamics (MdMD); this approach combines docking with simulations for exploration of both ligand and protein flexibility (Caulfield, 2012; Caulfield et al., 2019; Caulfield and Devkota, 2012; Caulfield and Medina-Franco, 2011; Caulfield et al., 2014; Kayode et al., 2016; von Roemeling et al., 2018). We then refined the drug-target interface our specific leader-like hit compounds using the quantum mechanics (QM)-based scoring within our MdMD matrix (Caulfield, 2012) to make our go/no-go assessment, which is particularly useful with NCEs and de novo compound design (DCDs). The protocol for library, structural modeling, dynamics, refinement, and hit identification as part of a pipeline is given (**Fig. 1B**).

### I. Modeling and Simulations for Improved Docking Outcome

To improve our docking outcome, we constructed x-ray structure-based models of Ace2 bound to S-protein, $M^{pro}$, and Tmprss2 in our molecular dynamics simulations (MDS) and virtual screening (**Fig. 1B,S1**). As S-protein interfaces with Ace2 at a distinct region from the active site (**Fig. S1A-D**), inhibition of the binding site by ligands may disrupt the Ace2/S-protein interaction. Canonical inhibitors of Ace2 bind at the active site where angiotensin interacts, whereas drugs directed at the structural region for S-protein binding are not overlapping with the binding site. The modulation of Ace2/S-protein interaction by canonical Ace2 inhibitors is likely allosteric and suboptimal. Therefore, directly targeting the interface of the interaction should increase efficacy of the approach and block COVID viral binding, precluding entry (**Fig. S1**). Additional investigation into the glycosylation sites of the S-protein demonstrated that the Ace2 binding site is mostly unaffected by these additions (**Fig. S2**).

### A. S-protein:Ace2 interaction (protein-protein inhibitor, PPI) requires dynamics to reveal binding site

To get the optimal interface for drug screening, we used our grid searching algorithms, as well as site mapping and protein-protein docking, to examine the protein-protein interactions surface using MDS (**Fig. 2-3,S1**) (Bhachoo and Beuming, 2017; Caulfield and Devkota, 2012; Caulfield et al., 2011; Caulfield and Harvey, 2007; Fernandez-Recio et al., 2005; Kozakov et al., 2006). The protein-protein inhibitor (PPI) interaction complex did not identify any immediate binding site on the surface of the PPI interfaces. Nevertheless, a small pore around one single beta-sheet in the center of the PPI interaction area could be exploited as a weak point that may perturb the interface equilibrium. Using UniProt, which contains information about a number of confirmed mutations, we determined the relative potencies of PPI binding residues, identifying those that would likely affect the integrity of the complex (**Fig. 2**). Residues K353 and Y41, which interact with D155 at the center of the PPI, are likely stabilizing its surface, potentially forming a useful “hot spot” for targeted druggability (**Fig. 2-3,S2**).

To check whether this is true and to understand how Ace2:S-protein cooperation functions, we performed two MD simulations, one with and one without the mutation of Y41A. This mutation causes strict inability to form the S-protein:Ace2 complex. Analysis of the trajectory of the wild-type protein, which possessed an intact complex, revealed the three most stable conformations of the “hot spot” region with expanded pores inside the triangle of residues K353, D155, Y41. Since it is impossible to determine which of these three conformations is the most stable, we ran three high-throughput screenings based on the donor-acceptor atoms and hydrophobic areas of the region. We then performed three MD simulations with top pose ligands. As demonstrated in **Figure 3K**, ligands failed binding within 10 ns, while docked ligands became leaders, as determined by energetic stability, during MD and interaction energy values (electrostatic – red, Van der Waals - blue) (**Fig. 3J/L)**.

**B. Identification of predicted inhibitors to interrupt S-protein:Ace2 PPI via docking**

To identify inhibitors of the S-protein:Ace2 interaction via docking, we used the best scoring compounds obtained after combination of molecular docking and molecular dynamics simulations, which feeds into the pipeline for constraint-based screening. The high-throughput screening (HTS) of a PPI library did not produce any results, since the PPI binding sites were weakly identified shallow regions (**Fig. 2,4A-D,S1**). Compounds that made good insertion into the sites situated between Ace2 and S-protein were able to perturb the association of S-protein with Ace2 via steric hindrance of S-protein association (**Fig. 3**). From the MDS, we detected compounds that decreased energy of stability between the Ace2:S-protein complex, which is desired in an inhibitor of protein-protein interaction. As a whole, this approach identified a deep and narrow binding site to disturb the S-protein interaction with Ace2 (**Fig. 3,4A-D**).

**C. Tmprss2 and $M^{pro}$ modeling requires dynamics to reveal optimal inhibitor binding**

To optimize the binding site of our inhibitors, we constructed a full-length (zymogen) model of Tmprss2 (epitheliasinogen), as well as a mature version of the protease (epitheliasin), as described in our method section (**Fig. 4E-G**). The mature protease model was used for MDS studies to generate a reference dynamical profile that can be used to assist in silico screening of Tmprss2 inhibitors. A control experiment was also completed with the uncleaved (non-catalytic) form of Tmprss2 to demonstrate the pocket's instability and poor ligand binding capacity (**Fig. S3**) (Ko et al., 2015; Lucas et al., 2014; Wilson et al., 2005). A full-length model of monomeric $M^{pro}$ was also constructed, as well as a homodimer (**Fig. 4H-K,S1**). The structure derived from PDB code 6Y2F with its ligand was used for a consensus virtual screen (Zhang et al., 2020). In addition, we used the dimer to generate a reference dynamical profile to assist with in silico screening and study its interdomain behavior.

**D. Tmprss2 inhibitors identified**

We acquired the dimer protein sequence from the UniProt database. BLAST search showed the highest identification values against factor XI, prothrombin, kallikrein proteases (~41-42%). However, we focused on ligands that could be active against active form of Tmprss2 protein. Thus, we found the ligand: (2s)-1-[(2r)-2-(Benzylsulfonylamino)-5-Guanidino-Pentanoyl]-N-[(4-Carbamimidoylphenyl)methyl]pyrrolidine-2-Carboxamide, contained within the ChemblDB repository (CHEMBL1229259) and active against Tmprss2, prothrombin, and Factor XI. Likewise, another docked model was recovered with macrocyclic ligand (CHEMBL3699198), called: Ethyl14-[[(E)-3-[5-chloro-2-(tetrazol-1-yl)phenyl]prop-2-enoyl]amino]-5-(methoxycarbonylamino)-17-oxo-8,16 diazatricyclo[13.3.1.02,7]nonadeca-1(18),2(7),3,5,15(19)-pentaene-9-carboxylate. We launched several molecular dynamics simulations (up to 75 ns of duration) to understand the interaction with the target protein-binding site. **Figure S3** shows the initial and stable/final states of our various models (**Fig. 4E-G**). The MD analysis provided useful results for selecting the appropriate model. After 15 ns MD, the putative binding site collapsed (**Fig. S3,4E-G**). Although the active form of thrombin was used for Tmprss2 modeling, as a negative control we also examined the region with prothrombin-based binding site for completeness of the docking study (**Fig. S3**). The overlay of the average homology model structure from MD and structure 3F68 (PDB code) was used as a template to compare protein-ligand interaction map and assign docking constraints (Baum et al., 2009). Two optimal inhibitors for Tmprss2 were selected for demonstration purpose in **Figure 5**. We also modeled Cathepsins L and K for preliminary work, since these can be implicated in late-endosomal entry of the virus (**Fig. S4**).

**E. $M^{pro}$ inhibitors identified**

For the viral main proteinase, $M^{pro}$, a key enzyme for coronavirus replication (SARS-CoV-2), and a potential target for anti-SARS drug development, several peptidomimetics synthetized in early 2012 against SARS-CoV-1 proteases were identified as selective. There is a high degree of sequence identity between the SARS-CoV-1 and SARS-CoV-2 $M^{pro}$. This means that SARS-focused ligands could form similar interaction map with $M^{pro}$ protein and offers good launching points for 3D-QSAR/Machine Learning-drive based drug design for future iterations. To perform the virtual screening, protein structure was taken from the PDB code 7BQY complex and significant attention was paid to the interaction between the crystallized ligand from the complex and protein-binding site (Jin et al., 2020) (**Fig. 6**). As the binding site is quite large (**Fig. 6A**) we used a set of additional crystal structures (PDB code 6Y2F and fragment-like compounds from https://www.diamond.ac.uk/) to narrow the source of possible conformations. The binding of the compounds inserted into this region demonstrated a very canonic and recurring interacting motif, represented with α-Keto amide group flanked with aliphatic or saturated rings. We then performed molecular dynamics of 75 ns for the ligand-free dimer structure of the $M^{pro}$ to evaluate and "catch" the most flexible elements of the binding site. Our simulation revealed that the extended binding pocket was not very stable, unlike its individual sub-pocket, which contains active cysteine (C145) residue (**Fig. 6B,6C**). We began our molecular docking after assigning several combinations of constraints that should define specific interactions with the protein-binding site. We performed several high-throughput screening procedures using the same set of features in different combinations of constraints by partial matching algorithm (**Fig. 6D-E**). We then ranged docking scores and compared obtained conformations inside the binding site with the co-crystalized ligands from 7BQY, 6Y2F structures to select the most potent compounds.

## II. Analysis of Identified Compounds

By disrupting the SARS-CoV-2 viral process in three different critical routes: Binding, Entry, and Replication with our virtual screening approaches against dynamic structures, we were able to identify 350 compounds (**Dataset S1**) and compile data reflecting physiochemical and chemoinformatic properties. An exemplar top hit from each target is summarized for docking score in **Table 1**. To classify the compounds and their chemical space, we completed various regression, K-means analyses and fingerprint measurements, and provide further details about their structures and properties, including commonly evaluated traits: MW, HBA, HBD, docking score, Rule of Three (Jorgensen), Rule of Five (Lipinksi), $logP_{o/w}$, and logS (**Dataset S2**). We focused on new compound searches. The MW for these initial screening compounds ranges from large fragment (~250 Da) to mature drug sized molecules (~500 Da) with only 10 of the 310 top scoring compounds being over 500 Da in size and the smallest fragment-based compound measured 178 Da. Overall the docking scores were very good with median around -7 kcal/mol using the Glide XP calculations. We also generated a list of most commonly related drugs and discuss some of our best hits to known and clinical trial drugs (**Dataset S3**). The general process for pruning the >30 million total chemical fragments and compounds from commercially available compounds for the initial round of virtual screening is described (**Fig. 1B**), which reduces the primary large set to 3 million per conformation of target.

**Table 1. Top 40 FDA predicted compounds for Ace2:S protein, $M^{Pro}$, and Tmprss2.**

| Drug | Synonyms | Predicted Protein | In Silico Score | Target | CAS |
|---|---|---|---|---|---|
| Metaproterenol sulfate | Orciprenaline Sulfate | Ace2 | -8.05 | Others | 5874-97-5 |
| Isoprenaline HCl | Isuprel, Isadrine, Euspiran, Proternol, NSC 37745, NSC 89747 | Ace2 | -7.44 | Adrenergic Receptor | 51-30-9 |
| Epinephrine HCl | N/A | Ace2 | -7.12 | Adrenergic Receptor | 55-31-2 |
| Levosulpiride | N/A | Ace2 | -6.87 | Dopamine Receptor | 23672-07-3 |
| Metaraminol bitartrate | Metaradrine Bitartrate | Ace2 | -6.84 | Others | 33402-03-8 |
| Valganciclovir HCl | N/A | Ace2 | -6.58 | Antifection (Anti-infection) | 175865-59-5 |
| Isoprenaline HCl | Isuprel, Isadrine, Euspiran, Proternol, NSC 37745, NSC 89747 | Ace2 | -6.45 | Adrenergic Receptor | 51-30-9 |
| S4817 Atenolol | Tenormin, Normiten, Blokium | Ace2 | -6.35 | β1 receptor, β2 receptor | 29122-68-7 |
| S3783 Echinacoside | N/A | Ace2 | -6.09 | Others | 82854-37-3 |
| Propafenone | Rythmol SR, Rytmonorm | Ace2 | -6.04 | Sodium Channel | 34183-22-7 |
| Amikacin sulfate | BB-K8 | Ace2 | -5.98 | Antifection | 39831-55-5 |
| Pro-chlorperazine dimaleate salt | Prochlorperazin, Compazine, Capazine, Stemetil | Ace2 | -5.79 | Dopamine Receptor | 30718 |
| Isoetharine mesylate | N/A | Ace2 | -5.47 | Others | 7279-75-6 |
| Levosulpiride | N/A | Ace2 | -6.87 | Dopamine Receptor | 23672-07-3 |
| S5023 Nadolol | Corgard, Solgol, Anabet | Ace2 | -5.16 | Androgen Receptor | 42200-33-9 |
| Benserazide HCl | Ro-4-4602 | Ace2 | -5.93 | Dopamine Receptor | 14919-77-8 |
| S3694 Glucosamine (HCl) | 2-Amino-2-deoxy-glucose HCl | Ace2 | -5.57 | Others | 66-84-2 |
| S4701 2-Deoxy-D-glucose | 2-deoxyglucose, NSC 15193 | Ace2 | -5.18 | Others | 154-17-6 |
| Inulin | N/A | Ace2 | -5.18 | Others | 9005-80-5 |
| Cephalexin | Alcephin, Cefablan, Keflex, Cefadin, Tepaxin | Ace2 | -5.11 | Antifection | 15686-71-2 |
| S4722 (+)-Catechin | Cianidanol, Catechinic acid, Catechuic acid | $M^{Pro}$ | -6.73 | Others | 154-23-4 |
| S4723 (-) Epicatechin | L-Epicatechin, (-)-Epicatechol | $M^{Pro}$ | -6.32 | Others | 490-46-0 |
| S5105 Proanthocyanidins | condensed tannins | $M^{Pro}$ | -6.19 | Others | 20347-71-1 |
| Carbenicillin disodium | N/A | $M^{Pro}$ | -5.78 | Antifection | 4800-94-6 |
| AG-120 (Ivosidenib) | N/A | $M^{Pro}$ | -5.52 | Dehydrogenase | 1448347-49-6 |
| Atorvastatin calcium | N/A | $M^{Pro}$ | -5.39 | HMG-CoA Reductase | 134523-03-8 |
| Bezafibrate | N/A | $M^{Pro}$ | -4.93 | PPAR | 41859-67-0 |
| PF299804 | N/A | $M^{Pro}$ | -4.34 | EGFR | 1110813-31-4 |
| Bumetanide | Bumex | Tmprss2 | -6.5 | Others | 28395-03-1 |
| Aloin | Barbaloin | Tmprss2 | -6.45 | Tyrosinase | 1415-73-2 |
| Salbutamol sulfate | Ventolin, Asthalin, Asthavent | Tmprss2 | -6.1 | Adrenergic Receptor | 51022-70-9 |
| S4953 Usnic acid | Usniacin | Tmprss2 | -5.8 | Others | 125-46-2 |
| Avanafil | N/A | Tmprss2 | -5.62 | PDE | 330784-47-9 |
| S3612 Rosmarinic acid | Rosemary acid | Tmprss2 | -5.6 | IKK-β | 20283-92-5 |
| S5105 Proantho-cyanidins | Condensed tannins | Tmprss2 | -5.51 | Others | 20347-71-1 |
| Ractopamine HCl | N/A | Tmprss2 | -5.22 | Others | 90274-24-1 |
| Neohesperidin dihydrochalcone | Neohesperidin dhc | Tmprss2 | -5.2 | Others | 20702-77-6 |
| Cidofovir | Vistide | Tmprss2 | -5.18 | Others | 113852-37-2 |
| Zidovudine | azidothymidine | Tmprss2 | -5.02 | Reverse Transcriptase | 30516-87-1 |

As an example, when examining some prototype compounds from our selected dataset of >300 NCEs screened from >10 million total compounds, we find the predicted interactions between drug and protein (**Table S2**) have some common binding modalities. When looking at the dynamical data for the drugs binding to the protein-protein site on Ace2, we find the RMSD, RMSF, and H-bond occupancy evidence strong binding capability, as calculated from three separate simulations of Ace2 with different ligands, referred to as *300*, *392*, and *488* (**Fig. 4,S1**). These observations can be applied to generate constraints for additional virtual screening to improve the performance at higher throughput. Based on these results, ligand 392 reduced the overall RMSD and per residue RMSF, while maintaining strong hydrogen bonds, as demonstrated by its greater occupancy during the simulation (**Table S1**). This information, particularly H-bond occupancy and modulation of interface residue RMSFs, can be used in conjunction with docking and other data to profile the compounds more thoroughly (**Fig. 4**). In some cases, where constraints were utilized, the docking score underrepresents the compound and testing is needed to get important single-point data to clarify actives from non-actives, as well as determine the real IC50s for the selected active compounds. We will enrich our dataset with the top compounds for future rounds of parallel chemical screening and eventual de novo chemical design for novel chemical entities. Current results of our approach are presented on all three targets (Ace2, Tmprss2, $M^{Pro}$).

**III. Screening FDA-approved drugs for repurposing to minimize delay towards clinical benefit**
For each of our targets, we screened for hits from a library of FDA-approved compounds alongside the more extensive library of NCEs. Our final result across all three targets identified a total of 350 specific compounds, with 167 against Ace2, 40 against Tmprss2, and 103 against $M^{pro}$. Among these are FDA-approved drugs that could be repurposed: 21 against Ace2, 11 against Tmprss2, and 8 against $M^{pro}$ (Supplemental Dataset **TableS1_topNCE-FDA-hits.xlsx**).

**A. Ace2 Repurposing Drugs (FDA set)**
Isoprenaline hydrochloride (isoprotenerol) is an adrenoreceptor agonist that can be repurposed as a vasopressor to augment cardiovascular function with a beta-receptor side benefit of bronchodilation to improve breathing function. Metaraminol bitartrate, a stereoisomer of meta-hydroxynorephedrine, is a potent sympathomimetic amine to raise blood pressure. Atenolol and nadolol are beta-receptor blocking agents used in chronic hypertension, a comorbid risk factor in COVID-19 patients. Propafenone is an anti-arrhythmic agent approved for patients with life-threatening ventricular tachycardia. Levosulpiride is an atypical antipsychotic medication with prokinetic function that can be used in patients with agitated delirium, and gut immotility. Valganciclovir hydrochloride is an antiviral agent used for cytomegalovirus (CMV), varicella zoster virus (VZV), and preventative medication in HIV patients (Wu et al., 2020b). Recent data shows COVID-19 deplete CD8 T helper cells similar to HIV (Zheng et al., 2020). Amikacin sulfate and cephalexin are antibiotic anti-bacterial drugs that can treat bacterial super-infection. Prochlorperazine dimaleate is a phenothiazine derivative prescribed in medicine for nausea. Isoetharine mesylate is a selective adrenergic beta-2 agonist and fast-acting aerosolized bronchodilator for COVID-19 respiratory distress. Benserazide hydrochloride is an aromatic L-amino acid decarboxylase (DOPA decarboxylase inhibitor) used with levodopa for the treatment of Parkinsonism. Glucosamine hydrochloride is constituent found in cartilage and used for osteoarthritis joint pains. S4701 or 2-Deoxy-D- glucose (2D-DG) compound can induce ketogenic state, a powerful pathway involved in reducing systemic inflammation. Inulin is a natural prebiotic agent that enhances GI function and digestion by increasing prebiotic GI homeostasis critical to stabilize downstream anti-inflammatory effects and prevent overgrowth of harmful bacteria. Metaproterenol is a bronchodilator (beta-2 receptor agonist) that is commonly used to treat a variety of respiratory disorders including asthma, COPD, bronchitis and wheezing associated with viral pneumonias in clinical practice. The novelty of this drug is that is aerosolized and can be given as a breathing treatment and similar reach the lungs, which have a tremendous surface area and enter the blood rapidly. By inhalation this drug acts rapidly and potentially with or in combination with other aerosolized drugs or oral or IV combination drugs. Its inhalational route of delivery also can reach alveolar type II cells which express Ace2 for dual synergism. Metaraminol bitartrate, a stereoisomer of meta-hydroxynorephedrine, is a potent sympathomimetic amine. This drug is used in patients with hypotension or low blood pressure. COVID-19 hospitalized patients in the intensive care unit (ICU) setting often need vasopressor agents to raise blood pressure in a condition called shock (dangerously low blood pressure) from COVID-19 disease or sepsis. Therefore, metaraminol has dual purpose of antiviral function at Ace2 docking site /entry as well as helping with systemic blood pressure in those acutely ill COVID-19 patients. This drug has immediate repurposing use in this patient population.

**B. $M^{pro}$ Repurposing Drugs (FDA set)**
Atorvastatin is a statin drug with anti-inflammatory, immunomodulatory (Diamantis et al., 2017) and endothelial benefits (Ackermann et al., 2020; Varga et al., 2020). Carbenicillin disodium is a penicillin derivative antibacterial antimicrobial agent. Catechins are derived from plants with many beneficial properties in human health including anticancer, anti-obesity, antidiabetic, anti-cardiovascular, anti-infectious, hepatoprotective, and neuroprotective effects (Isemura, 2019). These substances fall outside FDA purview since supplements and generally have a wide safety margin that will be tested on the multidrug platform. Epicatechine S5105 is a naturally occurring flavonoid found in chocolate with anti-sarcopenic effects on skeletal muscle (Gutierrez-Salmean et al., 2014). Ivosidenib is an experimental drug for treatment of several forms of cancer. Bezafibrate is a fibrate lipid-lowering drug, which creates a favorable anti-inflammatory ratio against cardiovascular diseases. PF299804 or dacomitinib is an EGFR inhibitor used in cancer therapeutics. Metaproterenol is a bronchodilator (beta-2 receptor agonist) that is commonly used to treat a variety of respiratory disorders with viral pneumonias in clinical practice. Carbenicillin disodium is a penicillin derivative antibacterial antimicrobial agent that as mentioned above can be used in conjunction with other anti-SARS-Cov-2 agents to shut down antiviral effects and used in combination with those COVID-19 patients with secondary super-infection with bacterial infection of lung, blood, or skin.

**C. Tmprss2 Repurposing Drugs (FDA set)**
Bumetanide is a loop-diuretic used to remove extra fluid in the body (edema) such as pulmonary edema. Aloin is an anthraquinone glycoside found naturally in aloe vera plants, a natural cathartic, and decreases 16s rRNA sequencing of dysbiosis-producing butyrate producing bacterial species via an emodin breakdown product (Gokulan et al., 2019). Emodin blocks Ace2 and viral docking (Ho et al., 2007). Salbutamol sulfate (albuterol) is a bronchodilator used in various breathing disorders. S4953 usnic acid is a naturally occurring dibenzofuran derivative found in lichen plant species, in some kombucha teas, with adrenergic function to raise blood pressure and potential bronchodilator. Usnic acid is an active ingredient in some and a preservative in others and has a wide array of antimicrobial action against human and plant pathogens with antiviral, antiprotozoal, antiproliferative, anti-inflammatory, and analgesic activity (Ingolfsdottir, 2002). Avanafil is a class of medications called phosphodiesterase (PDE) inhibitors, which are pulmonary artery and circulation dilators. S3612 Rosmarinic acid is a naturally occurring compound found in plants (rosemary and sage), which has broad range of antimicrobial activity including antiviral activity including HIV (Shekarchi et al., 2012). Ractopamine is a beta-agonist function used for bronchodilatation. Neohesperidin dihydrochalcone (NHDC) is a naturally derived plant sweetener (bitter orange) with anti-Tmprss2 effects. Cidofovir and zidovudine (ZDV) are both antiviral drugs used in HIV patients.

## DISCUSSION

### Clinical Unmet Need for COVID-19 Acute Therapeutics
There is a critical unmet patient need for therapeutics to treat the acute phase of COVID-19 disease now and for the future. Efforts to create and trial a vaccine are underway, but 11.6 million patients are confirmed infected globally (>540K deaths) with 25% infected within the United States and we are just at the midpoint of 2020. Therefore, there is an urgent need to rapidly speed drug discovery from the bench to the bedside. In order to accelerate drug discovery, translation and human application, a design funnel using high-powered artificial intelligence is needed to screen millions of compounds against macromolecular mechanistic targets against the virus. At the back end of this funnel 40 drug candidates emerged, many of which may represent repurposing candidates for use in humans due to known safety and tolerability profiles. However, the approach with the highest probability of overall clinical therapeutic success may be not a single drug therapy for this viral RNA disease but rather a multi-pronged drug approach gleaned from decades of HIV-AIDS epidemic research. A multidrug approach for HIV has improved survival, markedly reduced viral loads, and vastly improved management of the disease by preventing AIDS end-stage fatal complications. We therefore suggest that a multifaceted drug approach for SARS-Cov-2 may prove superior by attacking 3 viral entry and replication cycle sites simultaneously: **Ace2** receptor docking site and entry, **Tmprss2** endosomal packaging, and **M$^{Pro}$** viral replication. Multiple drug targets for each of the 3 sites also allow permutations and optimization for combinatorial success.

### Comparison of FDA compounds identified from other recent screening

A recent study that screened commercially available >10,000 clinical-staged and FDA-approved small molecules against SARS-CoV-2 in a cell-based assay (Riva et al., 2020) identified interesting compounds for alternative targets that complement our results. These FDA approved compounds included MDL-28170, a selective Cathepsin B inhibitor; VBY-825, a non-specific Cathepsin B, L, S, V inhibitor; Apilimod, an inhibitor of production of the interleukins IL-12 and IL-23; Z-LVG-CHN2, a tri-peptide derivative inhibitor for cysteine proteinases; ONO 5334, a selective Cathepsin K inhibitor; and SL-11128, a polyamine analogs designed against *E. cuniculi*, a antimicrobial agents used as an adjuvant treatment for opportunistic AIDS-associated infections. Overall these compounds are Cathepsin-centric or antibiotic in nature, with little to no effect on our intended targets (Tmprss2, Ace2, M$^{Pro}$). Additional top hits identified by Riva et al. include: **AMG-2674**, an AMGEN compound inhibitor of TRPV-1 (Vanilloid Receptor); **SB-616234-A** that possesses high affinity for human 5-HT1B receptors; **SDZ 62-434** that strongly inhibited various inflammatory responses induced by lipopolysaccharide (LPS) or function-activating antibody to CD29; **Hafangchin A** (also called "**Tetrandrine**"), a bis-benzylisoquinoline alkaloid, which acts as a calcium channel blocker; Elopiprazole an antipsychotic drug of the phenylpiperazine class (antagonist at dopamine D2 and D3 receptors and an agonist at serotonin1A receptors) that was never marketed; **YH-1238,** which inhibits dipeptidyl peptidase IV (DPP-IV) enzyme prolonging the action of the incretin hormones, glucagon-like peptide-1 (GLP-1) and glucose-dependent insulinotropic polypeptide (GIP); **KW-8232**, an anti-osteoporotic agent that can reduce the biosynthesis of PGE2; **Astemizole**, an antihistamine; N-tert-butyl Isoquine (also called "**GSK369796**"), an antimalarial drug candidate; and **Remdesivir**, a broad-spectrum antiviral medication developed by the biopharmaceutical company Gilead Sciences. Again, none of these compounds were geared toward targeting Tmprss2 or Mpro, and are also not specific to Ace2. While the lack of overlap may be surprising, results generated by Riva and colleagues are not in opposition to our findings and both approaches can complement each other. Most importantly, these approved FDA compounds can be combined with our set of identified NCE (310 compounds) that have been demonstrated to have low toxicity issues based on our chemoinformatics filtering (**Fig. 1B**). All NCE compounds identified were chemical moieties that do not overlap any FDA drugs. Altogether, the data presented here complements previously generated data and should help prioritize and rapidly identify safe treatments for COVID-19. Future work will rely on advanced 3D-QSAR, fragment-based drug design principles for de novo drug optimization.

### Selective AI-SARS-Cov-2 Targeting and Drug Repurposing Data - Ace2, Tmprss2, M$^{pro}$

Among millions of potential COVID-19 drugs screened the majority of the final 40 drug candidates have known medical use and/or FDA approval for a primary indication (e.g., hypertension, cardiac indication, hyperlipidemia) with well-established patient safety and tolerability profiles from large phase III human trials and post- market (Phase IV) analyses. These large human data provide both a clinically *significant* and scientifically *innovative* window of opportunity to test 40 compounds on the multidrug platform, and, in conjunction, observe longitudinal human survival outcomes of COVID-19 patients on these drugs for comparative effectiveness within established and ongoing patient registries. An emerging example of this important parallel is Ace2 pathway drugs (Ace inhibitors [AceI] and angiotensin receptor blocking drugs [ARB]), which are increasingly observed in humans with COVID-19 to be associated with improved survival advantage (Jarcho et al., 2020; Mancia et al., 2020; Mehta et al., 2020; Patel and Verma, 2020; Vaduganathan et al., 2020). However, there is a scientific knowledge gap within human registries data regarding a scientifically robust and testable translational platform to test mechanistic effects of these different molecular compounds. Therefore, creation of a “pandemic platform” using newer technology of compound AI drug throughput screening combined with animal multi-drug screening models creates an early Phase I/II safety, tolerability and early efficacy platform which is rapidly needed to expedite bedside human use for COVID-19 pandemic, and as a platform that can be used in future pandemics.

**NCE set of compounds**

A flurry of activity to identify compounds for SARS CoV-2 targets has been underway by academic labs globally. Here in our approach we introduce our novel Maxwell's demon molecular dynamics method for screening flexibility required to get rare and essential conformational transitions and pathways to find the most likely druggable state. We also used our quantum docking technique (QM-driven adaptive molecular dynamics scanning docking) (Caulfield, 2012) to identify compounds effective for targeting Ace2, Tmprss2 and $M^{pro}$. The compounds identified by our large-scale in silico platform can next be experimentally validated as binders for intended targets and for efficacy in models of the disease, evaluated for EC50/safety-toxicity data, and carried into hit-to-lead and lead optimization in a drug development pipeline. Structural studies such as X-ray crystallography will also be important to generate structural SAR data for these efforts.

In sum, our leading edge in silico methods incorporating structural dynamics have produced a set of 350 candidate compounds suitable for screening in biological disease models. Among these, 40 FDA-approved compounds are eligible for rapid clinical trial testing. Additionally, our results bring forward 310 NCEs predicted to possess potency and specificity for viral or human accessory target proteins to lower the viral load. Moreover, this resource offers the community a set of chemical tools to probe the behavior of these enzymes essential for SARS-CoV-2 progression, namely, *binding*, *entry* and *replication*. As SARS-CoV-2 is already endemic, the rapid identification of effective antivirals remains a paramount focus until we have an efficient vaccine to provide long-lasting protection.

**STAR*METHODS**

## I. General Modeling Methods

In general, COOT was used for building in missing residues and regularizing geometry (Emsley and Cowtan, 2004; Emsley et al., 2010). More details for the preparation of each model are given in the respective subsections. Since these structures were all used in downstream computational studies, a uniform structural preparation was implemented. The full-length structures are comprised of all residues and side chains. We added missing atoms in rotamers and de-clashed atoms, added missing residues for chain continuity, and removed extraneous molecules/atoms (e.g. artifacts of crystallography or alternative conformations of residues were removed (keeping the highest occupancy)), and the B-factors were set to isotropic. The PDBePISA server was used to data mine the interface between Ace2 and S-protein (Krissinel and Henrick, 2007). Surface interactions data is provided (Supplemental).
Calculations on molecular dynamics trajectories including RMSD, RMSF, and H-bonds were performed using VMD and internal tools thereof (RMSD trajectory tool and Tk Console). Prior to calculations, the backbone (CONC$\alpha$) atoms of each frame of the trajectories were aligned to the first frame as a reference, to remove the effect of random rotation/translation. After alignment, the per residue average of RMSF or RMSD per frame in Å across the entire MDS trajectory is given. For the Ace2-ligand simulations, the number of hydrogen bonds between the protein and ligand were recorded for each frame, and the occupancy of each specific H-bond is defined as the percentage of frames the bond is present. RMSD, RMSF, and H-bond data were plotted in 2D format in Excel. The RMSF was also appended to the beta column of the PDB and heat-mapped to the structure using a custom Tcl/Tk script and PyMOL. All molecular graphics were generated in PyMOL (Mooers, 2016).

## II. General Dynamics Conditions

Molecular Dynamics and Monte Carlo simulations were performed on the protein to allow local regional changes for full-length structure for all acids of each structure.

The X-ray refinement for Monte Carlo was built using YASARA SSP/PSSM Method (Altschul et al., 1997; Hooft et al., 1996a; Hooft et al., 1996b; King and Sternberg, 1996; Krieger et al., 2009; Qiu and Elber, 2006). The structure was relaxed to the YASARA/Amber force field using knowledge-based potentials within YASARA. The side chains and rotamers were adjusted with knowledge-based potentials, simulated annealing with explicit solvent, and small equilibration simulations using YASARA's refinement protocol (Laskowski RA, 1993). The entire full-length structure was modeled, filling in any gaps or unresolved portions from the X-ray.

Refinement of the finalized models was completed using either Schrodinger's LC-MOD Monte Carlo-based module or NAMD2 protocols. These refinements started with YASARA generated initial refinement of Tmprss2 (Altschul et al., 1997; Hooft et al., 1996a; Hooft et al., 1996b; Krieger et al., 2009). The superposition and subsequent refinement of each protein regions yields a complete model. The final structures were subjected to energy optimization with PR conjugate gradient with an R-dependent dielectric.

Atom consistency was checked for all amino acids of the full-length wild-type structure, verifying correctness of chain name, dihedrals, angles, torsions, non-bonds, electrostatics, atom typing, and parameters. Model was exported to the following formats: Maestro (MAE), YASARA (PDB). Model manipulation was done with Maestro (Macromodel, version 9.8, Schrodinger, LLC, New York, NY, 2010), or Visual Molecular Dynamics (VMD) (Humphrey et al., 1996).

MDS and MC searching were completed on each model for conformational sampling, using methods previously described in the literature (Caulfield and Devkota, 2012; Caulfield and Medina-Franco, 2011; Caulfield, 2011; Caulfield et al., 2011). Briefly, each protein system was minimized with relaxed restraints using either Steepest Descent or Conjugate Gradient PR, then allowed to undergo the MC search criteria, as shown in the literature (Caulfield and Devkota, 2012; Caulfield and Medina-Franco, 2011; Caulfield, 2011; Caulfield et al., 2011). The primary purpose of MC, in this scenario, is examining any conformational variability that may occur with each protein.

**III. Structural modeling Ace2/S-protein**

For Ace2/S-protein, PDB code 6VW1 was used to construct the model (Shang et al., 2020). While the structure was mostly complete, chain F (S-protein) was missing more residues, though it had residue Ala522. Chain E (S-protein) was only missing residue 522. Residue Ala522 was built into chain E using COOT and where the extraneous molecules (solvent/cryoprotectant) and chains were deleted to leave only the heterodimer Ace2/S-protein, which was processed to be used for computational studies, not to generate a de novo model or complete structure with missing atoms and sections.

All information about the protein was found on the corresponding Uniprot page. After identifying the hot spot residues using SiteMap or protein-protein interfaces, we used MD to find out how the Y41A mutation can affect of PPI inhibition. We performed MD for wild type and mutated protein. Residual mutation was also performed using PyMol's built-in tools. Gromacs 2018 and amber99 force field were used to conduct MD and further analysis of the results (Baugh et al., 2011; Dilip et al., 2016; Janson et al., 2017; Makarewicz and Kazmierkiewicz, 2013, 2016; Mooers, 2016). Visual inspection of every 10 frames allowed us to determine some tendency of structural deformation in a certain place on the protein surface. According to the literature data and our finding, we focused on the predicted binding site. Then, each trajectory was analyzed via the built-in clustering tool based on the RMSD distribution. Three the most stable conformations of the binding site were chosen for the docking studies. All received docking poses from each docking study were evaluated based on the docking scores, interaction diagrams and solvent exposure. To make some prediction regarding the binding method, we carried out another molecular dynamics simulation for the upper poses of each docking. After such a confirmation of our assumptions, we selected the most powerful and accurate compounds from the results of docking.

**IV. Structural modeling Tmprss2**

A homology model was constructed on the basis of prothrombin crystal structure in complex with the ligand analog (PDB code 3F68) (Baum et al., 2009). We modeled the 492 amino acid Tmprss2 protein two different ways: YASARA based and SwissModel server based (Krieger et al., 2002; Waterhouse et al., 2018; Zoete et al., 2011). First, the YASARA based model begins with the FASTA sequence:
MALNSGSPPAIGPYYENHGYQPENPYPAQPTVVPTVYEVHPAQYYPSPVPQYAPRVLTQASNPVVCT
QPKSPSGTVCTSKTKKALCITLTLGTFLVGAALAAGLLWKFMGSKCSNSGIECDSSGTCINPSNWCDG
VSHCPGGEDENRCVRLYGPNFILQVYSSQRKSWHPVCQDDWNENYGRAACRDMGYKNNFYSSQGI
VDDSGSTSFMKLNTSAGNVDIYKKLYHSDACSSKAVVSLRCIACGVNLNSSRQSRIVGGESALPGAWP
WQVSLHVQNVHVCGGSIITPEWIVTAAHCVEKPLNNPWHWTAFAGILRQSFMFYGAGYQVEKVISHPN
YDSKTKNNDIALMKLQKPLTFNDLVKPVCLPNPGMMLQPEQLCWISGWGATEEKGKTSEVLNAAKVLL
IETQRCNSRYVYDNLITPAMICAGFLQGNVDSCQGDSGGPLVTSKNNIWWLIGDTSWGSGCAKAYRP
GVYGNVMVFTDWIYRQMRADG. Topological domains have the following characteristics: residues 1 – 84 forms the cytoplasmic sequence; residues 85 – 105 form the transmembrane domain region (helical 21 aa); and residues 106 – 492 form the Signal-anchor for type II membrane protein (extracellular), where the protein as two main chains: non-catalytic chain (Met1-Arg225) and catalytic chain (Ile256-Gly492), where each domain modeled as a separate unit built together in composite. Disulfide bonds exist between several residues (113 ↔ 126), (120 ↔ 139), (133 ↔ 148), (172 ↔ 231), (185 ↔ 241), (244 ↔ 365), (281 ↔ 297), (410 ↔ 426), (437 ↔ 465), which can be informative for building the structure. Glycosylation sites are also possible at residues N213 and N249. Cleavage site (active) exists between Arg255 and Ile256 (see refinement section).

The second method, homological modeling was performed using the SwissModel server after performing a BLAST search on available protein structures in the RCSB database. Molecular dynamics simulations of 100 ns of both, suggested and re-modeled protein structures, was performed with GROMACS 2018 (Makarewicz and Kazmierkiewicz, 2013, 2016). Based on the structural analysis and the generated Connolly surfaces, we identified critical changes in the binding site of the proposed model and began creating a mesh for the binding site of the new homology model. Since our model was based on the structure of thrombin, we used its co-crystallized ligand as a template for assigning constraints and ensured we built the catalytically active state.

## V. Structural modeling M$^{pro}$

For M$^{pro}$ (PDB 6Y2F) co-crystallization with tert-butyl (1-((S)-1-(((S)-4-(benzylamino)-3,4-dioxo-1-((S)-2-oxopyrrolidin-3-yl)butan-2-yl)amino)-3-cyclopropyl-1-oxopropan-2-yl)-2-oxo-1,2-dihydropyridin-3-yl)carbamate (also referred to as alpha-ketoamide 13b) was used; the structure was also mostly complete. Residues E47 and D48 were built in using COOT, where the other preparations previously described were also performed. To build the missing residues, the coordinates and structure factors were downloaded, generated 2mFo-DFc and FEM maps, and real space refine zone/regularize zone were used to fit to electron density and optimize local geometry. The ligand (alpha-ketoamide 13b) was left for usage as a cognate ligand for virtual screening.

The protein structure was initially studied using MD to find out if the binding site is cruel enough or can break down without a ligand molecule during the simulation. Simulation of the dimeric complex for 100 ns was sufficient to compare conformational changes from different MD states. A set of positional and hydrogen bonds were assigned based on the available peptidomimetic structure. Thus, two screenings were conducted with an emphasis on positional constraints or interactions of hydrogen bonds.

## VI. Structure-refinement of Ace2 (S-protein:Ace2), Tmprss2, and M$^{pro}$ Models

Using MDS and MC refinement with Schrodinger and/or YASARA SSP/PSSM methods (Altschul et al., 1997; Hooft et al., 1996a; Hooft et al., 1996b; King and Sternberg, 1996; Krieger et al., 2009; Qiu and Elber, 2006), each structure was relaxed to the YASARA/Amber force field using knowledge-based potentials within YASARA. The side chains and rotamers were adjusted with knowledge-based potentials, simulated annealing with explicit solvent, and small equilibration simulations using YASARA's refinement protocol (Laskowski RA, 1993). The entire full-length structure was modeled, filling in any gaps or unresolved portions from the X-ray structure.

Refinement of the finalized models was completed using either Schrodinger's Monte Carlo-based module or in-house protocols. These refinements started with generated initial refinement for each independent structure (Altschul et al., 1997; Hooft et al., 1996a; Hooft et al., 1996b; Krieger et al., 2009). The superposition and subsequent refinement of the overlapping regions yields a complete model for all four proteins. The final structures were subjected to energy optimization with PR conjugate gradient with an R-dependent dielectric.

Atom consistency was checked for all amino acids (and atoms) of the full-length wild-type model, verifying correctness of chain name, dihedrals, angles, torsions, non-bonds, electrostatics, atom typing, and parameters. A multimeric-complex model is predicted, including cofactors and ions. All of the models were exported in the following formats Maestro (MAE), YASARA (PDB). Model manipulation was done with Maestro (Macromodel, version 9.8, Schrodinger, LLC, New York, NY, 2010), or Visual Molecular Dynamics (VMD) (Humphrey et al., 1996). Analyses were emphasized on the protein-protein interaction regions containing.

Monte Carlo dynamics searching (MC-search) was completed on each model for additional conformational sampling, using methods previously described in the literature (Caulfield and Devkota, 2012; Caulfield, 2011; Caulfield et al., 2011). Briefly, each protein system was minimized with relaxed restraints using either Steepest Descent or Conjugate Gradient PR, then allowed to undergo the MC search criteria, as shown in the literature (Caulfield and Devkota, 2012; Caulfield, 2011; Caulfield et al., 2011). The primary purpose of MC, in this scenario, is examining any conformational variability that may occur with different orientations in the region near to protein-protein interfaces.

## VII. MD Simulation Protocol

The total atomic force field was used to minimize the energy of the system, namely, the descent algorithm for 20,000 steps with an iteration interval of 2 fs. The equilibrium of the solvent was carried out using positional restrictions imposed on the atoms of protein structures, while the solvent molecules remained mobile for all 100 ps. Each system was placed in a box in which the layer of the TIP3P water molecule was 10 Å. The final systems were neutralized by the addition of Na + and Cl– ions to a concentration of 150 mM. All simulations were performed under periodic boundary conditions using the V-Rescale Thermostat algorithm to maintain temperature (310 K) and the Parrinello-Rahman Barostat algorithm for constant pressure (1 bar) (Bussi et al., 2007; Parrinello and Rahman, 1981). Long-range unrelated interactions were calculated using the Particle-Mesh-Ewald (PME) method (Abraham and Gready, 2011). All molecules were relaxed with a molecular dynamics simulation of 100 ns. Ligand topologies were created using the antechamber module from the AmberTools18 package (Case et al., 2005).

## VIII. DOCKING METHODS

### A. Site Mapping on Proteins

We used SiteMapper (Bhachoo and Beuming, 2017) to identify possible binding sites for docking affinity with the proteins Ace2 (allosteric site), Tmprss2, and $M^{pro}$. We also used our novel MDS biasing technique algorithm, Maxwell's demon MD, for searching within these sites for potential flexible zones that would have beneficial peptide interactions, which served as a reductive filter limiting the total number of possible sites screened on the proteins to those with adequately deep binding grooves (Caulfield, 2011; Kayode et al., 2016) or interesting insertion sites (Ace2).

### B. Glide Docking

Prior to the docking with the Ace2 (allosteric site), Tmprss2, and $M^{pro}$, we had completed rigorous molecular dynamics simulations (MDS) and Monte Carlo (MC) conformational searching for each model for additional conformational sampling, using methods previously described in the literature (Caulfield and Devkota, 2012; Caulfield, 2011; Caulfield et al., 2011). The primary purpose of MC, in this scenario, is examining any conformational variability that may occur with different orientations in the region near to protein-protein interfaces.

Over three million compounds were docked to each site using the Glide XP docking program (Bhachoo and Beuming, 2017). All compounds were accounted for using OPLS3 within Maestro program (Maestro-9.4, 2014). Using our published docking protocols on each identified site, we reductively scanned from 100s to the top 10 poses from each docking and then did cross-comparisons of docking scores to retain only the top binding pose of each compound from each site in a winner-takes-all strategy.

### C. Other Docking (positional constraints)

Each compound has been converted into a set of energy minimized three-dimensional shapes with the Ligprep module. Without protein preparation, it was used for the correct distribution of protonation and post-minimization in the OPLS3 force field. In the case of assigning restrictions based on ligands ($M^{pro}$, Tmprss2), we tried to cover the most important and strong interactions. In the case of Ace2, a set of constraints was generated in sufficient quantities to generate combinations of possible interactions. Positional constrains (1.8 A radius) and h-bond constraints were generated in the Schrodinger Glide module, namely in the mesh generation tool. Aromatic and hydrophobic features were represented with short SMARTS. A partial matching protocol for applying constraints has also been used to improve process accuracy. A high throughput screening protocol with regulated ligand flexibility was applied.

### D. Docking Parameters

Each compound has been converted into a set of energy minimized three-dimensional shapes with the Ligprep module. Without protein preparation, it was used for the correct distribution of protonation and post-minimization in the OPLS3 force field. In the case of assigning restrictions based on ligands ($M^{pro}$, Tmprss2), we tried to cover the most important and strong interactions. In the case of Ace2, a set of constraints was generated in sufficient quantities to generate combinations of possible interactions. Positional constrains (1.8 A radius) and h-bond constraints were generated in the Schrodinger Glide module, namely in the mesh generation tool. Aromatic and hydrophobic features were represented with short SMARTS. A partial matching protocol for applying constraints has also been used to improve process accuracy. A high throughput screening protocol with regulated ligand flexibility was applied.

Conformations of compound orientations were generated using our standard protocols (Bhachoo and Beuming, 2017; Kalid et al., 2012; Unger et al., 2015). The starting conformation of relaxed protein structures was first obtained by the method of Polak-Ribière conjugate gradient (PRCG) energy minimization with the Optimized Potentials for Liquid Simulations (OPLS) 2005 force field (Jorgensen, 2004; Jorgensen and Tiradorives, 1988) for 5000 steps, or until the energy difference between subsequent structures was less than 0.001 kJ/mol-Å units. Our docking methodology has been described previously (Caulfield and Devkota, 2012; Friesner et al., 2006; Loving et al., 2009; Vivoli et al., 2012).

Briefly, compounds were docked within the Schrödinger software suite (Mohamadi et al., 1990) using a virtual screening workflow (VSW) (Bhachoo and Beuming, 2017; Friesner et al., 2006; Jacobson et al., 2002; Kalid et al., 2012; Kozakov et al., 2006). Alternative docking methods were also employed, including in-house software techniques for top leads for SAR elucidation. The top seeded poses were ranked and unfavorable scoring poses were discarded. Top favorable scores from initial dockings yielded hundreds of poses with the top five poses retained. Molecular interactions of the ligand-protein interfaces were used to help determine the optimal binding set, which included descriptors were used to obtain atomic energy terms like hydrogen bond interaction, electrostatic interaction, hydrophobic enclosure and π-π stacking interaction that result during the docking run. Molecular modeling for importing and refining the proteins was completed (Maestro-9.4, 2014).

Examinations of structure stability were examined for all proteins investigated, S-protein:Ace2, Tmprss2, and $M^{pro}$, respectively (Caulfield and Devkota, 2012; Caulfield and Medina-Franco, 2011; Caulfield, 2011; Reumers et al., 2005; Schymkowitz et al., 2005; Zhang et al., 2013). Object stability was used to determine if any changes in structure that were deleterious to function from immediate inspection, which the FoldX algorithm can provide, prior to docking studies. Thus, we examined the local residues around the docking site and determined an electrostatic calculation may be useful to explain the change in function. The molecular model for the full structure and its truncated form are given (**Fig. S1**) using our state of the art methods, which have been established (Abdul-Hay et al., 2013; Ando et al., 2017; Caulfield and Devkota, 2012; Caulfield and Medina-Franco, 2011; Caulfield, 2011; Caulfield et al., 2011; Caulfield et al., 2014; Caulfield et al., 2015; Fiesel et al., 2015a; Fiesel et al., 2015b; Puschmann et al., 2017; Zhang et al., 2013).

Local residues within the 12Å cutoff near docking sites were analyzed (**Fig. S1-S2**). Any interactions requiring inducible fit, or Threonine/Serine hydroxyl rotation or other docking parameter (π-stacking/halogen-directionality) were also included. Mapping electrostatics was accomplished using the Poisson-Boltzmann calculation for solvation on all amino acids for each docked structure (Caulfield and Devkota, 2012; Caulfield and Medina-Franco, 2011; Caulfield, 2011; Reumers et al., 2005; Schymkowitz et al., 2005; Zhang et al., 2013)

**E. Libraries used**

Compounds were derived from either a set of all FDA approved and clinical tested compounds, bioactive set of compounds, or a large multi-million compound set from ZINC database. In the all cases the libraries were prepared using LigPrep described above. The ZINC database was pruned using parameters for better drug-like profile and removal of reactive functional groups and poor chemoinformatics properties delivering a large set suitable for screening on all targets across dynamic time points from MDS.

## SUPPLEMENTAL INFORMATION

Supplemental Information can be found online at: XXX

## ACKNOWLEDGEMENTS

## AUTHOR CONTRIBUTIONS

T.C. designed and conducted most experiments, analyzed data, and wrote the manuscript with inputs from J.M., K.L., M.C. and E.R.; T.C. and M.C. performed MDS, docking, and generated analyses; T.C. and M.C. performed post simulation analyses; C.B. and M.C. performed bioinformatics analysis; T.C. supervised M.C.; T.C. provided expertise on data analysis; E.R. provided expertise and insight interpreting experimental structures and homology models; and T.C. proposed the project to K.L., whom helped with formatting, detailing analysis and edited the manuscript.

## DECLARATIONS OF INTERESTS

The authors declare no competing interest.

## FIGURE LEGENDS

**Figure 1. Flowchart for drug pipeline for attacking COVID-19 via polypharma small molecule approach using in silico screening and advanced simulation biasing.** (A) Biological infection of SARS-CoV-2 from initial binding, entry and replication for virus proliferation. (B) Overview of COVID-19 Drug Discovery Pipeline.

**Figure 2. Protein-Protein Interaction (PPI) region on the surface of Ace2 identifies key residues.** (A) PPI region (yellow) on the surface of Ace2 is shown with important residues K353, D155, Y41, K31 highlighted in yellow. (B) Zoomed in detail panel shows beta sheet secondary structure and H-bond interactions targeted for disruption by docked small molecules.

**Figure 3. Ace2 protein docked with exemplar ligands during MD simulations and used as basis for large-scale constraint-based screening.** (A) protein and its final state of MD (B), which differs from Y41A mutant due to significant surface changes (C). (D-E) examination of the binding pockets change in shape as during MD simulations with the tested ligands bound with key interaction residues in red. (G-I) Surfaces removed and zoom into the ligands docked at the site (inserted versus slipping out). (J-L) Energy of the ligand lowers system (more stable versus slippage, where no effect observed).

**Figure 4. Modeling requires molecular dynamics to reflect optimal inhibitor binding sites.** (A-D) Ace2:S protein stabilization and effect of ligand binding at allosteric site. (A) Number of hydrogen bonds for each ligand with Ace2 against each frame of the simulation. Blue is ligand 300, orange is ligand 392, green is ligand 488. (B) RMSD of Ace2 across every frame in the simulation, bound to different ligands. (C) RMSF per residue of Ace2 in each MDS bound to different ligands. (D) RMSF heat-mapped onto Ace2 and ligand 300. A call-out box shows a close-up of ligand and binding site. Ligand and binding site residues represented as sticks with labels and interaction distances. The scale is a BWR gradient from 0 to 2.0 Å RMSF. (E-G) Tmprss2 dynamics reveal the catalytically active form suitable for inhibition. (E) RMSD in Å across the 25 ns MDS trajectory mapped as a 2D plot. (F) Per residue average RMSF in Å across the trajectory mapped as a 2D plot. Disulfide bonds and catalytic triad are represented as sticks. The scale is a BWR gradient from 0 to 2.0 Å RMSF. (G) Post-cleavage (mature protease) extracellular domain of Tmprss2. Call-out box shows close-up of canonical serine protease catalytic triad of mature Tmprss2, with distances of polar contacts. (H_K) Model refinement for Mpro reveals ligand binding sites suitable for docking. (H) Average RMSF per residue heat-mapped onto the $M^{pro}$ structure. The scale is a BWR gradient from 0 to 2.0 Å RMSF. (I) RMSD of Mpro for each frame of the simulation. (J) Average RMSF per residue of Mpro (each chain measured separately). (K) Mpro (orange) with small molecule inhibitor (cyan).

**Figure 5. Modelled catalytically active form of Tmprss2 bound to inhibitors.** (A) Homology model of TMPRSS2 based on crystal structure of thrombin (3F68) is shown docked with 1-(2-Fluoro-5-methylphenyl)-N-[2-(4-fluorophenyl)-2-hydroxypropyl]-4-hydroxy-1H-pyrazole-3-carboxamide (B). A proposed macrocycle-bound structure (C) and docked N-(2-4-[3-(2-Carbamoylphenyl)propanoyl]-1,1-dioxido-2-thiomorpholinyl}ethyl)-2-oxo-2,3-dihydro-1H-benzimidazole-4-carboxamide (D) as further exemplars for inhibition of Tmprss2.

**Figure 6. Druggability of $M^{pro}$ is demonstrated with detailed analysis of α-Keto amide group binding using MD simulations.** (A) The alignment of two $M^{pro}$ crystal structures (7BQY/cyan and $M^{pro}$-x0434/purple from diamond.ac.uk) bound to compounds containing an α-Keto amide group flanked by hydrophobic groups is shown. Sufficient structural stability of the binding site is demonstrated via comparative visualization of initial (B) and final (C) states of MD. Binding site retains its geometry and shape across the MD. (D) Two bound states of hit compounds from the large library of compounds give further exemplars: Z1609752806 (D) and Z1143050660 (E) in complex with $M^{pro}$ protein.

FIGURES

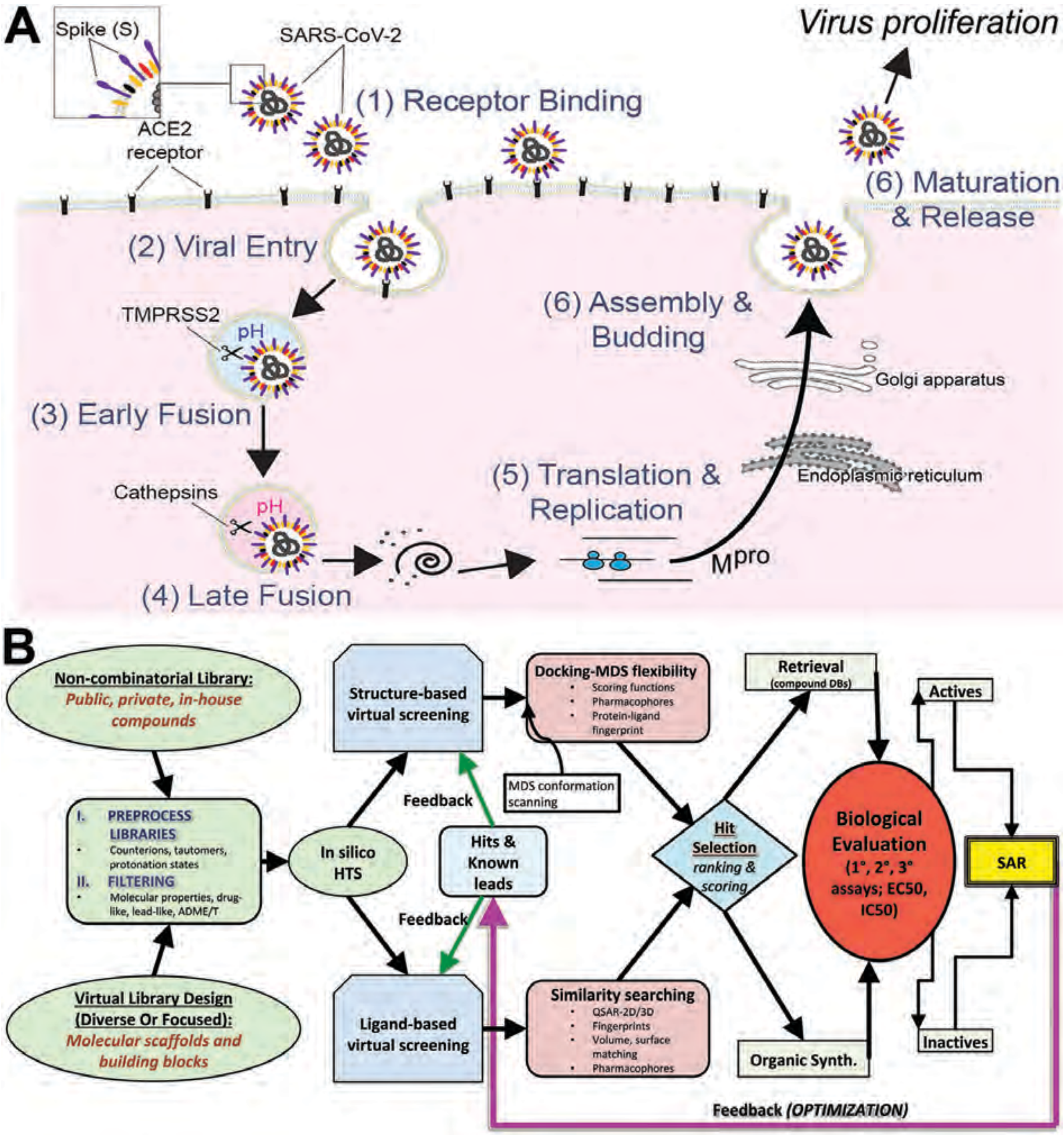

**Figure 1**

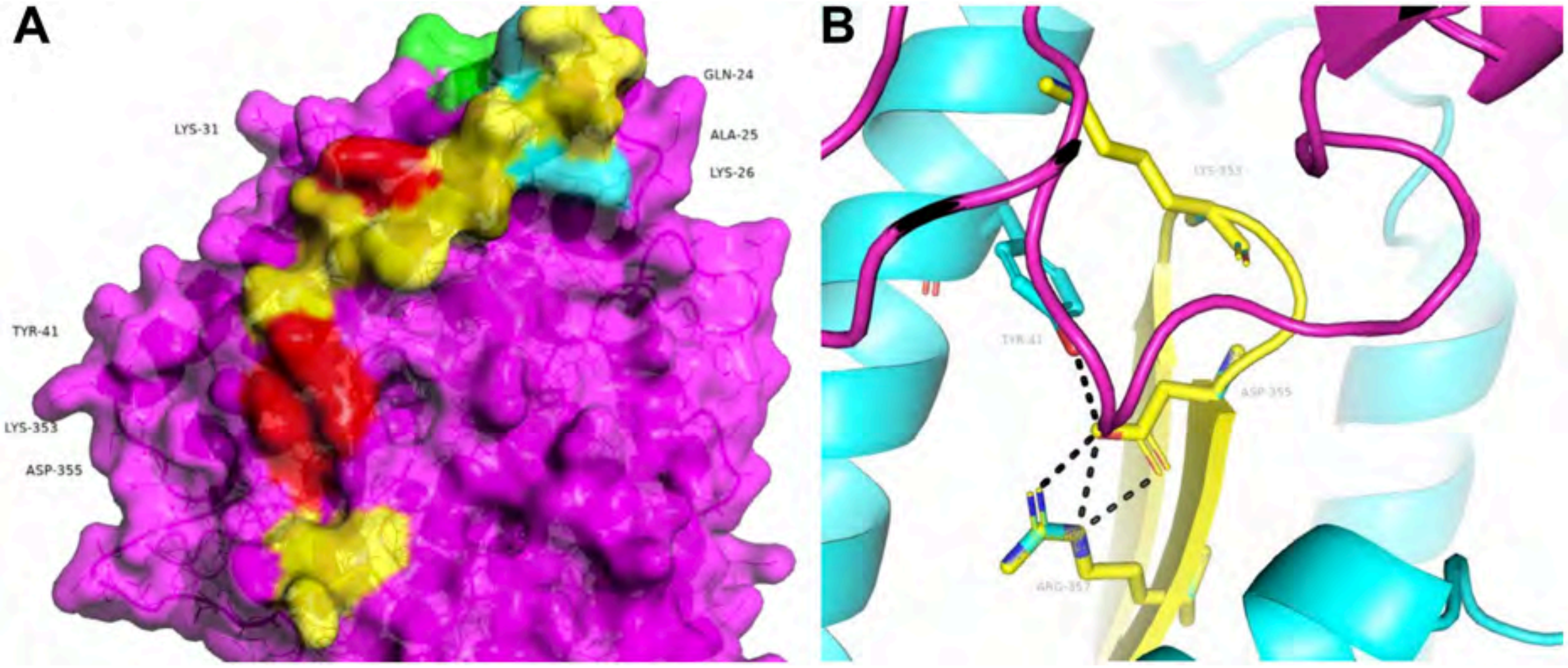

**Figure 2**

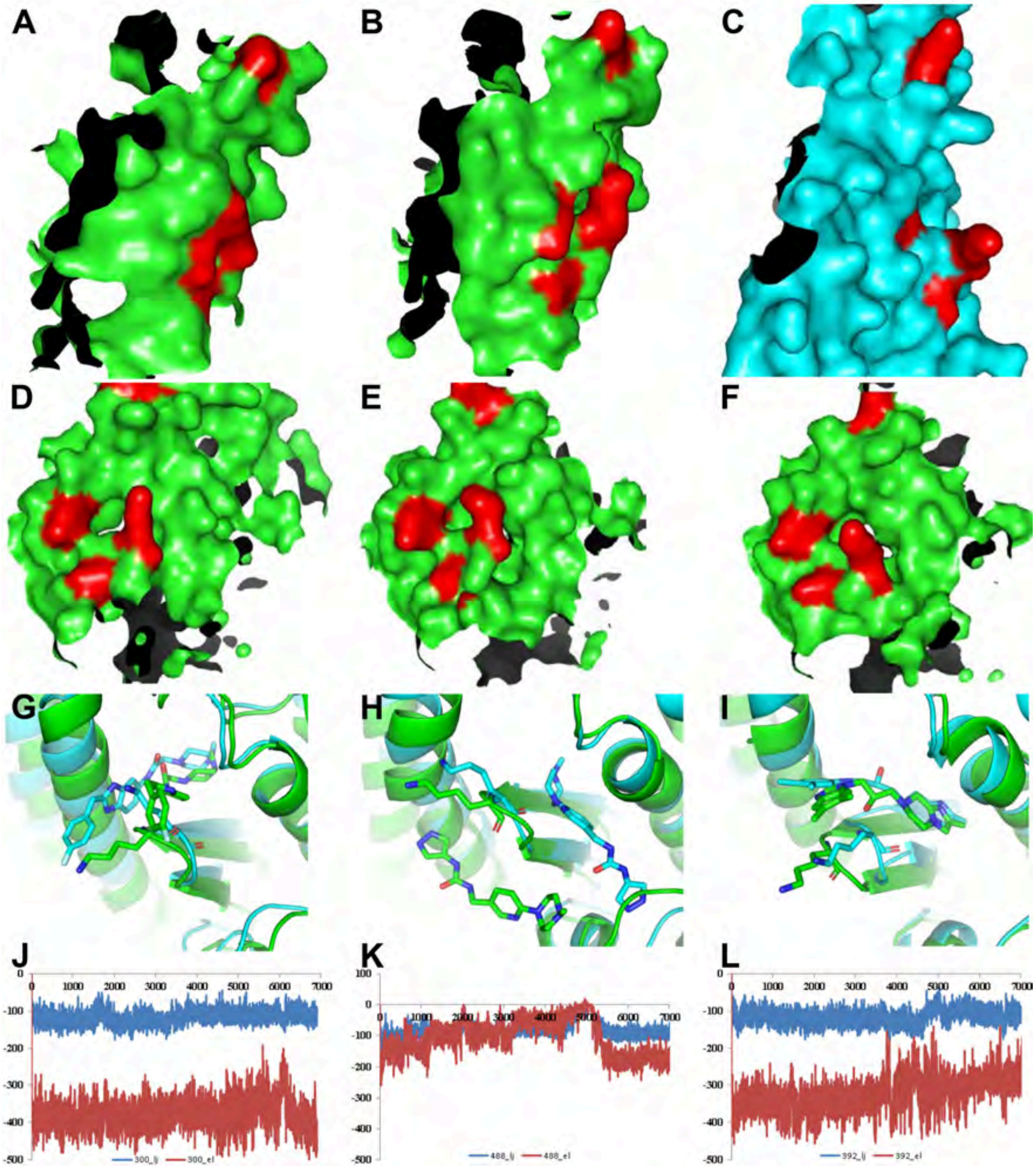

**Figure 3**

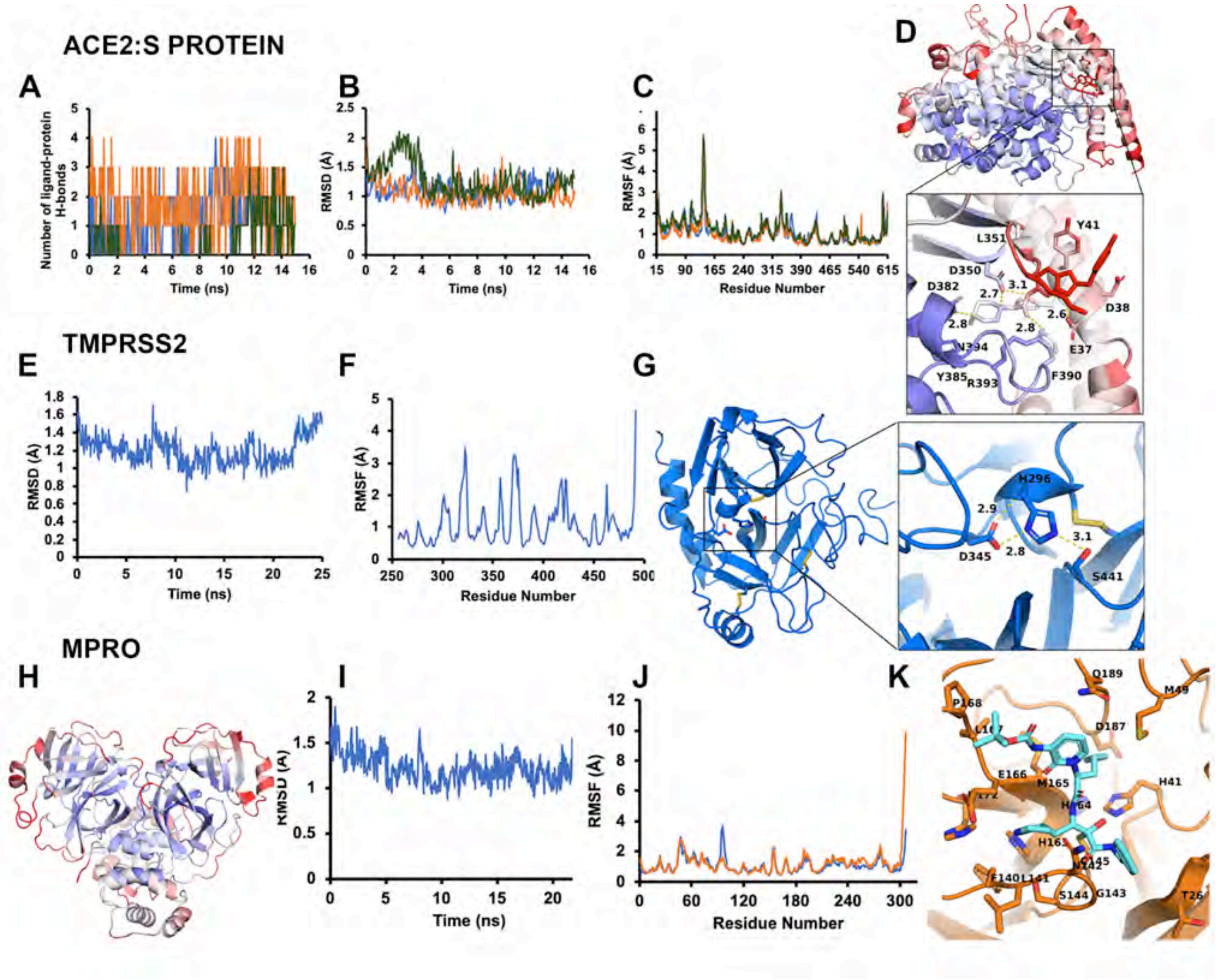

Figure 4

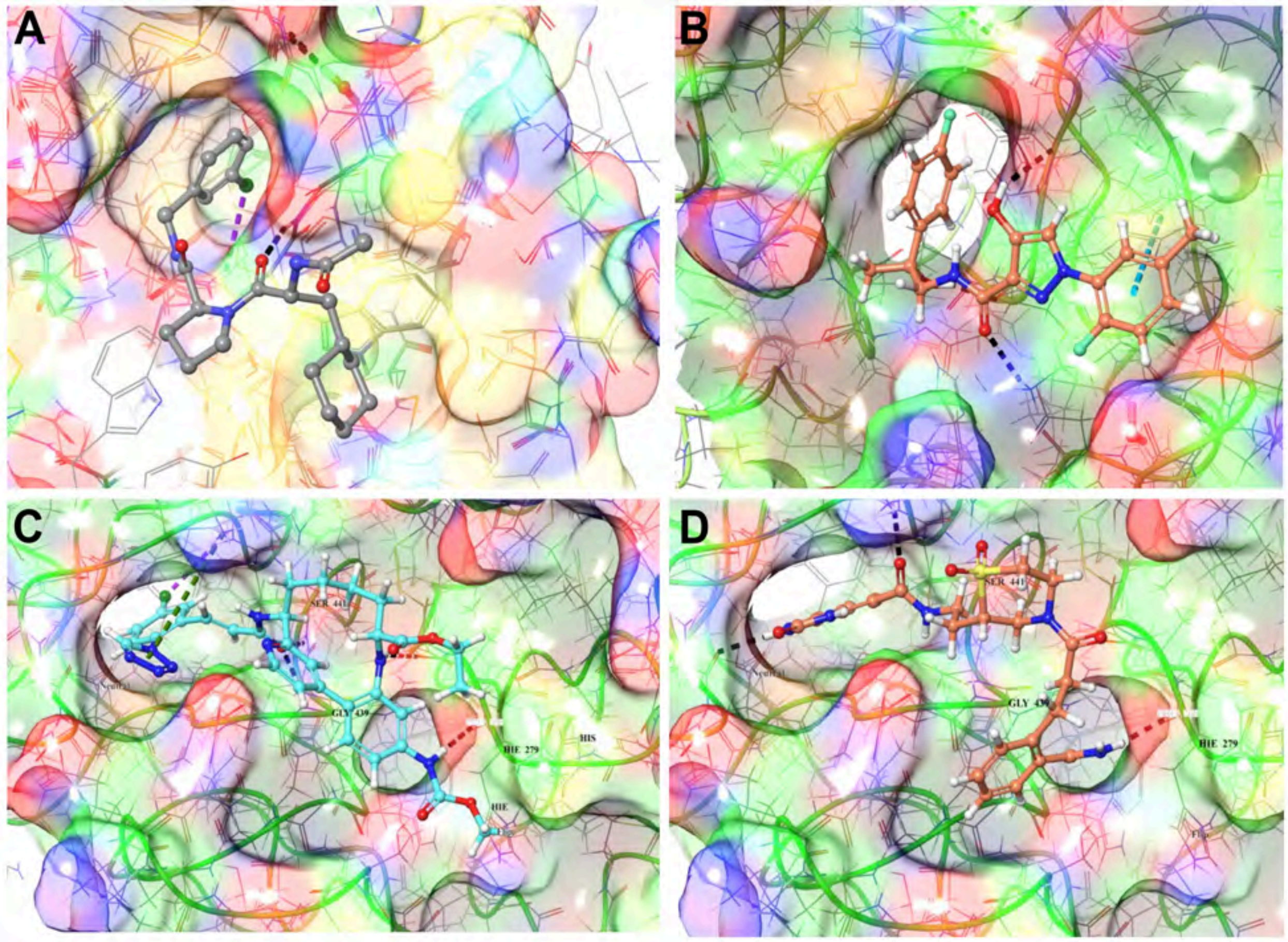

**Figure 5**

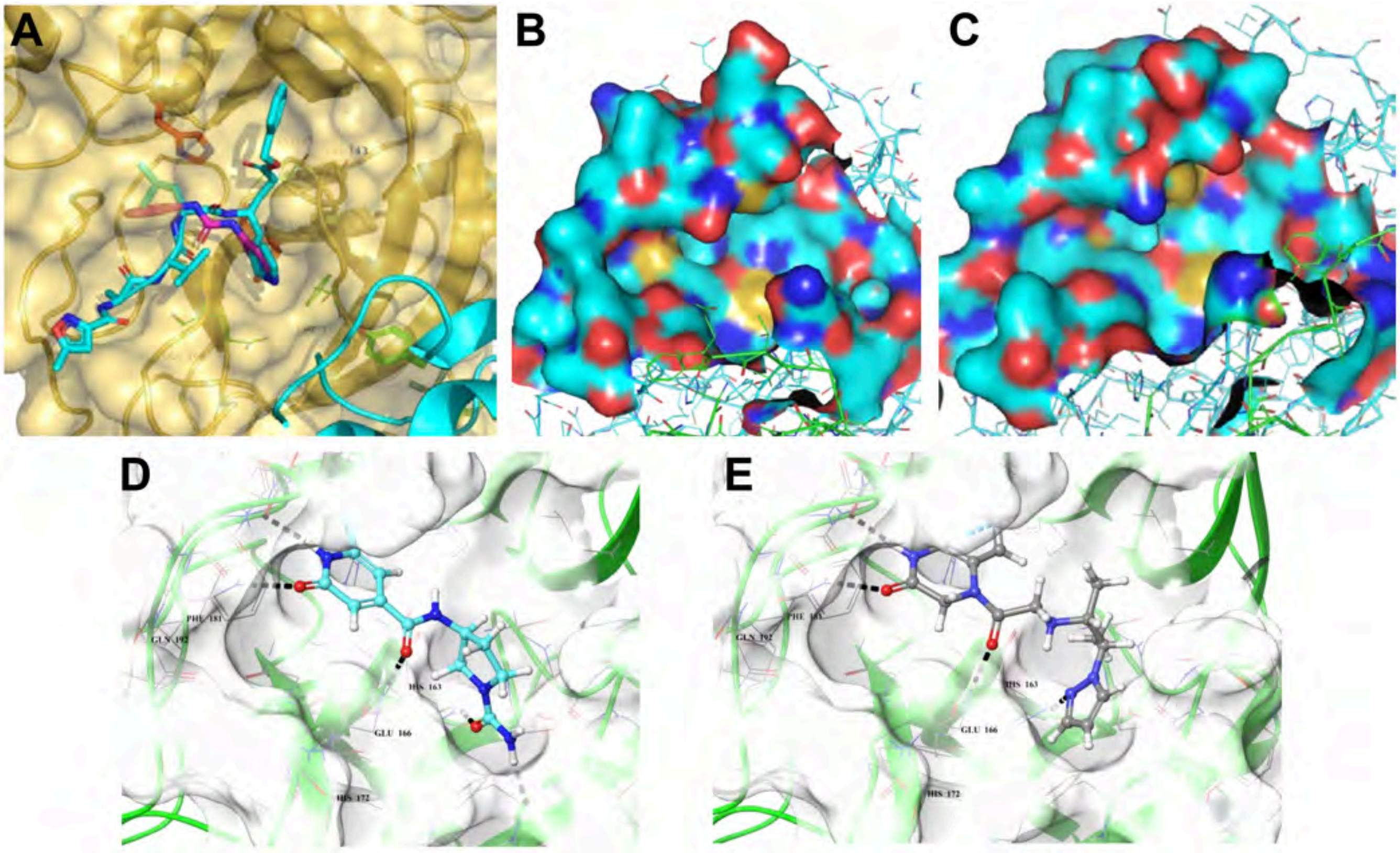

**Figure 6**

**SUPPLEMENTAL SECTION**

# Attacking COVID-19 Progression using Multi-Drug Therapy for Synergetic Target Engagement

Mathew Coban[1], Juliet Morrison PhD[2], William D. Freeman MD[3], Evette Radisky PhD[1], Karine G. Le Roch PhD[4], Thomas R. Caulfield, PhD[1,5-8]

**Affiliations**

[1] *Department of Cancer Biology, Mayo Clinic, 4500 San Pablo Road South, Jacksonville, FL, 32224 USA*

[2] *Department* of Microbiology and Plant Pathology*, University of California, Riverside, 900 University, Riverside, CA, 92521 USA*

[3] *Department of Neurology, Mayo Clinic, 4500 San Pablo South, Jacksonville, FL, 32224 USA*

[4] *Department of Molecular, Cell and Systems Biology, University of California, Riverside, 900 University, Riverside, CA, 92521 USA*

[5] *Department of Neuroscience, Mayo Clinic, 4500 San Pablo South, Jacksonville, FL, 32224 USA*

[6] *Department of Neurosurgery, Mayo Clinic, 4500 San Pablo South, Jacksonville, FL, 32224 USA*

[7] *Department of Health Science Research (BSI), Mayo Clinic, 4500 San Pablo South, Jacksonville, FL, 32224 USA*

[8] *Department of Clinical Genomics (Enterprise), Mayo Clinic, Rochester, MN 55905, USA*

**Correspondence to:**
Thomas R. Caulfield, PhD,
Dept of Neuroscience, Cancer Biology, Neurosurgery, Health Science Research, & Clinical Genomics
Mayo Clinic, 4500 San Pablo Road South
Jacksonville, FL 32224
Telephone: +1 904-953-6072,
E-mail: caulfield.thomas@mayo.edu

**Table S1. Hydrogen bond occupancy over 15 ns MDS trajectory for each ligand with ACE2.**

| Ligand | Ligand 300 w/ ACE H-bonds | | | Ligand 392 w/ ACE H-bonds | | | Ligand 488 w/ ACE H-bonds | | |
|---|---|---|---|---|---|---|---|---|---|
| Kind / % | donor | acceptor | occupancy | donor | acceptor | occupancy | donor | acceptor | occupancy |
| Type | drug | D350-Side | 31.41% | drug | D350-Side | 67.77% | drug | Q325-Main | 11.96% |
| Type | R393-Side | drug | 30.32% | D350-Side | drug | 42.52% | drug | D355-Side | 11.63% |
| Type | drug | D350-Side | 28.16% | drug | D350-Side | 33.55% | drug | Y41-Main | 9.97% |
| Type | drug | E37-Side | 24.91% | drug | D382-Side | 23.59% | drug | D350-Side | 6.64% |
| Type | drug | D382-Side | 19.13% | drug | G352-Main | 4.65% | T324-Side | drug | 1.00% |
| Type | drug | E37-Main | 2.53% | drug | D38-Side | 4.32% | drug | N322-Main | 0.66% |
| Type | | | | drug | E37-Main | 0.66% | drug | K353-Main | 0.66% |
| Type | | | | | | | drug | M383-Main | 0.33% |
| Type | | | | | | | Q325-Main | drug | 0.33% |
| Type | | | | | | | Q325-Side | drug | 0.33% |

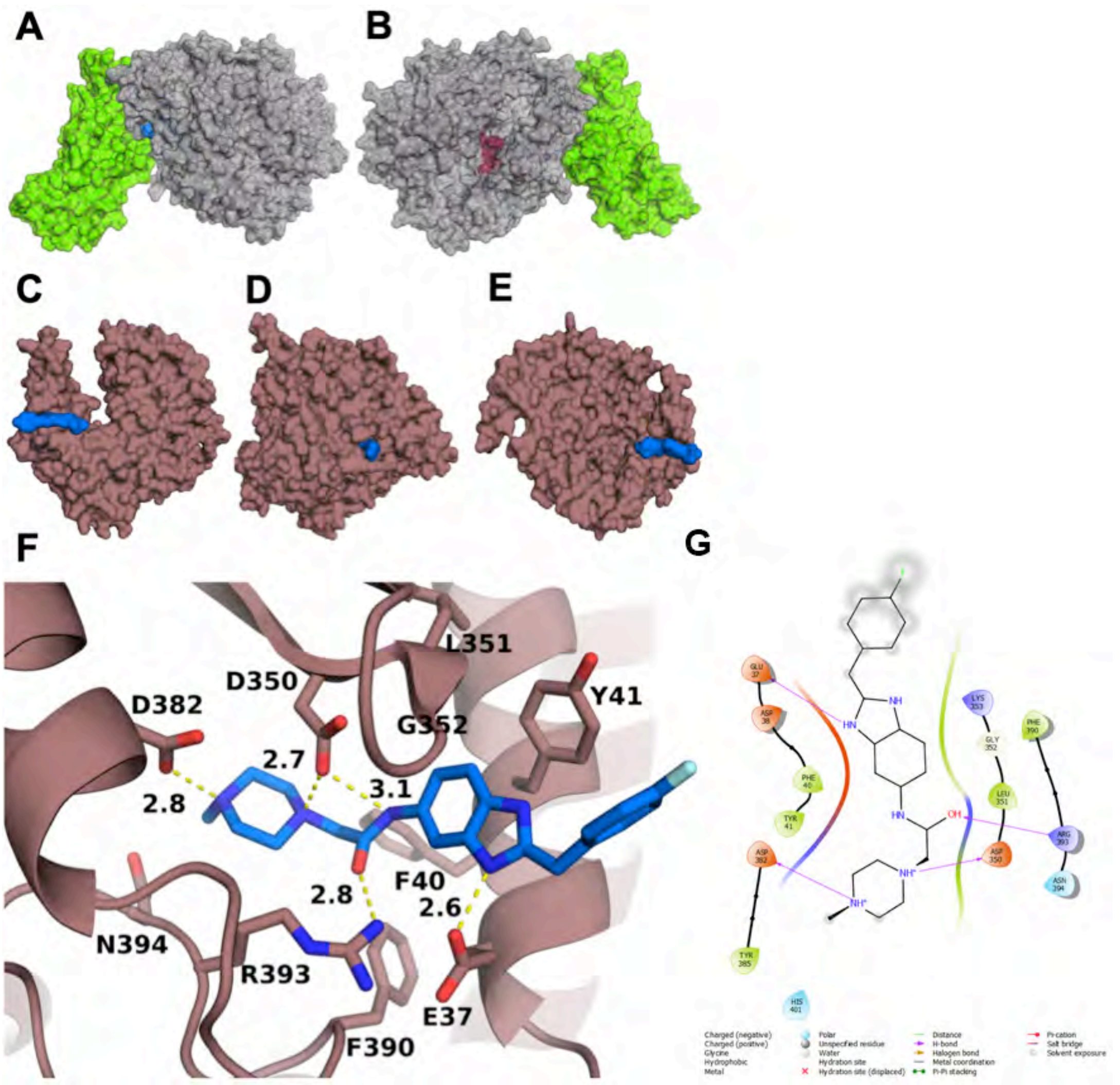

**Figure S1: Ace2:S protein interface and indication of allosteric site relative to active binding site.** **(A)** Sagittal view of ACE2 (gray) interface with RBD of COVID19 S-protein (green); the blue surface highlights the binding site for ligands that disrupt the interface between the two proteins. **(B)** ACE2 (gray) and S-protein (green) sagittal view. In maroon is the active site of ACE2. **(C)** ACE2 (salmon) with ligand 300 (blue) rendered as surfaces. 50% left side slab to examine deep insertion in more detail. **(D)** Full surface view of ACE2 and ligand 300. **(E)** 50% right side slab to examine deep insertion in more detail. **(F)** Example of docked compound that disrupts interface between ACE2 and S-protein. Close-up of binding site of ACE2 (salmon) and ligand 300 (blue) with residues and polar contact distances labeled. **(G)** Ligand Interaction Diagram rendered with Maestro for ACE2 with ligand 300 at the allosteric site impacting S-protein binding from SAR-CoV2. This 2D "flat" representation shows the interactions at this particular compounds interface on Ace2 that would interfere with S protein binding. In particular, extending from deeply inserted to superficial, the interactions are described in the subsequent sentences. D382 and D350 are hydrogen bond acceptors (side chains) from the opposite NH+ on the piperazine-like ring deeply inserted into the binding pocket. R393 is a hydrogen bond donor (side chain) to the alcohol group connecting the piperazine-like ring to the fused ring. E37 is a hydrogen bond acceptor (side chain) to one of the NH on the fused ring. The fluorocyclohexane group is entirely solvent-exposed at the mouth of the binding pocket.

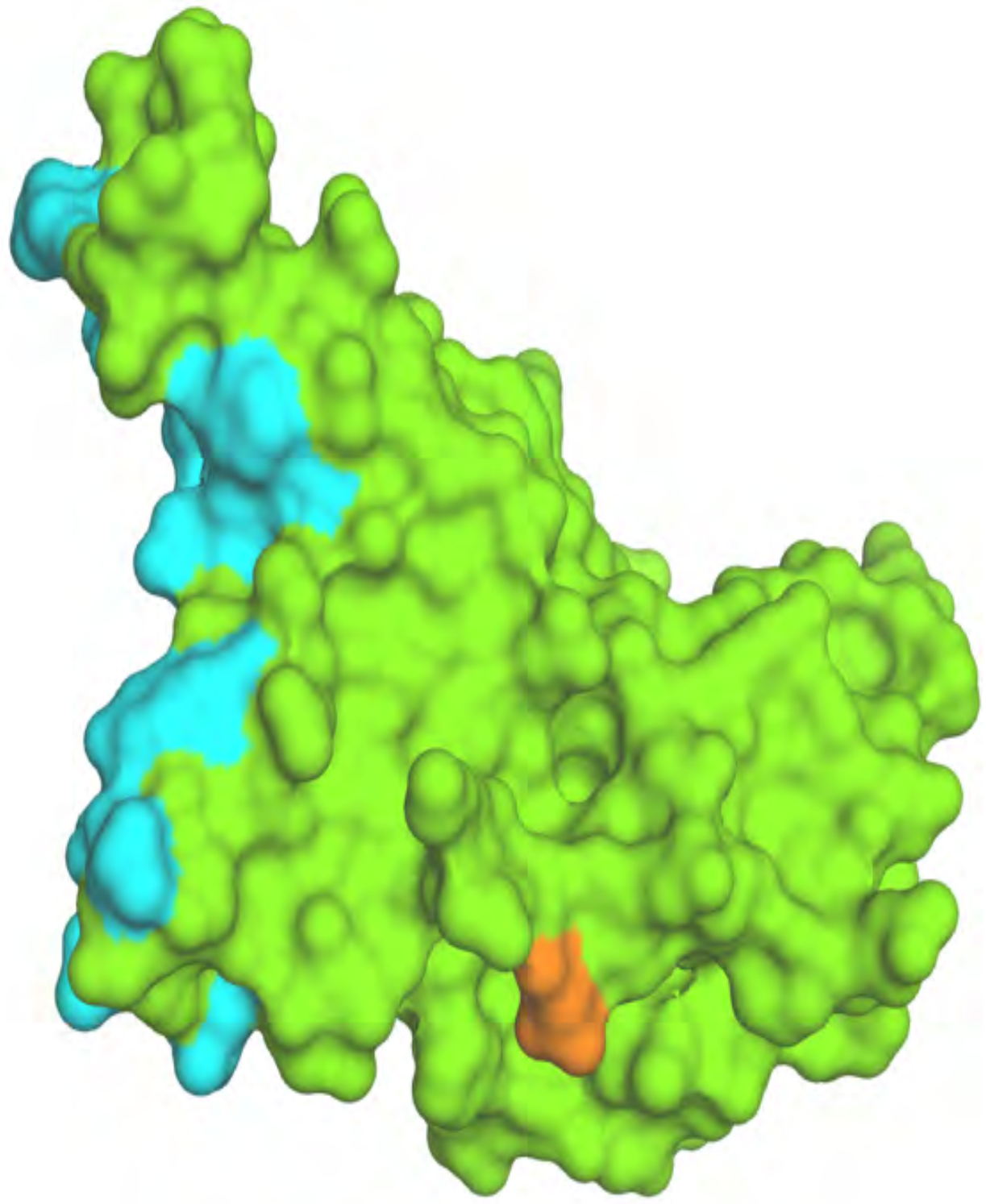

**Figure S2 Glycosylation sites of Ace2 protein (D616G highlighted red).** Although glycosylation sites at residues N165, N234, N343 from S-protein (PDB code 6VSB), are nearby the ACE2:S-protein binding interface, they do not overlap and interfere with the protein-protein interface, offering an adjacent site is readily available for PPI docking (S-prot glycosylation analysis: DOI: 10.1126/science.abb9983; 10.1101/2020.04.29.069054}. The majority of glycosylation sites are not on the RBD (Fig. S2), the glycosylation site that is actually present on the RBD, N343, is not in 3D proximity to the binding interface. Recently, a variant of the S-protein, D614G, was identified to possess enhanced transmissibility and resistance to contemporary interventions and this site is not present on the RBD. Neither the glycosylation sites, nor the enhanced transmissibility variant D614G, are within the 3D proximity to the drug binding site for our targeted protein-protein interface disrupting therapeutics for Ace2.

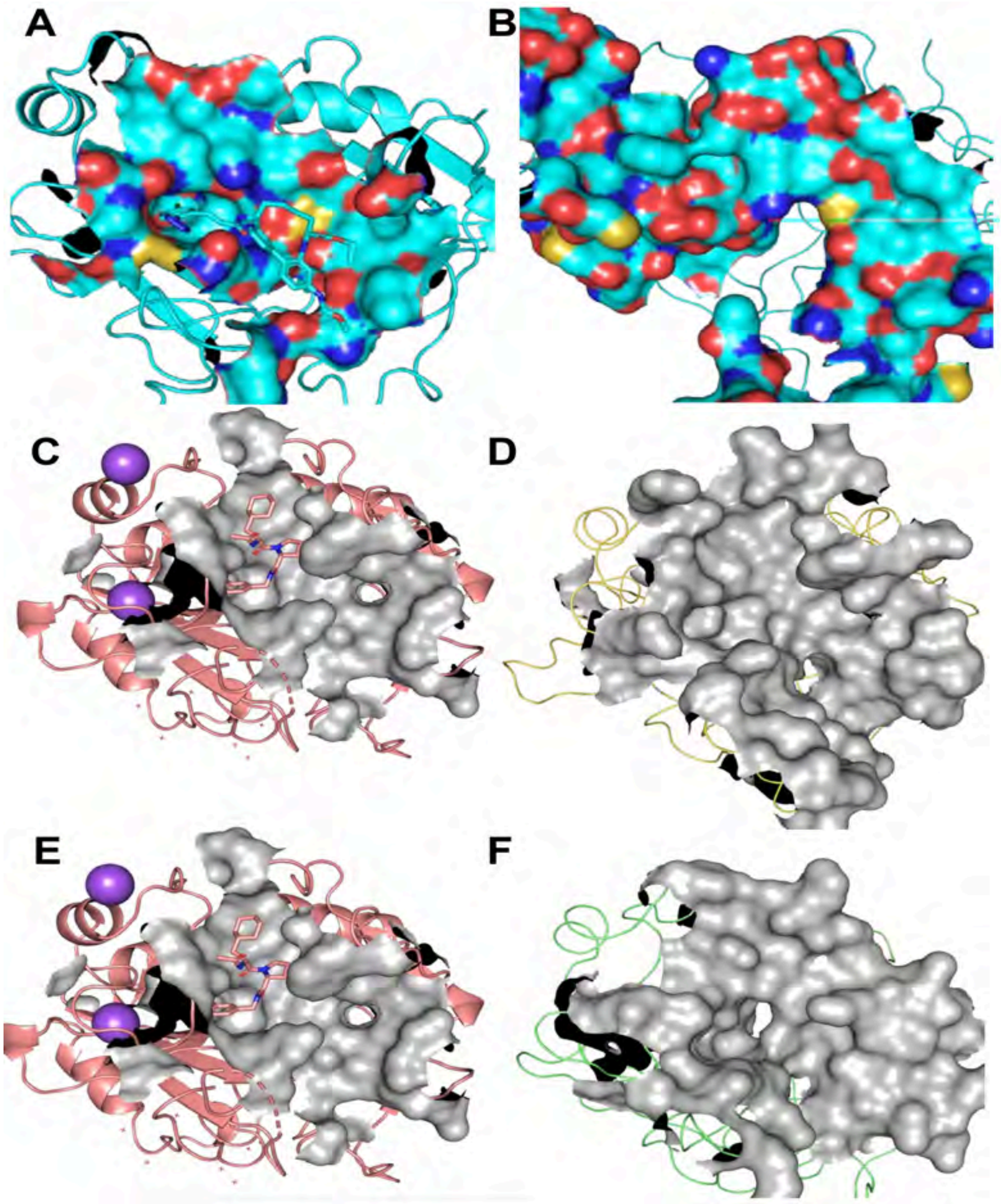

**Figure S3. Gradually crumbling binding site.** (A) Initital and (B) Final states - of the protein model, while catalytically active state has better preserved binding site. Prothrombin binding site (PDB 3F68) with its inhibitor (C) and the final state of Prothrombin (D) are shown. Again, Prothrombin binding site (3F68) with its inhibitor (E) and proposed structural model – a prothrombin-based homology model of TRPMSS2 (F), which looks more accurate then previous (B) model structure. This version maintains structural stability and is good candidate for drug docking with ligands. Purple spheres are constraints used to impose good relative positioning.

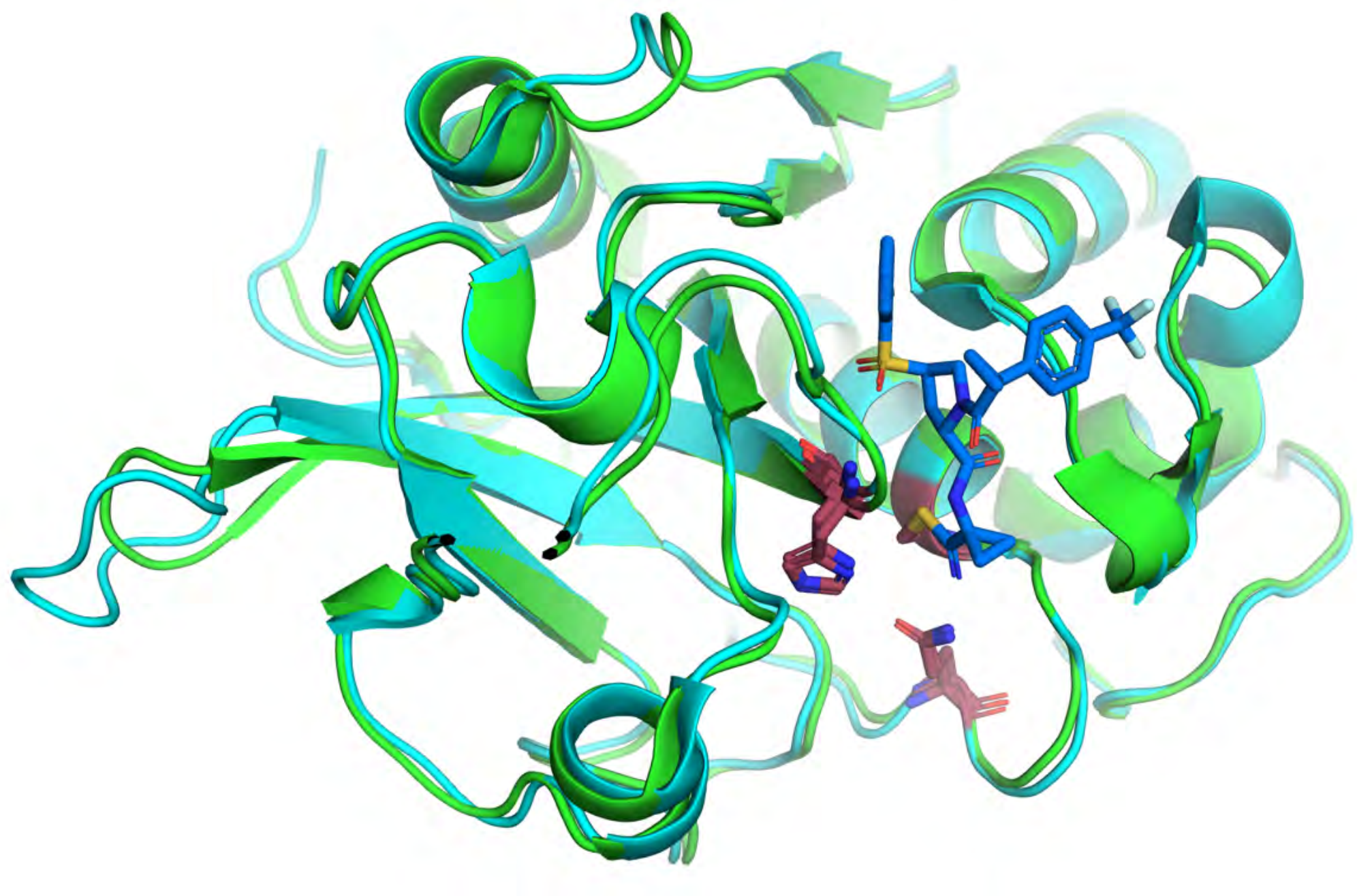

**Figure S4. Proteases cathepsin L and K can be also used in blocking ENTRY of COVID-19 during late-endosome progression. Top panel**. Structures for cathepsin K (PDB code: 4N8W, green) and cathepsin L (PDB code: 2YJB, cyan), shown with an inhibitor (blue). The active site residues are colored maroon. These represent additional host protease targets at another stage of the viral entry cycle. **Bottom panel**. Same as top but rotated 180°. Cathepsins K and L represent additional host protease targets at another stage of the viral entry cycle. Future efforts and alternative methods on our part may involve discovering effective compounds to exploit this point of intervention in synergy with our other therapeutics. In anticipation of this, we have already constructed models of both of these cathepsins, which exhibit remarkable structural homology with each other. For cathepsin K, 4N8W.pdb {PMID: 25422423} was used as a base from which to construct the model, and 2YJB.pdb {PMID: 21898833} was used for cathepsin L.

Coban et al., CHM 2020 -- Resource ARTICLE -- Dataset Table S2

**TABLE S2. Top 310 NCE compounds docked with Ace2, TMRPSS2, and Mpro (from >10million compounds on all targets)**

| 2D Structure | Compound Name | ENZYME | Docking Score | Smile |
|---|---|---|---|---|
| 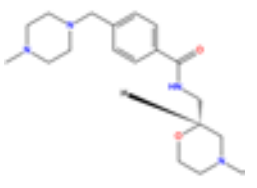 | N-[(4-methylmorpholin-2-yl)methyl]-4-[(4-methylpiperazin-1-yl)methyl]benzamide | ACE2 | -7.415595 | CN1CCN(Cc2ccc(cc2)C(=O)NCC2CN(C)CCO2)CC1 |
| 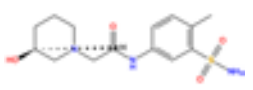 | 2-(3-hydroxypiperidin-1-yl)-N-(4-methyl-3-sulfamoylphenyl)acetamide | ACE2 | -7.377284 | Cc1ccc(NC(=O)CN2CCCC(O)C2)cc1S(N)(=O)=O |
| 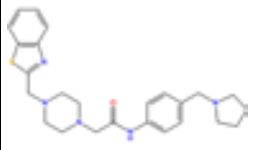 | 2-{4-[(1,3-benzothiazol-2-yl)methyl]piperazin-1-yl}-N-{4-[(pyrrolidin-1-yl)methyl]phenyl}acetamide | ACE2 | -7.733448 | O=C(CN1CCN(Cc2nc3ccccc3s2)CC1)Nc1ccc(CN2CCCC2)cc1 |
| 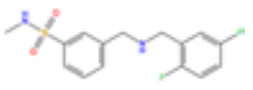 | 3-({[(2,5-difluorophenyl)methyl]amino}methyl)-N-methylbenzene-1-sulfonamide | ACE2 | -7.527599 | CNS(=O)(=O)c1cccc(CNCc2cc(F)ccc2F)c1 |
| 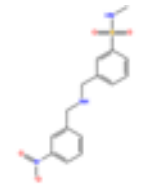 | N-methyl-3-({[(3-nitrophenyl)methyl]amino}methyl)benzene-1-sulfonamide | ACE2 | -7.486144 | CNS(=O)(=O)c1cccc(CNCc2cccc(c2)[N+]([O-])=O)c1 |
| 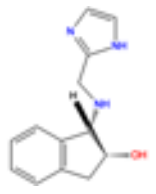 | (1S,2R)-1-{[(1H-imidazol-2-yl)methyl]amino}-2,3-dihydro-1H-inden-2-ol | ACE2 | -7.818057 | O[C@@H]1Cc2ccccc2[C@@H]1NCc1ncc[nH]1 |
| 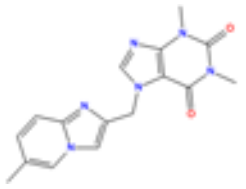 | 1,3-dimethyl-7-({6-methylimidazo[1,2-a]pyridin-2-yl}methyl)-2,3,6,7-tetrahydro-1H-purine-2,6-dione | ACE2 | -7.813326 | Cc1ccc2nc(Cn3cnc4n(C)c(=O)n(C)c(=O)c34)cn2c1 |
| 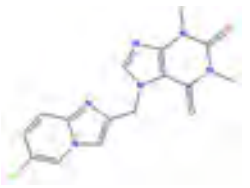 | 7-({6-chloroimidazo[1,2-a]pyridin-2-yl}methyl)-1,3-dimethyl-2,3,6,7-tetrahydro-1H-purine-2,6-dione | ACE2 | -7.787989 | Cn1c2ncn(Cc3cn4cc(Cl)ccc4n3)c2c(=O)n(C)c1=O |
| 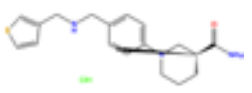 | 1-[4-({[(thiophen-3-yl)methyl]amino}methyl)phenyl]piperidine-3-carboxamide hydrochloride | ACE2 | -7.346939 | Cl.NC(=O)C1CCCN(C1)c1ccc(CNCc2ccsc2)cc1 |
| 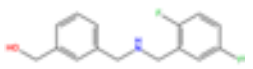 | [3-({[(2,5-difluorophenyl)methyl]amino}methyl)phenyl]methanol | ACE2 | -7.273845 | OCc1cccc(CNCc2cc(F)ccc2F)c1 |
| 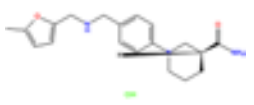 | 1-[4-({[(5-methylfuran-2-yl)methyl]amino}methyl)phenyl]piperidine-3-carboxamide hydrochloride | ACE2 | -7.368915 | Cl.Cc1ccc(CNCc2ccc(cc2)N2CCCC(C2)C(N)=O)o1 |
| 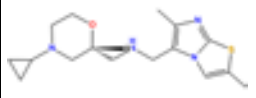 | [(4-cyclopropylmorpholin-2-yl)methyl]({2,6-dimethylimidazo[2,1-b][1,3]thiazol-5-yl}methyl)amine | ACE2 | -7.838944 | Cc1cn2c(CNCC3CN(CCO3)C3CC3)c(C)nc2s1 |
| 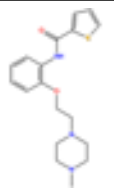 | N-{2-[2-(4-methylpiperazin-1-yl)ethoxy]phenyl}thiophene-2-carboxamide | ACE2 | -7.649969 | CN1CCN(CCOc2ccccc2NC(=O)c2cccs2)CC1 |
| 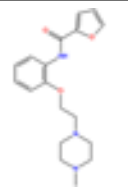 | N-{2-[2-(4-methylpiperazin-1-yl)ethoxy]phenyl}furan-2-carboxamide | **ACE2** | **-7.32206** | CN1CCN(CCOc2ccccc2NC(=O)c2ccco2)CC1 |
| 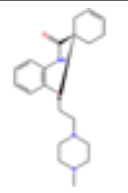 | N-{2-[2-(4-methylpiperazin-1-yl)ethoxy]phenyl}cyclohex-3-ene-1-carboxamide | **ACE2** | **-7.790159** | CN1CCN(CCOc2ccccc2NC(=O)C2CCC=CC2)CC1 |
| 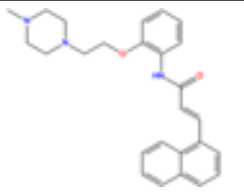 | N-{2-[2-(4-methylpiperazin-1-yl)ethoxy]phenyl}-3-(naphthalen-1-yl)prop-2-enamide | **ACE2** | **-7.25658** | CN1CCN(CCOc2ccccc2NC(=O)C=Cc2cccc3ccccc23)CC1 |
| 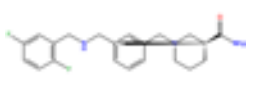 | 1-{[3-({[(2,5-difluorophenyl)methyl]amino}methyl)phenyl]methyl}piperidine-3-carboxamide | **ACE2** | **-7.835397** | NC(=O)C1CCCN(Cc2cccc(CNCc3cc(F)ccc3F)c2)C1 |
| 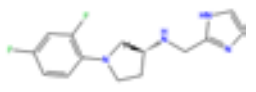 | 1-(2,4-difluorophenyl)-N-[(1H-imidazol-2-yl)methyl]pyrrolidin-3-amine | **ACE2** | **-7.890993** | Fc1ccc(N2CCC(C2)NCc2ncc[nH]2)c(F)c1 |
| 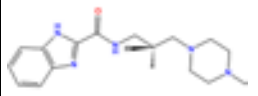 | N-[2-methyl-3-(4-methylpiperazin-1-yl)propyl]-1H-1,3-benzodiazole-2-carboxamide | **ACE2** | **-7.198821** | CC(CNC(=O)c1nc2ccccc2[nH]1)CN1CCN(C)CC1 |
| 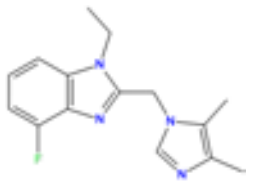 | 2-[(4,5-dimethyl-1H-imidazol-1-yl)methyl]-1-ethyl-4-fluoro-1H-1,3-benzodiazole | **ACE2** | **-7.938425** | CCn1c(Cn2cnc(C)c2C)nc2c(F)cccc12 |
| 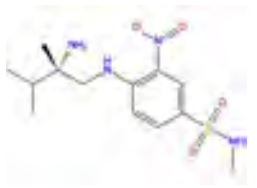 | 4-[(2-amino-2,3-dimethylbutyl)amino]-N-methyl-3-nitrobenzene-1-sulfonamide | **ACE2** | **-7.442683** | CNS(=O)(=O)c1ccc(NCC(C)(N)C(C)C)c(c1)[N+]([O-])=O |
| 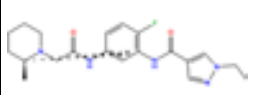 | 1-ethyl-N-{2-fluoro-5-[2-(2-methylpiperidin-1-yl)acetamido]phenyl}-1H-pyrazole-4-carboxamide | **ACE2** | **-7.589474** | CCn1cc(cn1)C(=O)Nc1cc(NC(=O)CN2CCCCC2C)ccc1F |
| 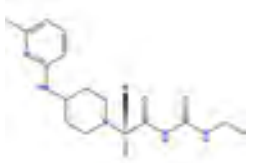 | 3-ethyl-1-(2-{4-[(6-methylpyridin-2-yl)amino]piperidin-1-yl}propanoyl)urea | **ACE2** | **-7.789715** | CCNC(=O)NC(=O)C(C)N1CCC(CC1)Nc1cccc(C)n1 |
| 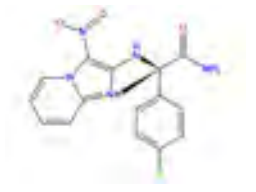 | 2-(4-chlorophenyl)-2-({3-nitroimidazo[1,2-a]pyridin-2-yl}amino)acetamide | **ACE2** | **-7.548985** | NC(=O)C(Nc1nc2ccccn2c1[N+]([O-])=O)c1ccc(Cl)cc1 |
| 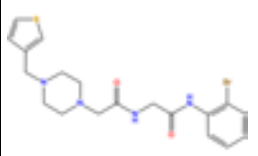 | N-{[(2-bromophenyl)carbamoyl]methyl}-2-{4-[(thiophen-3-yl)methyl]piperazin-1-yl}acetamide | **ACE2** | **-7.056257** | Brc1ccccc1NC(=O)CNC(=O)CN1CCN(Cc2ccsc2)CC1 |
| 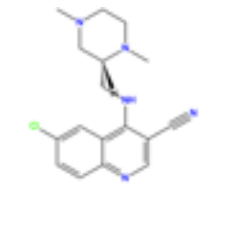 | 6-chloro-4-{[(1,4-dimethylpiperazin-2-yl)methyl]amino}quinoline-3-carbonitrile | **ACE2** | **-7.690945** | CN1CCN(C)C(CNc2c(cnc3ccc(Cl)cc23)C#N)C1 |
| 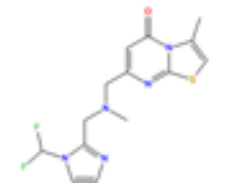 | 7-[({[1-(difluoromethyl)-1H-imidazol-2-yl]methyl}(methyl)amino)methyl]-3-methyl-5H-[1,3]thiazolo[3,2-a]pyrimidin-5-one | **ACE2** | **-7.368598** | CN(Cc1nccn1C(F)F)Cc1cc(=O)n2c(C)csc2n1 |
| 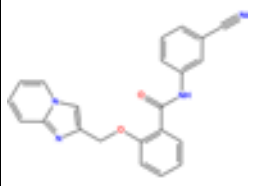 | N-(3-cyanophenyl)-2-({imidazo[1,2-a]pyridin-2-yl}methoxy)benzamide | **ACE2** | **-7.272896** | O=C(Nc1cccc(c1)C#N)c1ccccc1OCc1cn2ccccc2n1 |
| 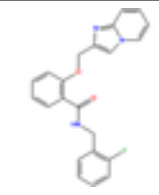 | N-[(2-fluorophenyl)methyl]-2-({imidazo[1,2-a]pyridin-2-yl}methoxy)benzamide | **ACE2** | **-7.504407** | Fc1ccccc1CNC(=O)c1ccccc1OCc1cn2ccccc2n1 |
| 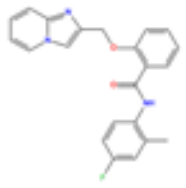 | N-(4-fluoro-2-methylphenyl)-2-({imidazo[1,2-a]pyridin-2-yl}methoxy)benzamide | **ACE2** | **-7.256458** | Cc1cc(F)ccc1NC(=O)c1ccccc1OCc1cn2ccccc2n1 |
| 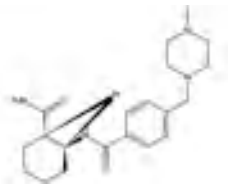 | N-(2-carbamoylcyclohexyl)-4-[(4-methylpiperazin-1-yl)methyl]benzamide | **ACE2** | **-7.2544** | CN1CCN(Cc2ccc(cc2)C(=O)NC2CCCCC2C(N)=O)CC1 |
| 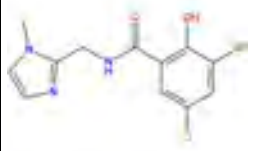 | 3-bromo-5-chloro-2-hydroxy-N-[(1-methyl-1H-imidazol-2-yl)methyl]benzamide | **ACE2** | **-7.396006** | Cn1ccnc1CNC(=O)c1cc(Cl)cc(Br)c1O |
| 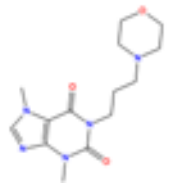 | 3,7-dimethyl-1-[3-(morpholin-4-yl)propyl]-2,3,6,7-tetrahydro-1H-purine-2,6-dione | **ACE2** | **-7.740514** | Cn1cnc2n(C)c(=O)n(CCCN3CCOCC3)c(=O)c12 |
| 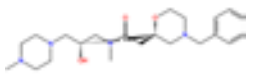 | 2-(4-benzylmorpholin-2-yl)-N-[2-hydroxy-3-(4-methylpiperazin-1-yl)propyl]-N-methylacetamide | **ACE2** | **-7.062327** | CN(CC(O)CN1CCN(C)CC1)C(=O)CC1CN(Cc2ccccc2)CCO1 |
| 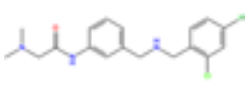 | N-[3-({[(2-chloro-4-fluorophenyl)methyl]amino}methyl)phenyl]-2-(dimethylamino)acetamide | **ACE2** | **-7.794252** | CN(C)CC(=O)Nc1cccc(CNCc2ccc(F)cc2Cl)c1 |
| 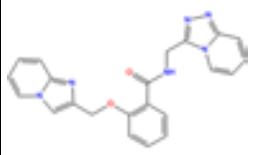 | 2-({imidazo[1,2-a]pyridin-2-yl}methoxy)-N-({[1,2,4]triazolo[4,3-a]pyridin-3-yl}methyl)benzamide | **ACE2** | **-7.670744** | O=C(NCc1nnc2ccccn12)c1ccccc1OCc1cn2ccccc2n1 |
| 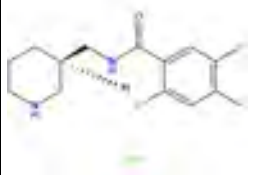 | 2,5-difluoro-4-methyl-N-[(piperidin-3-yl)methyl]benzamide hydrochloride | **ACE2** | **-7.244727** | Cl.Cc1cc(F)c(cc1F)C(=O)NCC1CCCNC1 |
| 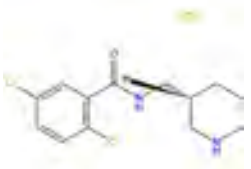 | 2,5-dichloro-N-[(piperidin-3-yl)methyl]benzamide hydrochloride | **ACE2** | **-7.847098** | Cl.Clc1ccc(Cl)c(c1)C(=O)NCC1CCCNC1 |
| 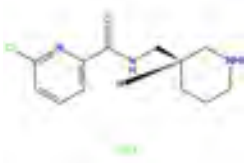 | 6-chloro-N-[(piperidin-3-yl)methyl]pyridine-2-carboxamide hydrochloride | **ACE2** | **-7.817936** | Cl.Clc1cccc(n1)C(=O)NCC1CCCNC1 |
| 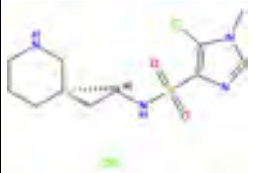 | 5-chloro-1-methyl-N-[2-(piperidin-3-yl)ethyl]-1H-imidazole-4-sulfonamide hydrochloride | **ACE2** | **-7.845325** | Cl.Cn1cnc(c1Cl)S(=O)(=O)NCCC1CCCNC1 |
| 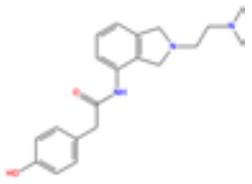 | N-{2-[2-(dimethylamino)ethyl]-2,3-dihydro-1H-isoindol-4-yl}-2-(4-hydroxyphenyl)acetamide | **ACE2** | **-7.443014** | CN(C)CCN1Cc2cccc(NC(=O)Cc3ccc(O)cc3)c2C1 |
| 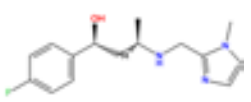 | 1-(4-fluorophenyl)-3-{[(1-methyl-1H-imidazol-2-yl)methyl]amino}butan-1-ol | **ACE2** | **-7.334131** | CC(CC(O)c1ccc(F)cc1)NCc1nccn1C |
| 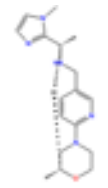 | [1-(1-methyl-1H-imidazol-2-yl)ethyl]({[6-(2-methylmorpholin-4-yl)pyridin-3-yl]methyl})amine | **ACE2** | **-7.83441** | CC(NCc1ccc(nc1)N1CCOC(C)C1)c1nccn1C |
| 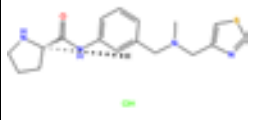 | N-[3-({methyl[(1,3-thiazol-4-yl)methyl]amino}methyl)phenyl]pyrrolidine-2-carboxamide hydrochloride | **ACE2** | **-7.740865** | Cl.CN(Cc1cscn1)Cc1cccc(NC(=O)C2CCCN2)c1 |
| 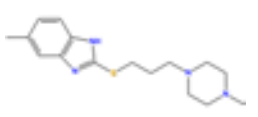 | 5-methyl-2-{[3-(4-methylpiperazin-1-yl)propyl]sulfanyl}-1H-1,3-benzodiazole | **ACE2** | **-7.876119** | CN1CCN(CCCSc2nc3cc(C)ccc3[nH]2)CC1 |
| 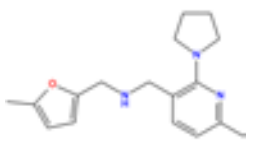 | {[6-methyl-2-(pyrrolidin-1-yl)pyridin-3-yl]methyl}[(5-methylfuran-2-yl)methyl]amine | **ACE2** | **-7.84725** | Cc1ccc(CNCc2ccc(C)nc2N2CCCC2)o1 |
| 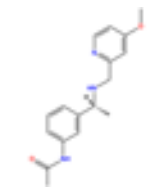 | N-[3-(1-{[(4-methoxypyridin-2-yl)methyl]amino}ethyl)phenyl]acetamide | **ACE2** | **-7.566234** | COc1ccnc(CNC(C)c2cccc(NC(C)=O)c2)c1 |
| 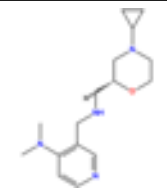 | 3-({[(4-cyclopropylmorpholin-2-yl)methyl]amino}methyl)-N,N-dimethylpyridin-4-amine | **ACE2** | **-7.891833** | CN(C)c1ccncc1CNCC1CN(CCO1)C1CC1 |
| 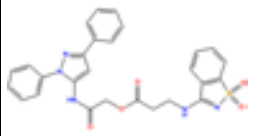 | [(1,3-diphenyl-1H-pyrazol-5-yl)carbamoyl]methyl 3-[(1,1-dioxo-1λ$^{6}$,2-benzothiazol-3-yl)amino]propanoate | **ACE2** | **-7.414** | O=C(COC(=O)CCNC1=NS(=O)(=O)c2ccccc12)Nc1cc(nn1-c1ccccc1)-c1ccccc1 |
| 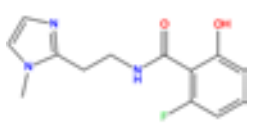 | 2-fluoro-6-hydroxy-N-[2-(1-methyl-1H-imidazol-2-yl)ethyl]benzamide | **ACE2** | **-7.366927** | Cn1ccnc1CCNC(=O)c1c(O)cccc1F |
| 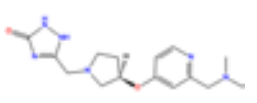 | 5-{[3-({2-[(dimethylamino)methyl]pyridin-4-yl}oxy)pyrrolidin-1-yl]methyl}-2,3-dihydro-1H-1,2,4-triazol-3-one | **ACE2** | **-7.785194** | CN(C)Cc1cc(OC2CCN(Cc3nc(=O)[nH][nH]3)C2)ccn1 |
| 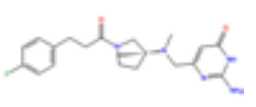 | 2-amino-6-[({1-[3-(4-fluorophenyl)propanoyl]pyrrolidin-3-yl}(methyl)amino)methyl]-3,4-dihydropyrimidin-4-one | **ACE2** | **-7.643808** | CN(Cc1cc(=O)[nH]c(N)n1)C1CCN(C1)C(=O)CCc1ccc(F)cc1 |
| 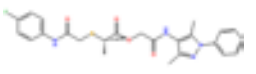 | [(3,5-dimethyl-1-phenyl-1H-pyrazol-4-yl)carbamoyl]methyl 2-({[(4-fluorophenyl)carbamoyl]methyl}sulfanyl)propanoate | **ACE2** | **-7.465323** | CC(SCC(=O)Nc1ccc(F)cc1)C(=O)OCC(=O)Nc1c(C)nn(c1C)-c1ccccc1 |
| 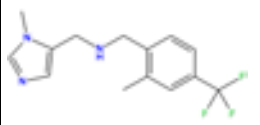 | [(1-methyl-1H-imidazol-5-yl)methyl]({[2-methyl-4-(trifluoromethyl)phenyl]methyl})amine | **ACE2** | **-7.915507** | Cc1cc(ccc1CNCc1cncn1C)C(F)(F)F |
| 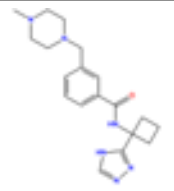 | 3-[(4-methylpiperazin-1-yl)methyl]-N-[1-(4H-1,2,4-triazol-3-yl)cyclobutyl]benzamide | **ACE2** | **-7.344405** | CN1CCN(Cc2cccc(c2)C(=O)NC2(CCC2)c2nnc[nH]2)CC1 |
| 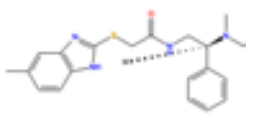 | N-[2-(dimethylamino)-2-phenylethyl]-2-[(5-methyl-1H-1,3-benzodiazol-2-yl)sulfanyl]acetamide | **ACE2** | **-7.040755** | CN(C)C(CNC(=O)CSc1nc2cc(C)ccc2[nH]1)c1ccccc1 |
| 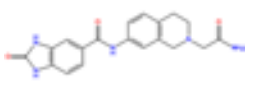 | N-[2-(carbamoylmethyl)-1,2,3,4-tetrahydroisoquinolin-7-yl]-2-oxo-2,3-dihydro-1H-1,3-benzodiazole-5-carboxamide | **ACE2** | **-7.125758** | NC(=O)CN1CCc2ccc(NC(=O)c3ccc4[nH]c(=O)[nH]c4c3)cc2C1 |
| 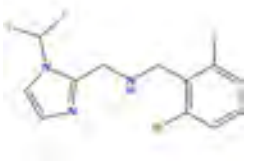 | [(2-bromo-6-fluorophenyl)methyl]({[1-(difluoromethyl)-1H-imidazol-2-yl]methyl})amine | **ACE2** | **-7.863935** | FC(F)n1ccnc1CNCc1c(F)cccc1Br |
| 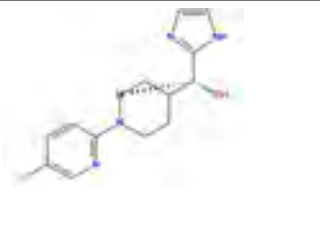 | [1-(5-chloropyridin-2-yl)piperidin-4-yl](1H-imidazol-2-yl)methanol | **ACE2** | **-7.910061** | OC(C1CCN(CC1)c1ccc(Cl)cn1)c1ncc[nH]1 |
| 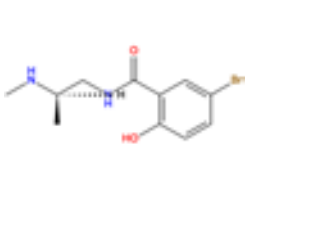 | 5-bromo-2-hydroxy-N-[2-(methylamino)propyl]benzamide | **ACE2** | **-7.621056** | CNC(C)CNC(=O)c1cc(Br)ccc1O |
| 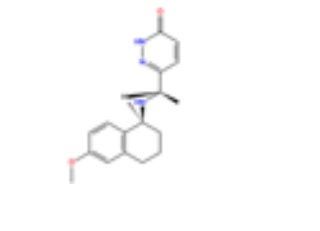 | 6-{1-[(6-methoxy-1,2,3,4-tetrahydronaphthalen-1-yl)amino]ethyl}-2,3-dihydropyridazin-3-one | **ACE2** | **-7.855813** | COc1ccc2C(CCCc2c1)NC(C)c1ccc(=O)[nH]n1 |
| 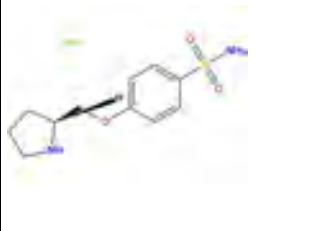 | 4-[(pyrrolidin-2-yl)methoxy]benzene-1-sulfonamide hydrochloride | **ACE2** | **-7.907147** | Cl.NS(=O)(=O)c1ccc(OCC2CCCN2)cc1 |
| 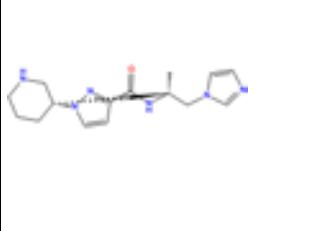 | N-[1-(1H-imidazol-1-yl)propan-2-yl]-1-(piperidin-3-yl)-1H-pyrazole-3-carboxamide | **ACE2** | **-7.895666** | CC(Cn1ccnc1)NC(=O)c1ccn(n1)C1CCCNC1 |
| 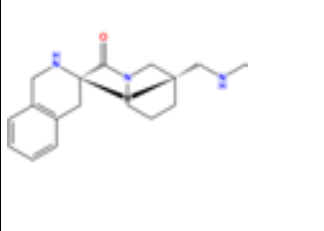 | methyl({[1-(1,2,3,4-tetrahydroisoquinoline-3-carbonyl)piperidin-3-yl]methyl})amine | **ACE2** | **-7.078639** | CNCC1CCCN(C1)C(=O)C1Cc2ccccc2CN1 |
| 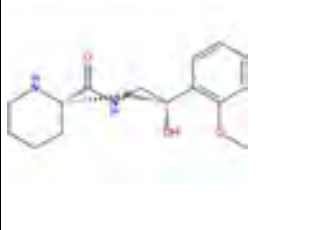 | N-[2-hydroxy-2-(2-methoxyphenyl)ethyl]piperidine-2-carboxamide | **ACE2** | **-7.278091** | COc1ccccc1C(O)CNC(=O)C1CCCCN1 |
| 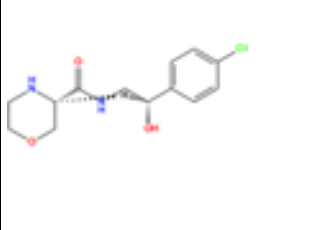 | N-[2-(4-chlorophenyl)-2-hydroxyethyl]morpholine-3-carboxamide | **ACE2** | **-7.905742** | OC(CNC(=O)C1COCCN1)c1ccc(Cl)cc1 |
| 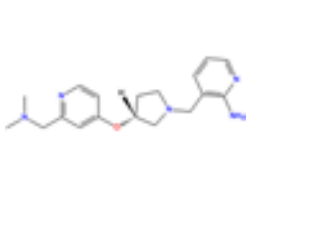 | 3-{[3-({2-[(dimethylamino)methyl]pyridin-4-yl}oxy)pyrrolidin-1-yl]methyl}pyridin-2-amine | **ACE2** | **-7.520437** | CN(C)Cc1cc(OC2CCN(Cc3cccnc3N)C2)ccn1 |
| 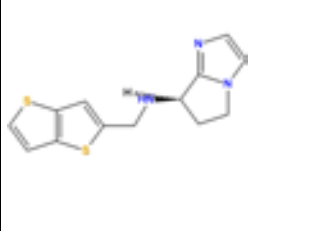 | N-({thieno[3,2-b]thiophen-2-yl}methyl)-5H,6H,7H-pyrrolo[1,2-a]imidazol-7-amine | **ACE2** | **-7.928268** | C(NC1CCn2ccnc12)c1cc2sccc2s1 |
| 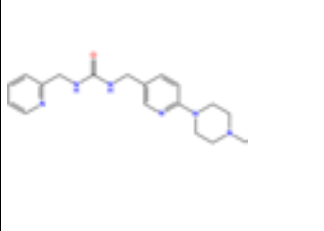 | 3-{[6-(4-methylpiperazin-1-yl)pyridin-3-yl]methyl}-1-[(pyridin-2-yl)methyl]urea | **ACE2** | **-7.875927** | CN1CCN(CC1)c1ccc(CNC(=O)NCc2ccccn2)cn1 |
| 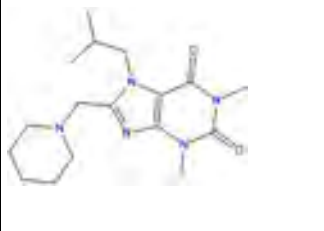 | 1,3-dimethyl-7-(2-methylpropyl)-8-[(piperidin-1-yl)methyl]-2,3,6,7-tetrahydro-1H-purine-2,6-dione | **ACE2** | **-7.933447** | CC(C)Cn1c(CN2CCCCC2)nc2n(C)c(=O)n(C)c(=O)c12 |
| 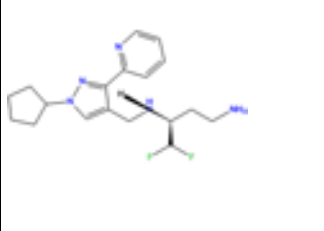 | (4-amino-1,1-difluorobutan-2-yl)({[1-cyclopentyl-3-(pyridin-2-yl)-1H-pyrazol-4-yl]methyl})amine | **ACE2** | **-7.875737** | NCCC(NCc1cn(nc1-c1ccccn1)C1CCCC1)C(F)F |
| 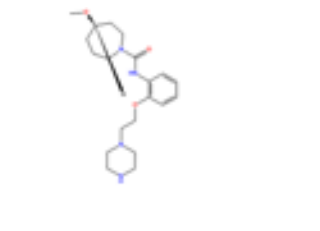 | 4-methoxy-N-{2-[2-(piperazin-1-yl)ethoxy]phenyl}azepane-1-carboxamide | **ACE2** | **-7.309969** | COC1CCCN(CC1)C(=O)Nc1ccccc1OCCN1CCNCC1 |
| 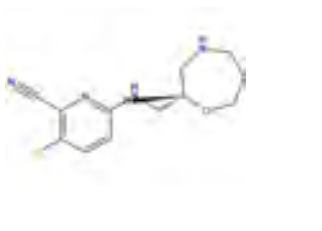 | 3-chloro-6-{[(1,4-oxazepan-2-yl)methyl]amino}pyridine-2-carbonitrile | **ACE2** | **-7.860086** | Clc1ccc(NCC2CNCCCO2)nc1C#N |
| 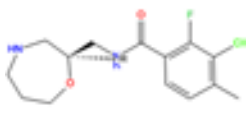 | 3-chloro-2-fluoro-4-methyl-N-[(1,4-oxazepan-2-yl)methyl]benzamide | **ACE2** | **-7.936678** | Cc1ccc(C(=O)NCC2CNCCCO2)c(F)c1Cl |
| 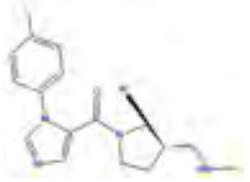 | ({1-[1-(4-fluorophenyl)-1H-imidazole-5-carbonyl]pyrrolidin-3-yl}methyl)(methyl)amine | **ACE2** | **-7.931379** | CNCC1CCN(C1)C(=O)c1cncn1-c1ccc(F)cc1 |
| 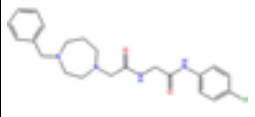 | 2-(4-benzyl-1,4-diazepan-1-yl)-N-{[(4-fluorophenyl)carbamoyl]methyl}acetamide | **ACE2** | **-7.271333** | Fc1ccc(NC(=O)CNC(=O)CN2CCCN(Cc3ccccc3)CC2)cc1 |
| 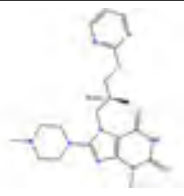 | 3-methyl-7-[2-methyl-3-(pyrimidin-2-ylsulfanyl)propyl]-8-(4-methylpiperazin-1-yl)-2,3,6,7-tetrahydro-1H-purine-2,6-dione | **ACE2** | **-7.751422** | CC(CSc1ncccn1)Cn1c(nc2n(C)c(=O)[nH]c(=O)c12)N1CCN(C)CC1 |
| 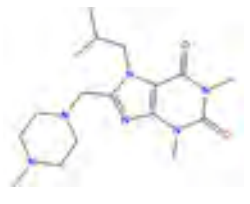 | 1,3-dimethyl-8-[(4-methylpiperazin-1-yl)methyl]-7-(2-methylpropyl)-2,3,6,7-tetrahydro-1H-purine-2,6-dione | **ACE2** | **-7.931598** | CC(C)Cn1c(CN2CCN(C)CC2)nc2n(C)c(=O)n(C)c(=O)c12 |
| 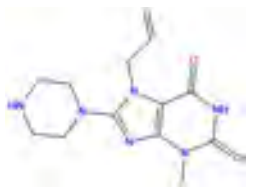 | 3-methyl-8-(piperazin-1-yl)-7-(prop-2-en-1-yl)-2,3,6,7-tetrahydro-1H-purine-2,6-dione | **ACE2** | **-7.460227** | Cn1c2nc(N3CCNCC3)n(CC=C)c2c(=O)[nH]c1=O |
| 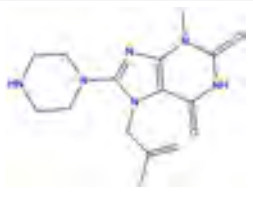 | 3-methyl-7-(2-methylprop-2-en-1-yl)-8-(piperazin-1-yl)-2,3,6,7-tetrahydro-1H-purine-2,6-dione | **ACE2** | **-7.33643** | CC(=C)Cn1c(nc2n(C)c(=O)[nH]c(=O)c12)N1CCNCC1 |
| 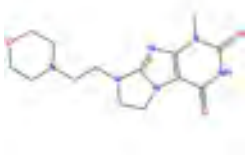 | 1-methyl-8-[2-(morpholin-4-yl)ethyl]-1H,2H,3H,4H,6H,7H,8H-imidazo[1,2-g]purine-2,4-dione | **ACE2** | **-7.330341** | Cn1c2nc3N(CCN4CCOCC4)CCn3c2c(=O)[nH]c1=O |
| 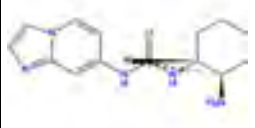 | rac-1-[(1R,2R)-2-aminocyclohexyl]-3-{imidazo[1,2-a]pyridin-7-yl}urea | **ACE2** | **-7.3361** | |
| 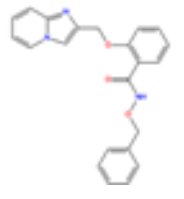 | N-(benzyloxy)-2-({imidazo[1,2-a]pyridin-2-yl}methoxy)benzamide | **ACE2** | **-7.671655** | O=C(NOCc1ccccc1)c1ccccc1OCc1cn2ccccc2n1 |
| 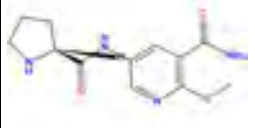 | 2-methoxy-5-(pyrrolidine-2-amido)pyridine-3-carboxamide | **ACE2** | **-7.783586** | COc1ncc(NC(=O)C2CCCN2)cc1C(N)=O |
| 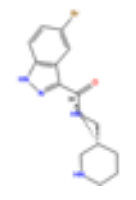 | 5-bromo-N-[(piperidin-3-yl)methyl]-1H-indazole-3-carboxamide | **ACE2** | **-7.822889** | Brc1ccc2[nH]nc(C(=O)NCC3CCCNC3)c2c1 |
| 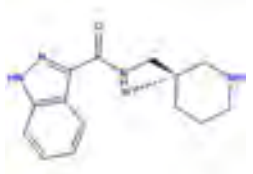 | N-[(piperidin-3-yl)methyl]-1H-indazole-3-carboxamide | **ACE2** | **-7.917339** | O=C(NCC1CCCNC1)c1n[nH]c2ccccc12 |
| 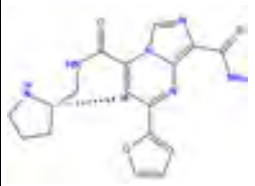 | 2-(furan-2-yl)-N4-[(pyrrolidin-2-yl)methyl]imidazo[1,5-a]pyrimidine-4,8-dicarboxamide | **ACE2** | **-7.250268** | NC(=O)c1ncn2c(cc(nc12)-c1ccco1)C(=O)NCC1CCCN1 |
| 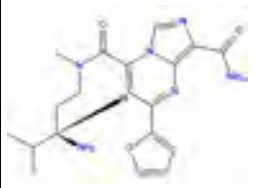 | N4-(3-amino-4-methylpentyl)-2-(furan-2-yl)-N4-methylimidazo[1,5-a]pyrimidine-4,8-dicarboxamide | **ACE2** | **-7.521512** | CC(C)C(N)CCN(C)C(=O)c1cc(nc2c(ncn12)C(N)=O)-c1ccco1 |
| 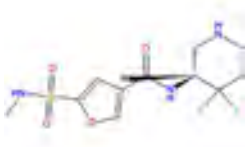 | N-(4,4-difluoropiperidin-3-yl)-5-(methylsulfamoyl)furan-3-carboxamide | **ACE2** | **-7.306415** | CNS(=O)(=O)c1cc(co1)C(=O)NC1CNCCC1(F)F |
| 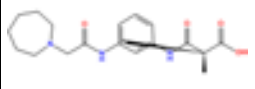 | 2-({3-[2-(azepan-1-yl)acetamido]phenyl}carbamoyl)-2-methylacetic acid | **ACE2** | **-7.696227** | CC(C(O)=O)C(=O)Nc1cccc(NC(=O)CN2CCCCCC2)c1 |
| 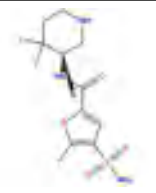 | N-(4,4-difluoropiperidin-3-yl)-5-methyl-4-sulfamoylfuran-2-carboxamide | **ACE2** | **-7.550276** | Cc1oc(cc1S(N)(=O)=O)C(=O)NC1CNCCC1(F)F |
| 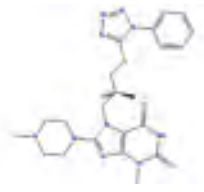 | 3-methyl-7-{2-methyl-3-[(1-phenyl-1H-1,2,3,4-tetrazol-5-yl)sulfanyl]propyl}-8-(4-methylpiperazin-1-yl)-2,3,6,7-tetrahydro-1H-purine-2,6-dione | **ACE2** | **-7.596339** | CC(CSc1nnnn1-c1ccccc1)Cn1c(nc2n(C)c(=O)[nH]c(=O)c12)N1CCN(C)CC1 |
| 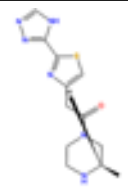 | 1-(3-methylpiperazin-1-yl)-2-[2-(4H-1,2,4-triazol-3-yl)-1,3-thiazol-4-yl]ethan-1-one | **ACE2** | **-7.906252** | CC1CN(CCN1)C(=O)Cc1csc(n1)-c1nnc[nH]1 |
| 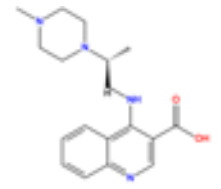 | 4-{[2-(4-methylpiperazin-1-yl)propyl]amino}quinoline-3-carboxylic acid | **ACE2** | **-7.251865** | CC(CNc1c(cnc2ccccc12)C(O)=O)N1CCN(C)CC1 |
| 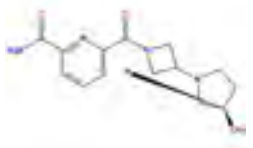 | 6-{3-[(3R)-3-hydroxypyrrolidin-1-yl]azetidine-1-carbonyl}pyridine-2-carboxamide | **ACE2** | **-7.815683** | NC(=O)c1cccc(n1)C(=O)N1CC(C1)N1CC[C@@H](O)C1 |
| 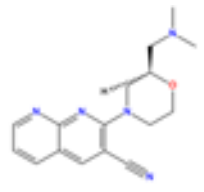 | 2-{2-[(dimethylamino)methyl]morpholin-4-yl}-1,8-naphthyridine-3-carbonitrile | **ACE2** | **-7.596166** | CN(C)CC1CN(CCO1)c1nc2ncccc2cc1C#N |
| 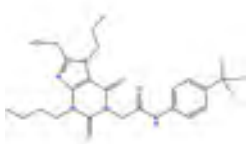 | 2-[3-butyl-8-(hydroxymethyl)-2,6-dioxo-7-propyl-2,3,6,7-tetrahydro-1H-purin-1-yl]-N-[4-(trifluoromethyl)phenyl]acetamide | **ACE2** | **-7.861967** | CCCCn1c2nc(CO)n(CCC)c2c(=O)n(CC(=O)Nc2ccc(cc2)C(F)(F)F)c1=O |
| 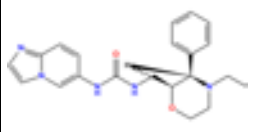 | rac-1-{[(2R,3R)-4-ethyl-3-phenylmorpholin-2-yl]methyl}-3-{imidazo[1,2-a]pyridin-6-yl}urea | **ACE2** | **-7.339682** | |
| 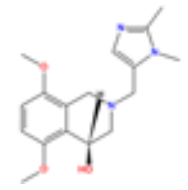 | 2-[(1,2-dimethyl-1H-imidazol-5-yl)methyl]-5,8-dimethoxy-1,2,3,4-tetrahydroisoquinolin-4-ol | **ACE2** | **-7.821377** | COc1ccc(OC)c2C(O)CN(Cc3cnc(C)n3C)Cc12 |
| 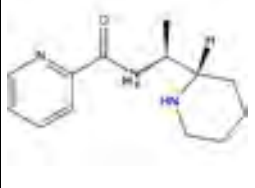 | N-[1-(piperidin-2-yl)ethyl]pyridine-2-carboxamide | **ACE2** | **-7.85541** | CC(NC(=O)c1ccccn1)C1CCCCN1 |
| 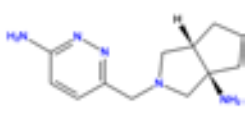 | 6-({3a-amino-octahydrocyclopenta[c]pyrrol-2-yl}methyl)pyridazin-3-amine | **ACE2** | **-7.279934** | Nc1ccc(CN2CC3CCCC3(N)C2)nn1 |
| 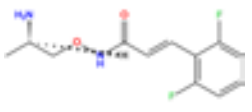 | N-(2-aminopropoxy)-3-(2,6-difluorophenyl)prop-2-enamide | **ACE2** | **-7.900215** | CC(N)CONC(=O)C=Cc1c(F)cccc1F |
| 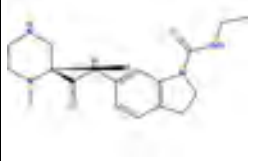 | N-ethyl-6-(1-methylpiperazine-2-amido)-2,3-dihydro-1H-indole-1-carboxamide | **ACE2** | **-7.505628** | CCNC(=O)N1CCc2ccc(NC(=O)C3CNCCN3C)cc12 |
| 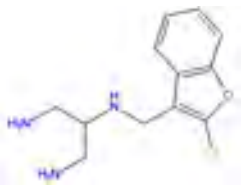 | [(2-chloro-1-benzofuran-3-yl)methyl](1,3-diaminopropan-2-yl)amine | **ACE2** | **-7.91669** | NCC(CN)NCc1c(Cl)oc2ccccc12 |
| 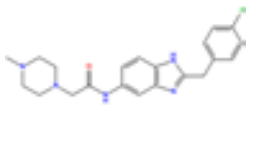 | N-{2-[(4-fluorophenyl)methyl]-1H-1,3-benzodiazol-5-yl}-2-(4-methylpiperazin-1-yl)acetamide | **ACE2** | **-7.82796** | CN1CCN(CC(=O)Nc2ccc3[nH]c(Cc4ccc(F)cc4)nc3c2)CC1 |
| 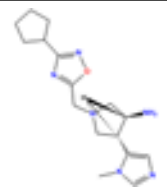 | rac-(3R,4S)-1-[(3-cyclopentyl-1,2,4-oxadiazol-5-yl)methyl]-4-(1-methyl-1H-imidazol-5-yl)pyrrolidin-3-amine | **ACE2** | **-7.82452** | |
| 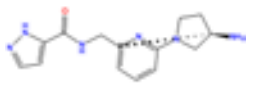 | N-{[6-(3-aminopyrrolidin-1-yl)pyridin-2-yl]methyl}-1H-pyrazole-5-carboxamide | **ACE2** | **-7.821383** | NC1CCN(C1)c1cccc(CNC(=O)c2ccn[nH]2)n1 |
| 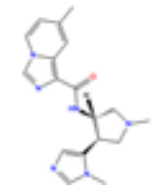 | 7-methyl-N-[(3S,4R)-1-methyl-4-(1-methyl-1H-imidazol-5-yl)pyrrolidin-3-yl]imidazo[1,5-a]pyridine-1-carboxamide | **ACE2** | **-7.856272** | CN1C[C@@H](NC(=O)c2ncn3ccc(C)cc23)[C@@H](C1)c1cncn1C |
| 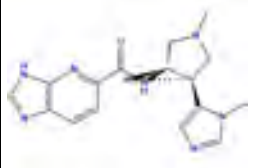 | N-[(3S,4R)-1-methyl-4-(1-methyl-1H-imidazol-5-yl)pyrrolidin-3-yl]-3H-imidazo[4,5-b]pyridine-5-carboxamide | **ACE2** | **-7.71168** | CN1C[C@@H](NC(=O)c2ccc3nc[nH]c3n2)[C@@H](C1)c1cncn1C |
| 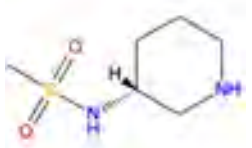 | N-(piperidin-3-yl)methanesulfonamide | **ACE2** | **-7.829874** | CS(=O)(=O)NC1CCCNC1 |
| 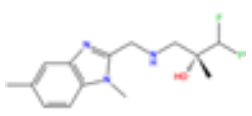 | 3-{[(1,5-dimethyl-1H-1,3-benzodiazol-2-yl)methyl]amino}-1,1-difluoro-2-methylpropan-2-ol | **ACE2** | **-7.784154** | Cc1ccc2n(C)c(CNCC(C)(O)C(F)F)nc2c1 |
| 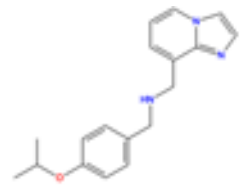 | ({imidazo[1,2-a]pyridin-8-yl}methyl)({[4-(propan-2-yloxy)phenyl]methyl})amine | **ACE2** | **-7.797499** | CC(C)Oc1ccc(CNCc2cccn3ccnc23)cc1 |
| 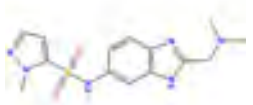 | N-{2-[(dimethylamino)methyl]-1H-1,3-benzodiazol-6-yl}-1-methyl-1H-pyrazole-5-sulfonamide | **ACE2** | **-7.921568** | CN(C)Cc1nc2ccc(NS(=O)(=O)c3ccnn3C)cc2[nH]1 |
| 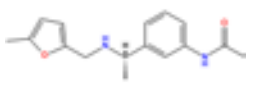 | N-[3-(1-{[(5-methylfuran-2-yl)methyl]amino}ethyl)phenyl]acetamide | **ACE2** | **-7.57716** | CC(NCc1ccc(C)o1)c1cccc(NC(C)=O)c1 |
| 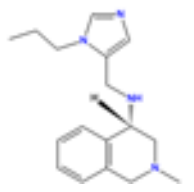 | 2-methyl-N-[(1-propyl-1H-imidazol-5-yl)methyl]-1,2,3,4-tetrahydroisoquinolin-4-amine | **ACE2** | **-7.915427** | CCCn1cncc1CNC1CN(C)Cc2ccccc12 |
| 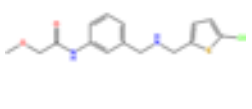 | N-[3-({[(5-chlorothiophen-2-yl)methyl]amino}methyl)phenyl]-2-methoxyacetamide | **ACE2** | **-7.445454** | COCC(=O)Nc1cccc(CNCc2ccc(Cl)s2)c1 |
| 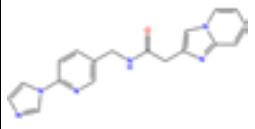 | 2-{imidazo[1,2-a]pyridin-2-yl}-N-{[6-(1H-imidazol-1-yl)pyridin-3-yl]methyl}acetamide | **ACE2** | **-7.33799** | O=C(Cc1cn2ccccc2n1)NCc1ccc(nc1)-n1ccnc1 |
| 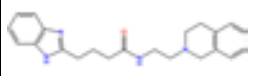 | 4-(1H-1,3-benzodiazol-2-yl)-N-[2-(1,2,3,4-tetrahydroisoquinolin-2-yl)ethyl]butanamide | **ACE2** | **-7.425942** | O=C(CCCc1nc2ccccc2[nH]1)NCCN1CCc2ccccc2C1 |
| 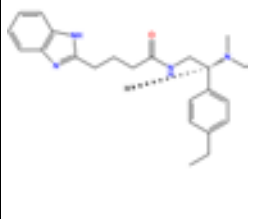 | 4-(1H-1,3-benzodiazol-2-yl)-N-[2-(dimethylamino)-2-(4-ethylphenyl)ethyl]butanamide | **ACE2** | **-7.538094** | CCc1ccc(cc1)C(CNC(=O)CCCc1nc2cccc c2[nH]1)N(C)C |
| 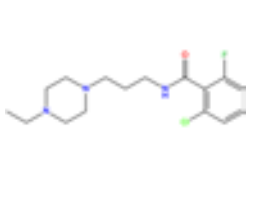 | 2-chloro-N-[3-(4-ethylpiperazin-1-yl)propyl]-6-fluorobenzamide | **ACE2** | **-7.878207** | CCN1CCN(CCCNC(=O)c2c(F)cccc2Cl)CC 1 |
| 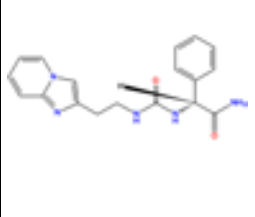 | 2-{[(2-{imidazo[1,2-a]pyridin-2-yl}ethyl)carbamoyl]amino}-2-phenylacetamide | **ACE2** | **-7.495692** | NC(=O)C(NC(=O)NCCc1cn2ccccc2n1)c 1ccccc1 |
| 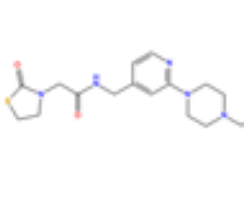 | N-{[2-(4-methylpiperazin-1-yl)pyridin-4-yl]methyl}-2-(2-oxo-1,3-thiazolidin-3-yl)acetamide | **ACE2** | **-7.935508** | CN1CCN(CC1)c1cc(CNC(=O)CN2CCSC2 =O)ccn1 |
| 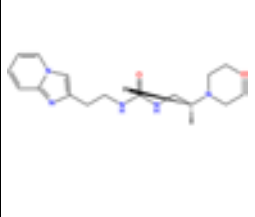 | 3-(2-{imidazo[1,2-a]pyridin-2-yl}ethyl)-1-[2-(morpholin-4-yl)propyl]urea | **ACE2** | **-7.478851** | CC(CNC(=O)NCCc1cn2ccccc2n1)N1CC OCC1 |
| 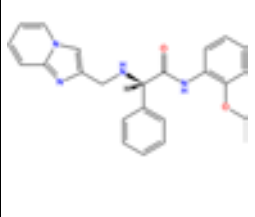 | N-(2-ethoxyphenyl)-2-[({imidazo[1,2-a]pyridin-2-yl}methyl)amino]-2-phenylacetamide | **ACE2** | **-7.382268** | CCOc1ccccc1NC(=O)C(NCc1cn2ccccc2 n1)c1ccccc1 |
| 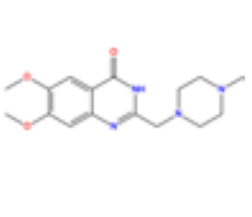 | 6,7-dimethoxy-2-[(4-methylpiperazin-1-yl)methyl]-3,4-dihydroquinazolin-4-one | **ACE2** | **-7.85665** | COc1cc2nc(CN3CCN(C)CC3)[nH]c(=O)c 2cc1OC |
| 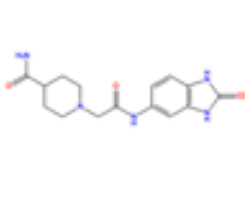 | 1-{[(2-oxo-2,3-dihydro-1H-1,3-benzodiazol-5-yl)carbamoyl]methyl}piperidine-4-carboxamide | **ACE2** | **-7.011648** | NC(=O)C1CCN(CC(=O)Nc2ccc3[nH]c(= O)[nH]c3c2)CC1 |
| 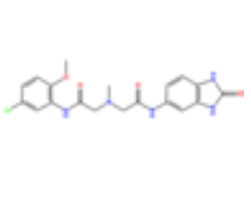 | 2-({[(5-chloro-2-methoxyphenyl)carbamoyl]methyl}(methyl)amino)-N-(2-oxo-2,3-dihydro-1H-1,3-benzodiazol-5-yl)acetamide | **ACE2** | **-7.666898** | COc1ccc(Cl)cc1NC(=O)CN(C)CC(=O)Nc 1ccc2[nH]c(=O)[nH]c2c1 |
| 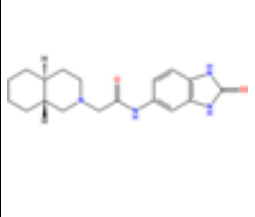 | 2-(decahydroisoquinolin-2-yl)-N-(2-oxo-2,3-dihydro-1H-1,3-benzodiazol-5-yl)acetamide | **ACE2** | **-7.392527** | O=C(CN1CCC2CCCCC2C1)Nc1ccc2[nH] c(=O)[nH]c2c1 |
| 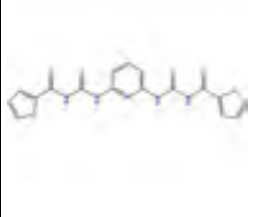 | 3-(furan-2-carbonyl)-1-[6-({[(furan-2-yl)formamido]methanethioyl}amino)pyridin-2-yl]thiourea | **ACE2** | **-7.848658** | O=C(NC(=S)Nc1cccc(NC(=S)NC(=O)c2c cco2)n1)c1ccco1 |
| 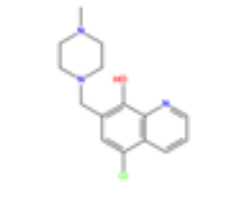 | 5-chloro-7-[(4-methylpiperazin-1-yl)methyl]quinolin-8-ol | **ACE2** | **-7.666059** | CN1CCN(Cc2cc(Cl)c3cccnc3c2O)CC1 |
| 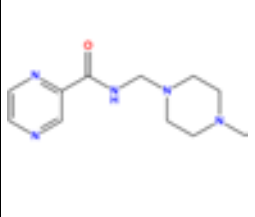 | N-[(4-methylpiperazin-1-yl)methyl]pyrazine-2-carboxamide | **ACE2** | **-7.566398** | CN1CCN(CNC(=O)c2cnccn2)CC1 |
| 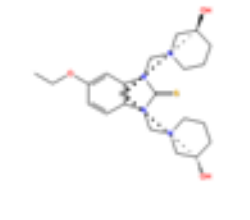 | 5-ethoxy-1,3-bis[(3-hydroxypiperidin-1-yl)methyl]-2,3-dihydro-1H-1,3-benzodiazole-2-thione | **ACE2** | **-7.6796** | CCOc1ccc2n(CN3CCCC(O)C3)c(=S)n(C N3CCCC(O)C3)c2c1 |
| 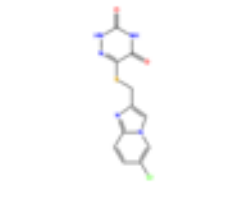 | 6-[({6-chloroimidazo[1,2-a]pyridin-2-yl}methyl)sulfanyl]-2,3,4,5-tetrahydro-1,2,4-triazine-3,5-dione | **ACE2** | **-7.577509** | Clc1ccc2nc(CSc3n[nH]c(=O)[nH]c3=O) cn2c1 |
| 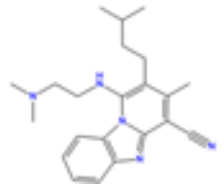 | 13-{[2-(dimethylamino)ethyl]amino}-11-methyl-12-(3-methylbutyl)-1,8-diazatricyclo$[7.4.0.0^{2,7}]$trideca-2,4,6,8,10,12-hexaene-10-carbonitrile | **ACE2** | **-7.305675** | CC(C)CCc1c(C)c(C#N)c2nc3ccccc3n2c1NCCN(C)C |
| 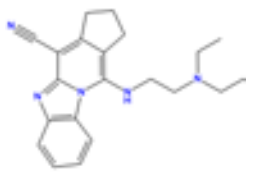 | 16-{[2-(diethylamino)ethyl]amino}-1,8-diazatetracyclo$[7.7.0.0^{2,7}.0^{11,15}]$hexadeca-2,4,6,8,10,15-hexaene-10-carbonitrile | **ACE2** | **-7.363593** | CCN(CC)CCNc1c2CCCc2c(C#N)c2nc3ccccc3n12 |
| 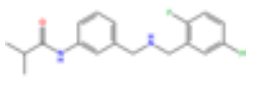 | N-[3-({[(2,5-difluorophenyl)methyl]amino}methyl)phenyl]-2-methylpropanamide | **ACE2** | **-7.412559** | CC(C)C(=O)Nc1cccc(CNCc2cc(F)ccc2F)c1 |
| 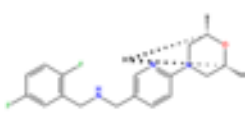 | [(2,5-difluorophenyl)methyl]({[6-(2,6-dimethylmorpholin-4-yl)pyridin-3-yl]methyl})amine | **ACE2** | **-7.830545** | CC1CN(CC(C)O1)c1ccc(CNCc2cc(F)ccc2F)cn1 |
| 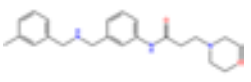 | N-[3-({[(3-methylphenyl)methyl]amino}methyl)phenyl]-3-(morpholin-4-yl)propanamide | **ACE2** | **-7.438191** | Cc1cccc(CNCc2cccc(NC(=O)CCN3CCOCC3)c2)c1 |
| 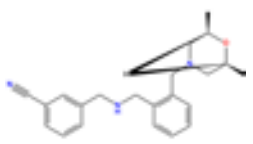 | 3-{[({2-[(2,6-dimethylmorpholin-4-yl)methyl]phenyl}methyl)amino]methyl}benzonitrile | **ACE2** | **-7.301904** | CC1CN(Cc2ccccc2CNCc2cccc(c2)C#N)CC(C)O1 |
| 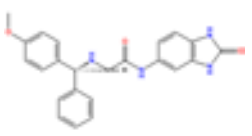 | 2-{[(4-methoxyphenyl)(phenyl)methyl]amino}-N-(2-oxo-2,3-dihydro-1H-1,3-benzodiazol-5-yl)acetamide | **ACE2** | **-7.302589** | COc1ccc(cc1)C(NCC(=O)Nc1ccc2[nH]c(=O)[nH]c2c1)c1ccccc1 |
| 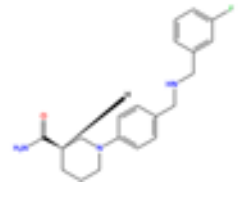 | 1-[4-({[(3-fluorophenyl)methyl]amino}methyl)phenyl]piperidine-3-carboxamide | **ACE2** | **-7.795565** | NC(=O)C1CCCN(C1)c1ccc(CNCc2cccc(F)c2)cc1 |
| 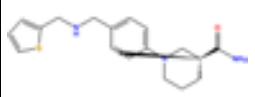 | 1-[4-({[(thiophen-2-yl)methyl]amino}methyl)phenyl]piperidine-3-carboxamide | **ACE2** | **-7.26034** | NC(=O)C1CCCN(C1)c1ccc(CNCc2cccs2)cc1 |
| 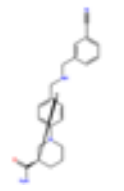 | 1-[4-({[(3-cyanophenyl)methyl]amino}methyl)phenyl]piperidine-3-carboxamide | **ACE2** | **-7.247289** | NC(=O)C1CCCN(C1)c1ccc(CNCc2cccc(c2)C#N)cc1 |
| 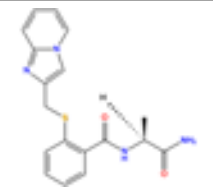 | 2-({2-[({imidazo[1,2-a]pyridin-2-yl}methyl)sulfanyl]phenyl}formamido)propanamide | **ACE2** | **-7.695058** | CC(NC(=O)c1ccccc1SCc1cn2ccccc2n1)C(N)=O |
| 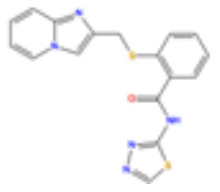 | 2-[({imidazo[1,2-a]pyridin-2-yl}methyl)sulfanyl]-N-(1,3,4-thiadiazol-2-yl)benzamide | **ACE2** | **-7.843258** | O=C(Nc1nncs1)c1ccccc1SCc1cn2ccccc2n1 |
| 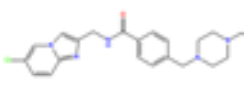 | N-({6-chloroimidazo[1,2-a]pyridin-2-yl}methyl)-4-[(4-methylpiperazin-1-yl)methyl]benzamide | **ACE2** | **-7.623186** | CN1CCN(Cc2ccc(cc2)C(=O)NCc2cn3cc(Cl)ccc3n2)CC1 |
| 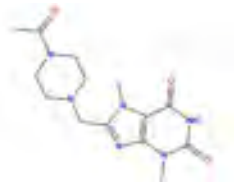 | 8-[(4-acetylpiperazin-1-yl)methyl]-3,7-dimethyl-2,3,6,7-tetrahydro-1H-purine-2,6-dione | **ACE2** | **-7.252306** | CC(=O)N1CCN(Cc2nc3n(C)c(=O)[nH]c(=O)c3n2C)CC1 |
| 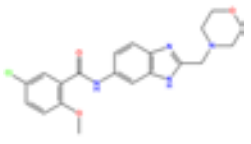 | 5-chloro-2-methoxy-N-{2-[(morpholin-4-yl)methyl]-1H-1,3-benzodiazol-6-yl}benzamide | **ACE2** | **-7.268946** | COc1ccc(Cl)cc1C(=O)Nc1ccc2nc(CN3CCOCC3)[nH]c2c1 |
| 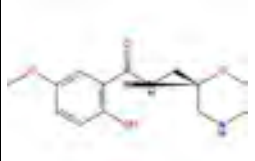 | 2-hydroxy-5-methoxy-N-[(morpholin-2-yl)methyl]benzamide | **ACE2** | **-7.411243** | COc1ccc(O)c(c1)C(=O)NCC1CNCCO1 |
| 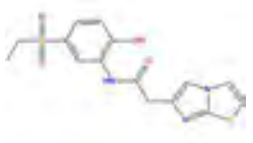 | N-[5-(ethanesulfonyl)-2-hydroxyphenyl]-2-{imidazo[2,1-b][1,3]thiazol-6-yl}acetamide | **ACE2** | **-7.304225** | CCS(=O)(=O)c1ccc(O)c(NC(=O)Cc2cn3ccsc3n2)c1 |
| 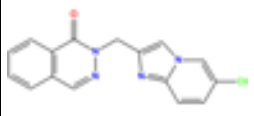 | 2-({6-chloroimidazo[1,2-a]pyridin-2-yl}methyl)-1,2-dihydrophthalazin-1-one | **ACE2** | **-7.361447** | Clc1ccc2nc(Cn3ncc4ccccc4c3=O)cn2c1 |
| 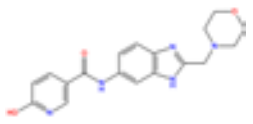 | 6-hydroxy-N-{2-[(morpholin-4-yl)methyl]-1H-1,3-benzodiazol-6-yl}pyridine-3-carboxamide | **ACE2** | **-7.398419** | Oc1ccc(cn1)C(=O)Nc1ccc2nc(CN3CCOCC3)[nH]c2c1 |
| 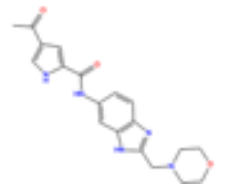 | 4-acetyl-N-{2-[(morpholin-4-yl)methyl]-1H-1,3-benzodiazol-6-yl}-1H-pyrrole-2-carboxamide | **ACE2** | **-7.627306** | CC(=O)c1c[nH]c(c1)C(=O)Nc1ccc2nc(CN3CCOCC3)[nH]c2c1 |
| 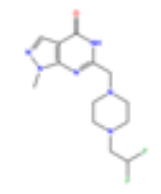 | 6-{[4-(2,2-difluoroethyl)piperazin-1-yl]methyl}-1-methyl-1H,4H,5H-pyrazolo[3,4-d]pyrimidin-4-one | **ACE2** | **-7.78928** | Cn1ncc2c1nc(CN1CCN(CC(F)F)CC1)[nH]c2=O |
| 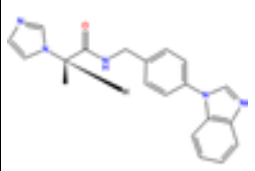 | N-{[4-(1H-1,3-benzodiazol-1-yl)phenyl]methyl}-2-(1H-imidazol-1-yl)propanamide | **ACE2** | **-7.800908** | CC(C(=O)NCc1ccc(cc1)-n1cnc2ccccc12)n1ccnc1 |
| 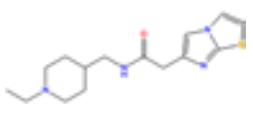 | N-[(1-ethylpiperidin-4-yl)methyl]-2-{imidazo[2,1-b][1,3]thiazol-6-yl}acetamide | **ACE2** | **-7.597419** | CCN1CCC(CNC(=O)Cc2cn3ccsc3n2)CC1 |
| 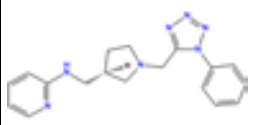 | N-({1-[(1-phenyl-1H-1,2,3,4-tetrazol-5-yl)methyl]pyrrolidin-3-yl}methyl)pyridin-2-amine | **ACE2** | **-7.576307** | C(Nc1ccccn1)C1CCN(Cc2nnnn2-c2ccccc2)C1 |
| 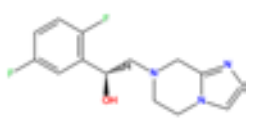 | 1-(2,5-difluorophenyl)-2-{5H,6H,7H,8H-imidazo[1,2-a]pyrazin-7-yl}ethan-1-ol | **ACE2** | **-7.881457** | OC(CN1CCn2ccnc2C1)c1cc(F)ccc1F |
| 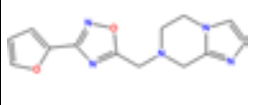 | 3-(furan-2-yl)-5-({5H,6H,7H,8H-imidazo[1,2-a]pyrazin-7-yl}methyl)-1,2,4-oxadiazole | **ACE2** | **-7.871089** | C(N1CCn2ccnc2C1)c1nc(no1)-c1ccco1 |
| 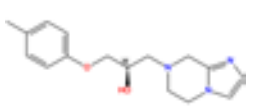 | 1-{5H,6H,7H,8H-imidazo[1,2-a]pyrazin-7-yl}-3-(4-methylphenoxy)propan-2-ol | **ACE2** | **-7.796023** | Cc1ccc(OCC(O)CN2CCn3ccnc3C2)cc1 |
| 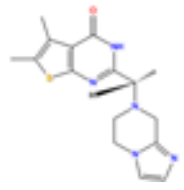 | 2-(1-{5H,6H,7H,8H-imidazo[1,2-a]pyrazin-7-yl}ethyl)-5,6-dimethyl-3H,4H-thieno[2,3-d]pyrimidin-4-one | **ACE2** | **-7.890335** | CC(N1CCn2ccnc2C1)c1nc2sc(C)c(C)c2c(=O)[nH]1 |
| 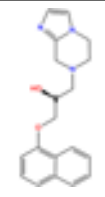 | 1-{5H,6H,7H,8H-imidazo[1,2-a]pyrazin-7-yl}-3-(naphthalen-1-yloxy)propan-2-ol | **ACE2** | **-7.909158** | OC(COc1cccc2ccccc12)CN1CCn2ccnc2C1 |
| 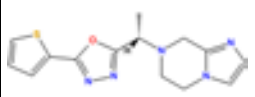 | 2-(1-{5H,6H,7H,8H-imidazo[1,2-a]pyrazin-7-yl}ethyl)-5-(thiophen-2-yl)-1,3,4-oxadiazole | **ACE2** | **-7.595696** | CC(N1CCn2ccnc2C1)c1nnc(o1)-c1cccs1 |
| 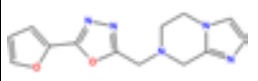 | 2-(furan-2-yl)-5-({5H,6H,7H,8H-imidazo[1,2-a]pyrazin-7-yl}methyl)-1,3,4-oxadiazole | **ACE2** | **-7.402563** | C(N1CCn2ccnc2C1)c1nnc(o1)-c1ccco1 |
| 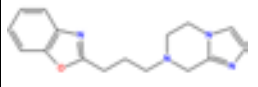 | 2-(3-{5H,6H,7H,8H-imidazo[1,2-a]pyrazin-7-yl}propyl)-1,3-benzoxazole | **ACE2** | **-7.644226** | C(CN1CCn2ccnc2C1)Cc1nc2ccccc2o1 |
| 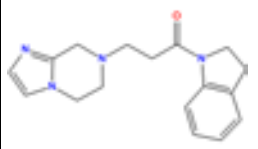 | 1-(2,3-dihydro-1H-indol-1-yl)-3-{5H,6H,7H,8H-imidazo[1,2-a]pyrazin-7-yl}propan-1-one | **ACE2** | **-7.778121** | O=C(CCN1CCn2ccnc2C1)N1CCc2ccccc12 |
| 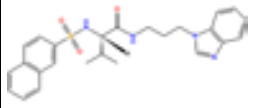 | N-[3-(1H-1,3-benzodiazol-1-yl)propyl]-3-methyl-2-(naphthalene-2-sulfonamido)butanamide | **MPRO** | **-7.638746** | CC(C)C(NS(=O)(=O)c1ccc2ccccc2c1)C(=O)NCCCn1cnc2ccccc12 |
| 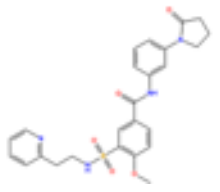 | 4-methoxy-N-[3-(2-oxopyrrolidin-1-yl)phenyl]-3-{[2-(pyridin-2-yl)ethyl]sulfamoyl}benzamide | **MPRO** | **-7.644347** | COc1ccc(cc1S(=O)(=O)NCCc1ccccn1)C(=O)Nc1cccc(c1)N1CCCC1=O |
| 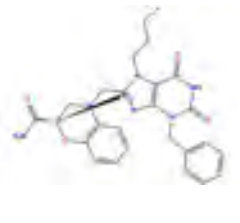 | 4-[(3-benzyl-7-butyl-2,6-dioxo-2,3,6,7-tetrahydro-1H-purin-8-yl)methyl]-3,4-dihydro-2H-1,4-benzoxazine-2-carboxamide | **MPRO** | **-7.759647** | CCCCn1c(CN2CC(Oc3ccccc23)C(N)=O)nc2n(Cc3ccccc3)c(=O)[nH]c(=O)c12 |
| 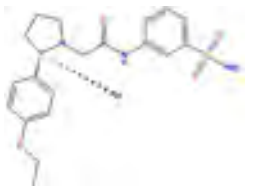 | 2-[2-(4-ethoxyphenyl)pyrrolidin-1-yl]-N-(3-sulfamoylphenyl)acetamide | **MPRO** | **-7.748233** | CCOc1ccc(cc1)C1CCCN1CC(=O)Nc1cccc(c1)S(N)(=O)=O |
| 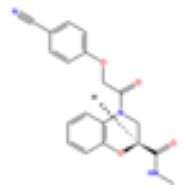 | 4-[2-(4-cyanophenoxy)acetyl]-N-methyl-3,4-dihydro-2H-1,4-benzoxazine-2-carboxamide | **MPRO** | **-7.673447** | CNC(=O)C1CN(C(=O)COc2ccc(cc2)C#N)c2ccccc2O1 |
| 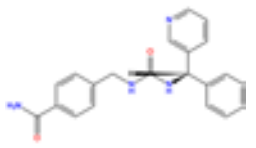 | 4-[({[phenyl(pyridin-3-yl)methyl]carbamoyl}amino)methyl]benzamide | **MPRO** | **-7.749195** | NC(=O)c1ccc(CNC(=O)NC(c2ccccc2)c2cccnc2)cc1 |
| 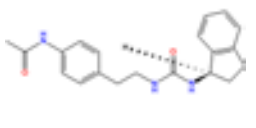 | N-[4-(2-{[(2,3-dihydro-1H-inden-1-yl)carbamoyl]amino}ethyl)phenyl]acetamide | **MPRO** | **-7.640749** | CC(=O)Nc1ccc(CCNC(=O)NC2CCc3ccccc23)cc1 |
| 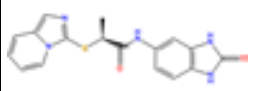 | 2-{imidazo[1,5-a]pyridin-3-ylsulfanyl}-N-(2-oxo-2,3-dihydro-1H-1,3-benzodiazol-5-yl)propanamide | **MPRO** | **-7.711375** | CC(Sc1ncc2ccccn12)C(=O)Nc1ccc2[nH]c(=O)[nH]c2c1 |
| 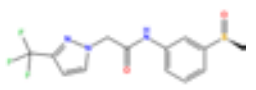 | N-(3-methanesulfinylphenyl)-2-[3-(trifluoromethyl)-1H-pyrazol-1-yl]acetamide | **MPRO** | **-7.618535** | CS(=O)c1cccc(NC(=O)Cn2ccc(n2)C(F)(F)F)c1 |
| 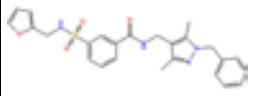 | N-[(1-benzyl-3,5-dimethyl-1H-pyrazol-4-yl)methyl]-3-{[(furan-2-yl)methyl]sulfamoyl}benzamide | **MPRO** | **-7.646959** | Cc1nn(Cc2ccccc2)c(C)c1CNC(=O)c1cccc(c1)S(=O)(=O)NCc1ccco1 |
| 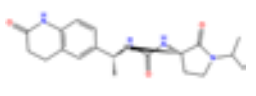 | 1-[1-(2-oxo-1,2,3,4-tetrahydroquinolin-6-yl)ethyl]-3-[2-oxo-1-(propan-2-yl)pyrrolidin-3-yl]urea | **MPRO** | **-7.6444** | CC(C)N1CCC(NC(=O)NC(C)c2ccc3NC(=O)CCc3c2)C1=O |
| 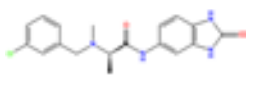 | 2-{[(3-chlorophenyl)methyl](methyl)amino}-N-(2-oxo-2,3-dihydro-1H-1,3-benzodiazol-5-yl)propanamide | **MPRO** | **-7.705485** | CC(N(C)Cc1cccc(Cl)c1)C(=O)Nc1ccc2[nH]c(=O)[nH]c2c1 |
| 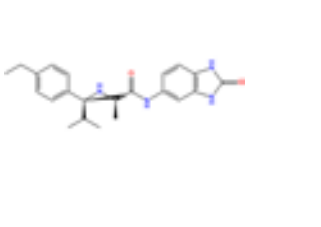 | 2-{[1-(4-ethylphenyl)-2-methylpropyl]amino}-N-(2-oxo-2,3-dihydro-1H-1,3-benzodiazol-5-yl)propanamide | **MPRO** | **-7.654702** | CCc1ccc(cc1)C(NC(C)C(=O)Nc1ccc2[nH]c(=O)[nH]c2c1)C(C)C |
| 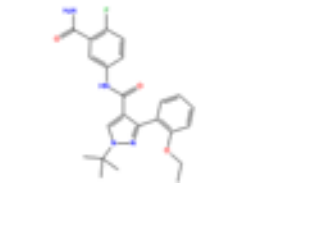 | 1-tert-butyl-N-(3-carbamoyl-4-fluorophenyl)-3-(2-ethoxyphenyl)-1H-pyrazole-4-carboxamide | **MPRO** | **-7.656709** | CCOc1ccccc1-c1nn(cc1C(=O)Nc1ccc(F)c(c1)C(N)=O)C(C)(C)C |
| 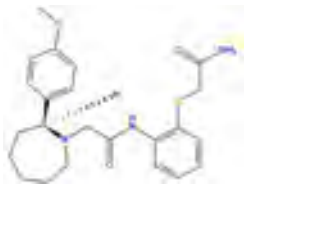 | 2-[(2-{2-[2-(4-methoxyphenyl)azepan-1-yl]acetamido}phenyl)sulfanyl]acetamide | **MPRO** | **-7.739322** | COc1ccc(cc1)C1CCCCCN1CC(=O)Nc1ccccc1SCC(N)=O |
| 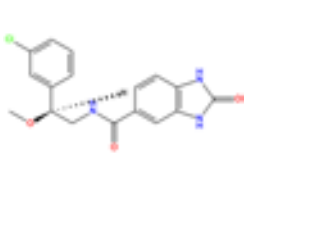 | N-[2-(3-chlorophenyl)-2-methoxyethyl]-2-oxo-2,3-dihydro-1H-1,3-benzodiazole-5-carboxamide | **MPRO** | **-7.666192** | COC(CNC(=O)c1ccc2[nH]c(=O)[nH]c2c1)c1cccc(Cl)c1 |
| 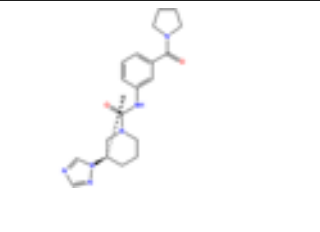 | N-[3-(pyrrolidine-1-carbonyl)phenyl]-3-(1H-1,2,4-triazol-1-yl)piperidine-1-carboxamide | **MPRO** | **-7.708583** | O=C(Nc1cccc(c1)C(=O)N1CCCC1)N1CCCC(C1)n1cncn1 |
| 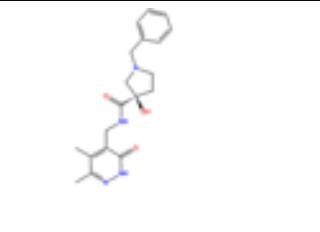 | 1-benzyl-3-hydroxy-N-[(3-hydroxy-5,6-dimethylpyridazin-4-yl)methyl]pyrrolidine-3-carboxamide | **MPRO** | **-7.768783** | Cc1nnc(O)c(CNC(=O)C2(O)CCN(Cc3ccccc3)C2)c1C |
| 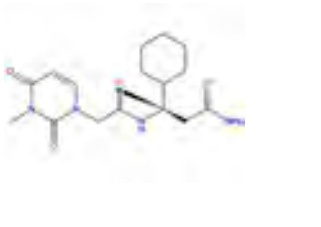 | 3-cyclohexyl-3-[2-(3-methyl-2,4-dioxo-1,2,3,4-tetrahydropyrimidin-1-yl)acetamido]propanamide | **MPRO** | **-7.651972** | Cn1c(=O)ccn(CC(=O)NC(CC(N)=O)C2CCCCC2)c1=O |
| 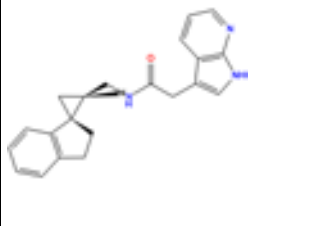 | N-({2',3'-dihydrospiro[cyclopropane-1,1'-inden]-3-yl}methyl)-2-{1H-pyrrolo[2,3-b]pyridin-3-yl}acetamide | **MPRO** | **-7.679692** | O=C(Cc1c[nH]c2ncccc12)NCC1CC11CCc2ccccc12 |
| 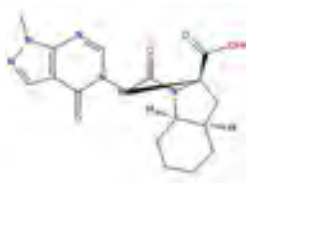 | 1-(2-{1-methyl-4-oxo-1H,4H,5H-pyrazolo[3,4-d]pyrimidin-5-yl}acetyl)-octahydro-1H-indole-2-carboxylic acid | **MPRO** | **-7.705968** | Cn1ncc2c1ncn(CC(=O)N1C3CCCCC3CC1C(O)=O)c2=O |
| 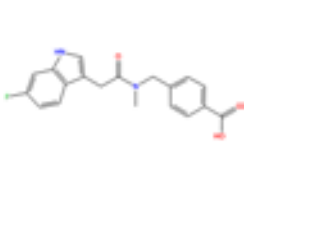 | 4-{[2-(6-fluoro-1H-indol-3-yl)-N-methylacetamido]methyl}benzoic acid | **MPRO** | **-7.737004** | CN(Cc1ccc(cc1)C(O)=O)C(=O)Cc1c[nH]c2cc(F)ccc12 |
| 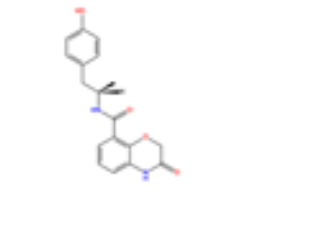 | N-[1-(4-hydroxyphenyl)propan-2-yl]-3-oxo-3,4-dihydro-2H-1,4-benzoxazine-8-carboxamide | **MPRO** | **-7.626542** | CC(Cc1ccc(O)cc1)NC(=O)c1cccc2NC(=O)COc12 |
| 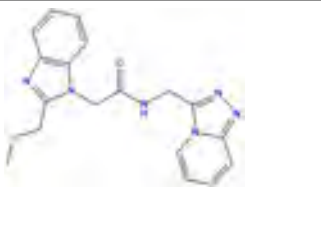 | 2-{2-[(methylsulfanyl)methyl]-1H-1,3-benzodiazol-1-yl}-N-({[1,2,4]triazolo[4,3-a]pyridin-3-yl}methyl)acetamide | **MPRO** | **-7.632684** | CSCc1nc2ccccc2n1CC(=O)NCc1nnc2ccccn12 |
| 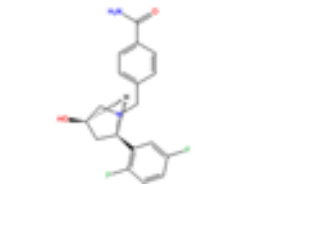 | 4-{[2-(2,5-difluorophenyl)-4-hydroxypyrrolidin-1-yl]methyl}benzamide | **MPRO** | **-7.676844** | NC(=O)c1ccc(CN2CC(O)CC2c2cc(F)ccc2F)cc1 |
| 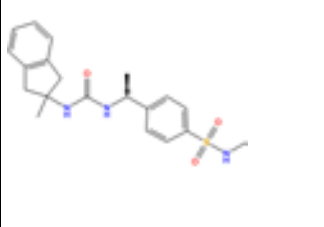 | 3-(2-methyl-2,3-dihydro-1H-inden-2-yl)-1-{1-[4-(methylsulfamoyl)phenyl]ethyl}urea | **MPRO** | **-7.75727** | CNS(=O)(=O)c1ccc(cc1)C(C)NC(=O)NC1(C)Cc2ccccc2C1 |
| 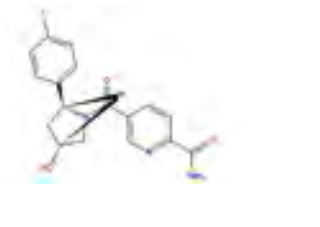 | 5-[2-(4-fluorophenyl)-4-hydroxypyrrolidine-1-carbonyl]pyridine-2-carboxamide | **MPRO** | **-7.718797** | NC(=O)c1ccc(cn1)C(=O)N1CC(O)CC1c1ccc(F)cc1 |
| 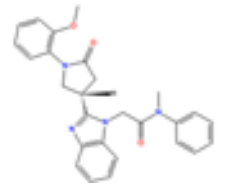 | 2-{2-[1-(2-methoxyphenyl)-5-oxopyrrolidin-3-yl]-1H-1,3-benzodiazol-1-yl}-N-methyl-N-phenylacetamide | **MPRO** | **-7.687013** | COc1ccccc1N1CC(CC1=O)c1nc2ccccc2n1CC(=O)N(C)c1ccccc1 |
| 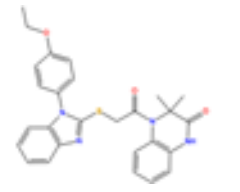 | 4-(2-{[1-(4-ethoxyphenyl)-1H-1,3-benzodiazol-2-yl]sulfanyl}acetyl)-3,3-dimethyl-1,2,3,4-tetrahydroquinoxalin-2-one | **MPRO** | **-7.629347** | CCOc1ccc(cc1)-n1c(SCC(=O)N2c3ccccc3NC(=O)C2(C)C)nc2ccccc12 |
| 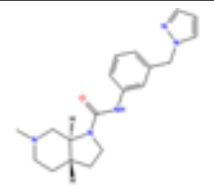 | 6-methyl-N-{3-[(1H-pyrazol-1-yl)methyl]phenyl}-octahydro-1H-pyrrolo[2,3-c]pyridine-1-carboxamide | **MPRO** | **-7.762516** | CN1CCC2CCN(C2C1)C(=O)Nc1cccc(Cn2cccn2)c1 |
| 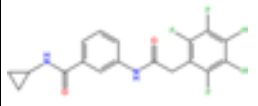 | N-cyclopropyl-3-[2-(2,3,4,5,6-pentafluorophenyl)acetamido]benzamide | **MPRO** | **-7.647161** | Fc1c(F)c(F)c(CC(=O)Nc2cccc(c2)C(=O)NC2CC2)c(F)c1F |
| 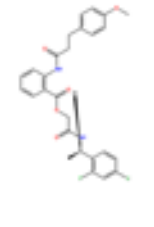 | {[1-(2,4-difluorophenyl)ethyl]carbamoyl}methyl 2-[3-(4-methoxyphenyl)propanamido]benzoate | **MPRO** | **-7.620529** | COc1ccc(CCC(=O)Nc2ccccc2C(=O)OCC(=O)NC(C)c2ccc(F)cc2F)cc1 |
| 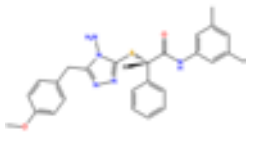 | 2-({4-amino-5-[(4-methoxyphenyl)methyl]-4H-1,2,4-triazol-3-yl}sulfanyl)-N-(3,5-dimethylphenyl)-2-phenylacetamide | **MPRO** | **-7.684616** | COc1ccc(Cc2nnc(SC(C(=O)Nc3cc(C)cc(C)c3)c3ccccc3)n2N)cc1 |
| 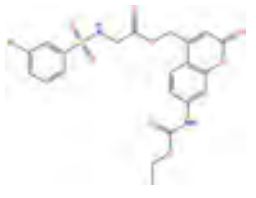 | {7-[(ethoxycarbonyl)amino]-2-oxo-2H-chromen-4-yl}methyl 2-(3-bromobenzenesulfonamido)acetate | **MPRO** | **-7.636072** | CCOC(=O)Nc1ccc2c(COC(=O)CNS(=O)(=O)c3cccc(Br)c3)cc(=O)oc2c1 |
| 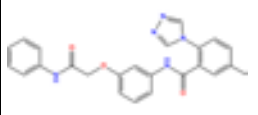 | 5-methyl-N-{3-[(phenylcarbamoyl)methoxy]phenyl}-2-(4H-1,2,4-triazol-4-yl)benzamide | **MPRO** | **-7.70219** | Cc1ccc(c(c1)C(=O)Nc1cccc(OCC(=O)Nc2ccccc2)c1)-n1cnnc1 |
| 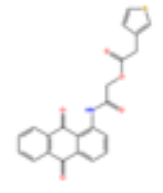 | [(9,10-dioxo-9,10-dihydroanthracen-1-yl)carbamoyl]methyl 2-(thiophen-3-yl)acetate | **MPRO** | **-7.757383** | O=C(COC(=O)Cc1ccsc1)Nc1cccc2C(=O)c3ccccc3C(=O)c12 |
| 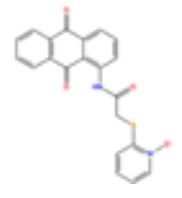 | 2-({[(9,10-dioxo-9,10-dihydroanthracen-1-yl)carbamoyl]methyl}sulfanyl)pyridin-1-ium-1-olate | **MPRO** | **-7.802135** | [O-][n+]1ccccc1SCC(=O)Nc1cccc2C(=O)c3ccccc3C(=O)c12 |
| 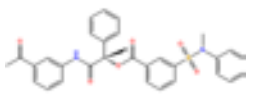 | [(3-acetylphenyl)carbamoyl](phenyl)methyl 3-[methyl(phenyl)sulfamoyl]benzoate | **MPRO** | **-7.784198** | CN(c1ccccc1)S(=O)(=O)c1cccc(c1)C(=O)OC(C(=O)Nc1cccc(c1)C(C)=O)c1ccccc1 |
| 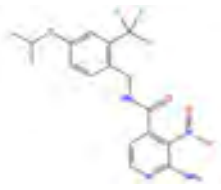 | 2-amino-3-nitro-N-{[4-(propan-2-yloxy)-2-(trifluoromethyl)phenyl]methyl}pyridine-4-carboxamide | **MPRO** | **-7.691484** | CC(C)Oc1ccc(CNC(=O)c2ccnc(N)c2[N+]([O-])=O)c(c1)C(F)(F)F |
| 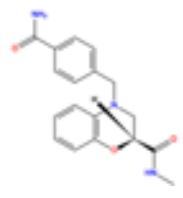 | 4-[(4-carbamoylphenyl)methyl]-N-methyl-3,4-dihydro-2H-1,4-benzoxazine-2-carboxamide | **MPRO** | **-7.658907** | CNC(=O)C1CN(Cc2ccc(cc2)C(N)=O)c2ccccc2O1 |
| 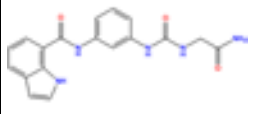 | N-(3-{[(carbamoylmethyl)carbamoyl]amino}phenyl)-1H-indole-7-carboxamide | **MPRO** | **-7.68726** | NC(=O)CNC(=O)Nc1cccc(NC(=O)c2cccc3cc[nH]c23)c1 |
| 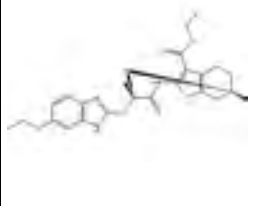 | ethyl 2-{2-[(6-ethoxy-1H-1,3-benzodiazol-2-yl)sulfanyl]propanamido}-6-methyl-4,5,6,7-tetrahydro-1-benzothiophene-3-carboxylate | **MPRO** | **-7.618485** | CCOC(=O)c1c(NC(=O)C(C)Sc2nc3ccc(OCC)cc3[nH]2)sc2CC(C)CCc12 |
| 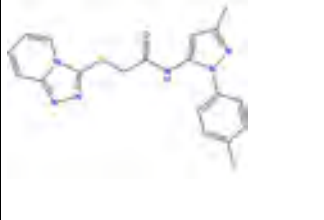 | N-[1-(4-fluorophenyl)-3-methyl-1H-pyrazol-5-yl]-2-{[1,2,4]triazolo[4,3-a]pyridin-3-ylsulfanyl}acetamide | **MPRO** | **-7.703882** | Cc1cc(NC(=O)CSc2nnc3ccccn23)n(n1)-c1ccc(F)cc1 |
| 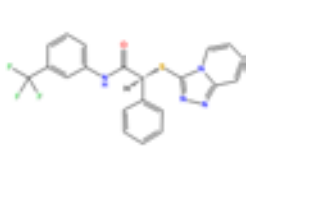 | 2-phenyl-2-{[1,2,4]triazolo[4,3-a]pyridin-3-ylsulfanyl}-N-[3-(trifluoromethyl)phenyl]acetamide | **MPRO** | **-7.618894** | FC(F)(F)c1cccc(NC(=O)C(Sc2nnc3ccccn23)c2ccccc2)c1 |
| 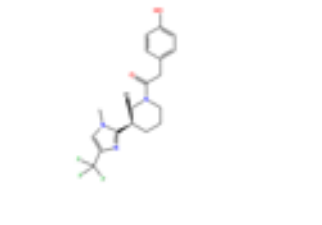 | 2-(4-hydroxyphenyl)-1-{3-[1-methyl-4-(trifluoromethyl)-1H-imidazol-2-yl]piperidin-1-yl}ethan-1-one | **MPRO** | **-7.722702** | Cn1cc(nc1C1CCCN(C1)C(=O)Cc1ccc(O)cc1)C(F)(F)F |
| 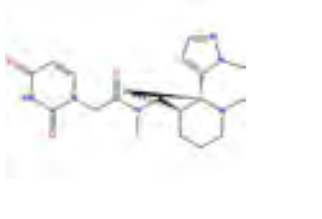 | 2-(2,4-dioxo-1,2,3,4-tetrahydropyrimidin-1-yl)-N-methyl-N-{[(2R,3S)-1-methyl-2-(1-methyl-1H-pyrazol-5-yl)piperidin-3-yl]methyl}acetamide | **MPRO** | **-7.712632** | CN(C[C@@H]1CCCN(C)[C@H]1c1ccnn1C)C(=O)Cn1ccc(=O)[nH]c1=O |
| 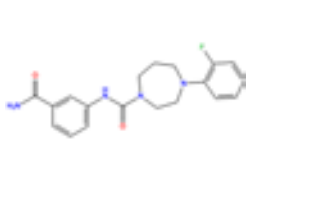 | N-(3-carbamoylphenyl)-4-(2-fluorophenyl)-1,4-diazepane-1-carboxamide | **MPRO** | **-7.652268** | NC(=O)c1cccc(NC(=O)N2CCCN(CC2)c2ccccc2F)c1 |
| 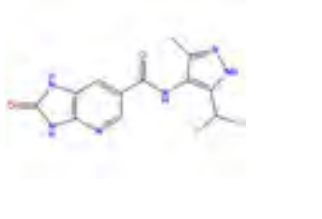 | N-[3-(difluoromethyl)-5-methyl-1H-pyrazol-4-yl]-2-oxo-1H,2H,3H-imidazo[4,5-b]pyridine-6-carboxamide | **MPRO** | **-7.796011** | Cc1[nH]nc(C(F)F)c1NC(=O)c1cnc2[nH]c(=O)[nH]c2c1 |
| 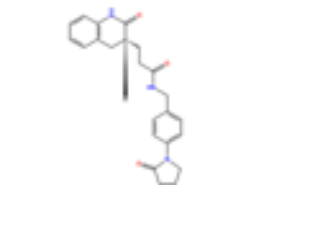 | 3-(2-oxo-1,2,3,4-tetrahydroquinolin-3-yl)-N-{[4-(2-oxopyrrolidin-1-yl)phenyl]methyl}propanamide | **MPRO** | **-7.747528** | O=C(CCC1Cc2ccccc2NC1=O)NCc1ccc(cc1)N1CCCC1=O |
| 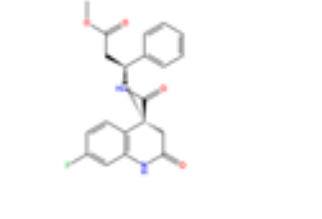 | methyl 3-[(7-fluoro-2-oxo-1,2,3,4-tetrahydroquinolin-4-yl)formamido]-3-phenylpropanoate | **MPRO** | **-7.771538** | COC(=O)CC(NC(=O)C1CC(=O)Nc2cc(F)ccc12)c1ccccc1 |
| 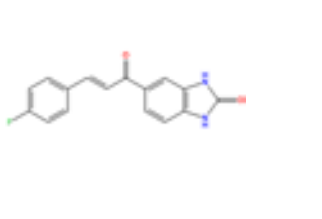 | 5-[3-(4-fluorophenyl)prop-2-enoyl]-2,3-dihydro-1H-1,3-benzodiazol-2-one | **MPRO** | **-7.719031** | Fc1ccc(C=CC(=O)c2ccc3[nH]c(=O)[nH]c3c2)cc1 |
| 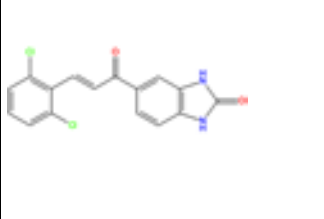 | 5-[3-(2,6-dichlorophenyl)prop-2-enoyl]-2,3-dihydro-1H-1,3-benzodiazol-2-one | **MPRO** | **-7.693111** | Clc1cccc(Cl)c1C=CC(=O)c1ccc2[nH]c(=O)[nH]c2c1 |
| 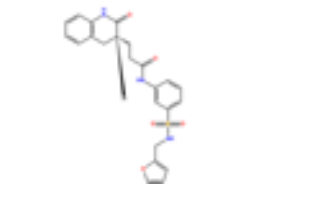 | N-(3-{[(furan-2-yl)methyl]sulfamoyl}phenyl)-3-(2-oxo-1,2,3,4-tetrahydroquinolin-3-yl)propanamide | **MPRO** | **-7.706308** | O=C(CCC1Cc2ccccc2NC1=O)Nc1cccc(c1)S(=O)(=O)NCc1ccco1 |
| 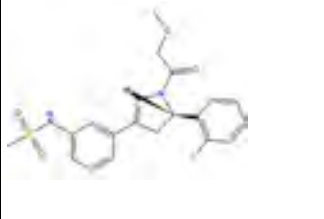 | N-{3-[5-(2-fluorophenyl)-1-(2-methoxyacetyl)-4,5-dihydro-1H-pyrazol-3-yl]phenyl}methanesulfonamide | **MPRO** | **-7.733552** | COCC(=O)N1N=C(CC1c1ccccc1F)c1cccc(NS(C)(=O)=O)c1 |
| 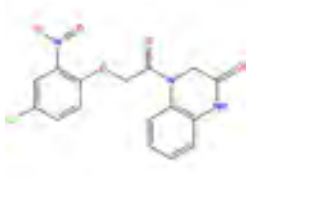 | 4-[2-(4-chloro-2-nitrophenoxy)acetyl]-1,2,3,4-tetrahydroquinoxalin-2-one | **MPRO** | **-7.749496** | [O-][N+](=O)c1cc(Cl)ccc1OCC(=O)N1CC(=O)Nc2ccccc12 |
| 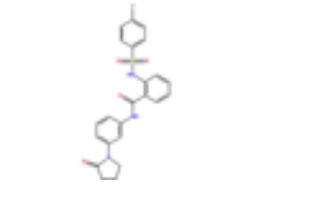 | 2-(4-fluorobenzenesulfonamido)-N-[3-(2-oxopyrrolidin-1-yl)phenyl]benzamide | **MPRO** | **-7.643129** | Fc1ccc(cc1)S(=O)(=O)Nc1ccccc1C(=O)Nc1cccc(c1)N1CCCC1=O |
| 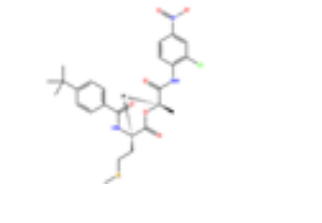 | 1-[(2-chloro-4-nitrophenyl)carbamoyl]ethyl 2-[(4-tert-butylphenyl)formamido]-4-(methylsulfanyl)butanoate | **MPRO** | **-7.697784** | CSCCC(NC(=O)c1ccc(cc1)C(C)(C)C)C(=O)OC(C)C(=O)Nc1ccc(cc1Cl)[N+]([O-])=O |
| 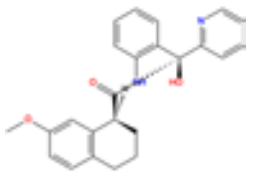 | N-{2-[hydroxy(pyridin-2-yl)methyl]phenyl}-7-methoxy-1,2,3,4-tetrahydronaphthalene-1-carboxamide | **MPRO** | **-7.680547** | COc1ccc2CCCC(C(=O)Nc3ccccc3C(O)c3ccccn3)c2c1 |
| 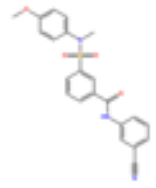 | N-(3-cyanophenyl)-3-[(4-methoxyphenyl)(methyl)sulfamoyl]benzamide | **MPRO** | **-7.608325** | COc1ccc(cc1)N(C)S(=O)(=O)c1cccc(c1)C(=O)Nc1cccc(c1)C#N |
| 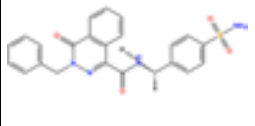 | 3-benzyl-4-oxo-N-[1-(4-sulfamoylphenyl)ethyl]-3,4-dihydrophthalazine-1-carboxamide | **MPRO** | **-7.71278** | CC(NC(=O)c1nn(Cc2ccccc2)c(=O)c2ccccc12)c1ccc(cc1)S(N)(=O)=O |
| 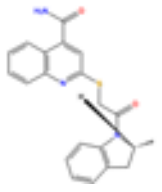 | 2-{[2-(2-methyl-2,3-dihydro-1H-indol-1-yl)-2-oxoethyl]sulfanyl}quinoline-4-carboxamide | **MPRO** | **-7.642768** | CC1Cc2ccccc2N1C(=O)CSc1cc(C(N)=O)c2ccccc2n1 |
| 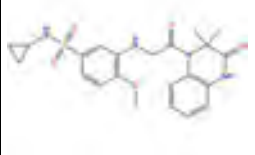 | N-cyclopropyl-3-{[2-(2,2-dimethyl-3-oxo-1,2,3,4-tetrahydroquinoxalin-1-yl)-2-oxoethyl]amino}-4-methoxybenzene-1-sulfonamide | **MPRO** | **-7.783271** | COc1ccc(cc1NCC(=O)N1c2ccccc2NC(=O)C1(C)C)S(=O)(=O)NC1CC1 |
| 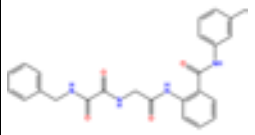 | N'-benzyl-N-[({2-[(3-methylphenyl)carbamoyl]phenyl}carbamoyl)methyl]ethanediamide | **MPRO** | **-7.640353** | Cc1cccc(NC(=O)c2ccccc2NC(=O)CNC(=O)C(=O)NCc2ccccc2)c1 |
| 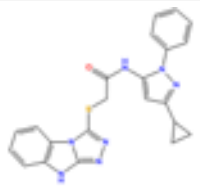 | N-(3-cyclopropyl-1-phenyl-1H-pyrazol-5-yl)-2-{2,4,5,7-tetraazatricyclo[6.4.0.0$^{2,6}$]dodeca-1(12),3,5,8,10-pentaen-3-ylsulfanyl}acetamide | **MPRO** | **-7.809379** | O=C(CSc1nnc2[nH]c3ccccc3n12)Nc1cc(nn1-c1ccccc1)C1CC1 |
| 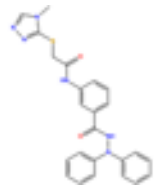 | N-[3-(N',N'-diphenylhydrazinecarbonyl)phenyl]-2-[(4-methyl-4H-1,2,4-triazol-3-yl)sulfanyl]acetamide | **MPRO** | **-7.719529** | Cn1cnnc1SCC(=O)Nc1cccc(c1)C(=O)NN(c1ccccc1)c1ccccc1 |
| 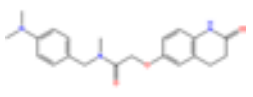 | N-{[4-(dimethylamino)phenyl]methyl}-N-methyl-2-[(2-oxo-1,2,3,4-tetrahydroquinolin-6-yl)oxy]acetamide | **MPRO** | **-7.769056** | CN(C)c1ccc(CN(C)C(=O)COc2ccc3NC(=O)CCc3c2)cc1 |
| 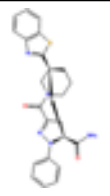 | 3-[3-(1,3-benzothiazol-2-yl)piperidine-1-carbonyl]-1-phenyl-4,5-dihydro-1H-pyrazole-5-carboxamide | **MPRO** | **-7.602391** | NC(=O)C1CC(=NN1c1ccccc1)C(=O)N1CCCC(C1)c1nc2ccccc2s1 |
| 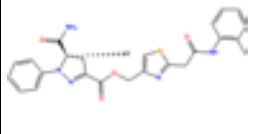 | (2-{[(2-methylphenyl)carbamoyl]methyl}-1,3-thiazol-4-yl)methyl 5-carbamoyl-1-phenyl-4,5-dihydro-1H-pyrazole-3-carboxylate | **MPRO** | **-7.69133** | Cc1ccccc1NC(=O)Cc1nc(COC(=O)C2=NN(C(C2)C(N)=O)c2ccccc2)cs1 |
| 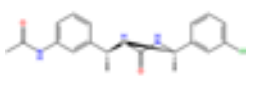 | N-{3-[1-({[1-(3-fluorophenyl)ethyl]carbamoyl}amino)ethyl]phenyl}acetamide | **MPRO** | **-7.802171** | CC(NC(=O)NC(C)c1cccc(NC(C)=O)c1)c1cccc(F)c1 |
| 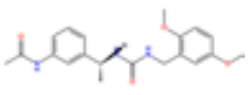 | N-{3-[1-({[(2,5-dimethoxyphenyl)methyl]carbamoyl}amino)ethyl]phenyl}acetamide | **MPRO** | **-7.721882** | COc1ccc(OC)c(CNC(=O)NC(C)c2cccc(NC(C)=O)c2)c1 |
| 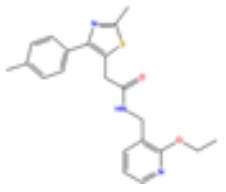 | N-[(2-ethoxypyridin-3-yl)methyl]-2-[2-methyl-4-(4-methylphenyl)-1,3-thiazol-5-yl]acetamide | **MPRO** | **-7.648012** | CCOc1ncccc1CNC(=O)Cc1sc(C)nc1-c1ccc(C)cc1 |
| 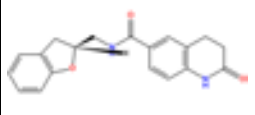 | N-[(2,3-dihydro-1-benzofuran-2-yl)methyl]-2-oxo-1,2,3,4-tetrahydroquinoline-6-carboxamide | **MPRO** | **-7.632513** | O=C(NCC1Cc2ccccc2O1)c1ccc2NC(=O)CCc2c1 |
| 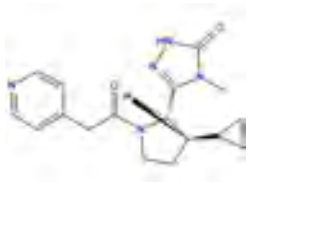 | 3-{3-cyclopropyl-1-[2-(pyridin-4-yl)acetyl]pyrrolidin-2-yl}-4-methyl-4,5-dihydro-1H-1,2,4-triazol-5-one | **MPRO** | **-7.675123** | Cn1c(n[nH]c1=O)C1C(CCN1C(=O)Cc1ccncc1)C1CC1 |
| 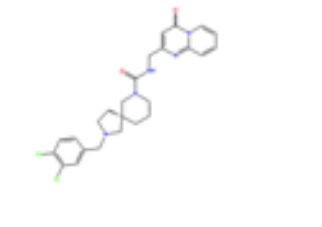 | 2-[(3,4-dichlorophenyl)methyl]-N-({4-oxo-4H-pyrido[1,2-a]pyrimidin-2-yl}methyl)-2,7-diazaspiro[4.5]decane-7-carboxamide | **MPRO** | **-7.628686** | Clc1ccc(CN2CCC3(C2)CCCN(C3)C(=O)NCc2cc(=O)n3ccccc3n2)cc1Cl |
| 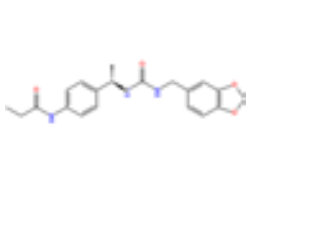 | N-{4-[1-({[(2H-1,3-benzodioxol-5-yl)methyl]carbamoyl}amino)ethyl]phenyl}propanamide | **MPRO** | **-7.656456** | CCC(=O)Nc1ccc(cc1)C(C)NC(=O)NCc1ccc2OCOc2c1 |
| 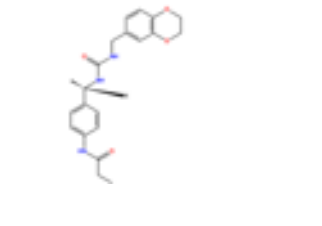 | N-{4-[1-({[(2,3-dihydro-1,4-benzodioxin-6-yl)methyl]carbamoyl}amino)ethyl]phenyl}propanamide | **MPRO** | **-7.740255** | CCC(=O)Nc1ccc(cc1)C(C)NC(=O)NCc1ccc2OCCOc2c1 |
| 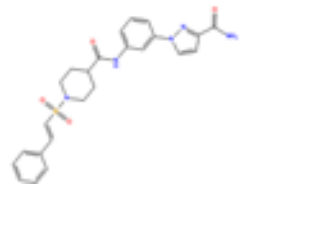 | N-[3-(3-carbamoyl-1H-pyrazol-1-yl)phenyl]-1-(2-phenylethenesulfonyl)piperidine-4-carboxamide | **MPRO** | **-7.715121** | NC(=O)c1ccn(n1)-c1cccc(NC(=O)C2CCN(CC2)S(=O)(=O)C=Cc2ccccc2)c1 |
| 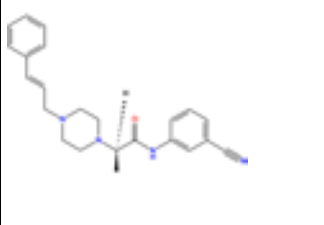 | N-(3-cyanophenyl)-2-[4-(3-phenylprop-2-en-1-yl)piperazin-1-yl]propanamide | **MPRO** | **-7.756551** | CC(N1CCN(CC=Cc2ccccc2)CC1)C(=O)Nc1cccc(c1)C#N |
| 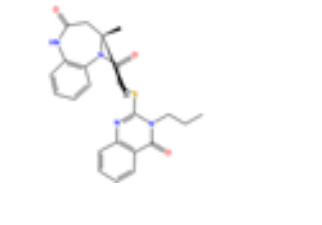 | 4-methyl-5-{2-[(4-oxo-3-propyl-3,4-dihydroquinazolin-2-yl)sulfanyl]acetyl}-2,3,4,5-tetrahydro-1H-1,5-benzodiazepin-2-one | **MPRO** | **-7.768639** | CCCn1c(SCC(=O)N2C(C)CC(=O)Nc3ccccc23)nc2ccccc2c1=O |
| 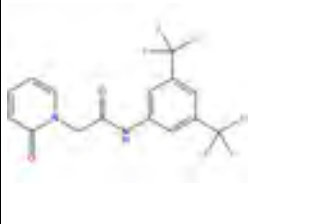 | N-[3,5-bis(trifluoromethyl)phenyl]-2-(2-oxo-1,2-dihydropyridin-1-yl)acetamide | **MPRO** | **-7.623712** | FC(F)(F)c1cc(NC(=O)Cn2ccccc2=O)cc(c1)C(F)(F)F |
| 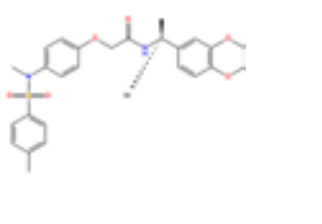 | N-[1-(2,3-dihydro-1,4-benzodioxin-6-yl)ethyl]-2-[4-(N-methyl4-methylbenzenesulfonamido)phenoxy]acetamide | **MPRO** | **-7.61314** | CC(NC(=O)COc1ccc(cc1)N(C)S(=O)(=O)c1ccc(C)cc1)c1ccc2OCCOc2c1 |
| 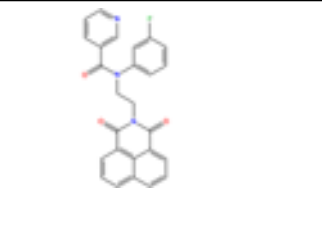 | N-(2-{2,4-dioxo-3-azatricyclo[7.3.1.0$^{5,13}$]trideca-1(13),5,7,9,11-pentaen-3-yl}ethyl)-N-(3-fluorophenyl)pyridine-3-carboxamide | **MPRO** | **-7.63636** | Fc1cccc(c1)N(CCN1C(=O)c2cccc3cccc(C1=O)c23)C(=O)c1cccnc1 |
| 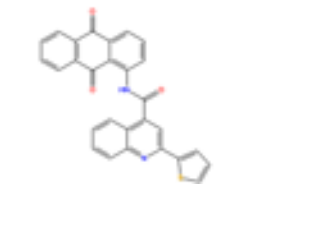 | N-(9,10-dioxo-9,10-dihydroanthracen-1-yl)-2-(thiophen-2-yl)quinoline-4-carboxamide | **MPRO** | **-7.800731** | O=C(Nc1cccc2C(=O)c3ccccc3C(=O)c12)c1cc(nc2ccccc12)-c1cccs1 |
| 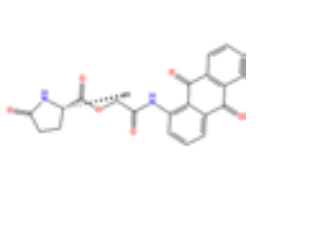 | [(9,10-dioxo-9,10-dihydroanthracen-1-yl)carbamoyl]methyl 5-oxopyrrolidine-2-carboxylate | **MPRO** | **-7.792261** | O=C(COC(=O)C1CCC(=O)N1)Nc1cccc2C(=O)c3ccccc3C(=O)c12 |
| 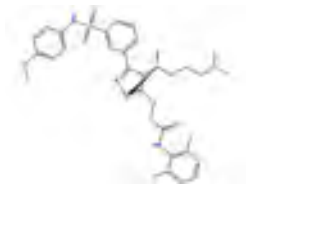 | N-(2,6-dimethylphenyl)-2-[(5-{3-[(4-methoxyphenyl)sulfamoyl]phenyl}-4-(6-methylheptan-2-yl)-4H-1,2,4-triazol-3-yl)sulfanyl]acetamide | **MPRO** | **-7.635464** | COc1ccc(NS(=O)(=O)c2cccc(c2)-c2nnc(SCC(=O)Nc3c(C)cccc3C)n2C(C)CCCC(C)C)cc1 |
| 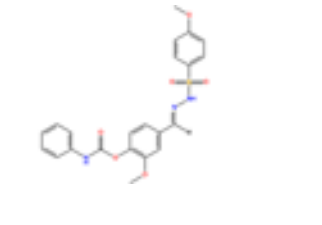 | 2-methoxy-4-{[(4-methoxybenzenesulfonamido)imino]methyl}phenyl N-phenylcarbamate | **MPRO** | **-7.623766** | COc1ccc(cc1)S(=O)(=O)NN=Cc1ccc(OC(=O)Nc2ccccc2)c(OC)c1 |
| 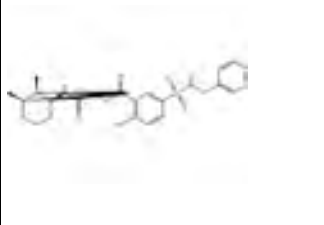 | [(2,3-dimethylcyclohexyl)carbamoyl]methyl 5-(benzylsulfamoyl)-2-hydroxybenzoate | **MPRO** | **-7.691032** | CC1CCCC(NC(=O)COC(=O)c2cc(ccc2O)S(=O)(=O)NCc2ccccc2)C1C |
| 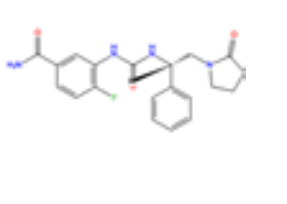 | 4-fluoro-3-({[2-(2-oxopyrrolidin-1-yl)-1-phenylethyl]carbamoyl}amino)benzamide | **MPRO** | **-7.779582** | NC(=O)c1ccc(F)c(NC(=O)NC(CN2CCCC2=O)c2ccccc2)c1 |
| 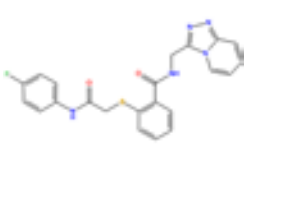 | 2-({[(4-fluorophenyl)carbamoyl]methyl}sulfanyl)-N-({[1,2,4]triazolo[4,3-a]pyridin-3-yl}methyl)benzamide | **MPRO** | **-7.678648** | Fc1ccc(NC(=O)CSc2ccccc2C(=O)NCc2nnc3ccccn23)cc1 |
| 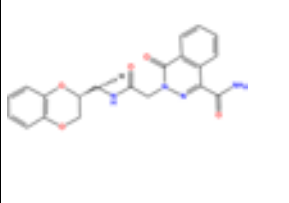 | 3-({[(2,3-dihydro-1,4-benzodioxin-2-yl)methyl]carbamoyl}methyl)-4-oxo-3,4-dihydrophthalazine-1-carboxamide | **MPRO** | **-7.677514** | NC(=O)c1nn(CC(=O)NCC2COc3ccccc3O2)c(=O)c2ccccc12 |
| 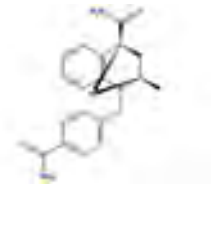 | 1-[(4-carbamoylphenyl)methyl]-2-methyl-1,2,3,4-tetrahydroquinoline-4-carboxamide | **MPRO** | **-7.686286** | CC1CC(C(N)=O)c2ccccc2N1Cc1ccc(cc1)C(N)=O |
| 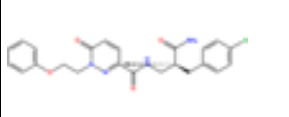 | 2-[(4-fluorophenyl)methyl]-3-{[6-oxo-1-(2-phenoxyethyl)-1,6-dihydropyridazin-3-yl]formamido}propanamide | **MPRO** | **-7.656212** | NC(=O)C(CNC(=O)c1ccc(=O)n(CCOc2ccccc2)n1)Cc1ccc(F)cc1 |
| 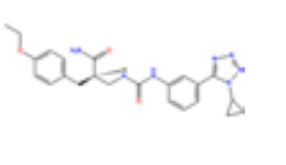 | 3-({[3-(1-cyclopropyl-1H-1,2,3,4-tetrazol-5-yl)phenyl]carbamoyl}amino)-2-[(4-ethoxyphenyl)methyl]propanamide | **MPRO** | **-7.739139** | CCOc1ccc(CC(CNC(=O)Nc2cccc(c2)-c2nnnn2C2CC2)C(N)=O)cc1 |
| 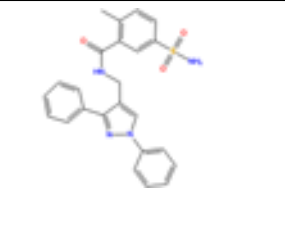 | N-[(1,3-diphenyl-1H-pyrazol-4-yl)methyl]-2-methyl-5-sulfamoylbenzamide | **MPRO** | **-7.629789** | Cc1ccc(cc1C(=O)NCc1cn(nc1-c1ccccc1)-c1ccccc1)S(N)(=O)=O |
| 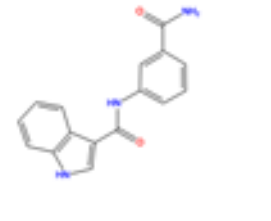 | N-(3-carbamoylphenyl)-1H-indole-3-carboxamide | **MPRO** | **-7.65138** | NC(=O)c1cccc(NC(=O)c2c[nH]c3ccccc23)c1 |
| 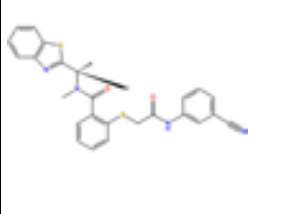 | N-[1-(1,3-benzothiazol-2-yl)ethyl]-2-({[(3-cyanophenyl)carbamoyl]methyl}sulfanyl)-N-methylbenzamide | **MPRO** | **-7.604902** | CC(N(C)C(=O)c1ccccc1SCC(=O)Nc1cccc(c1)C#N)c1nc2ccccc2s1 |
| 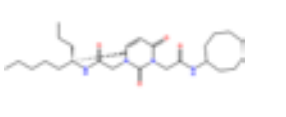 | 2-{3-[(cyclooctylcarbamoyl)methyl]-2,4-dioxo-1,2,3,4-tetrahydropyrimidin-1-yl}-N-(nonan-4-yl)acetamide | **MPRO** | **-7.685229** | CCCCCC(CCC)NC(=O)Cn1ccc(=O)n(CC(=O)NC2CCCCCCC2)c1=O |
| 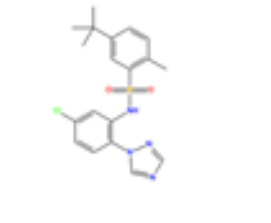 | 5-tert-butyl-N-[5-chloro-2-(1H-1,2,4-triazol-1-yl)phenyl]-2-methylbenzene-1-sulfonamide | **MPRO** | **-7.628456** | Cc1ccc(cc1S(=O)(=O)Nc1cc(Cl)ccc1-n1cncn1)C(C)(C)C |
| 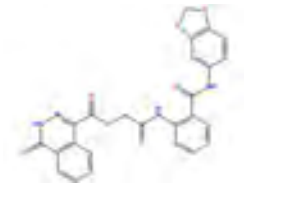 | ({2-[(2H-1,3-benzodioxol-5-yl)carbamoyl]phenyl}carbamoyl)methyl 4-oxo-3,4-dihydrophthalazine-1-carboxylate | **MPRO** | **-7.796552** | O=C(COC(=O)c1n[nH]c(=O)c2ccccc12)Nc1ccccc1C(=O)Nc1ccc2OCOc2c1 |
| 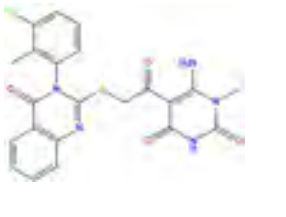 | 6-amino-5-(2-{[3-(3-chloro-2-methylphenyl)-4-oxo-3,4-dihydroquinazolin-2-yl]sulfanyl}acetyl)-1-methyl-1,2,3,4-tetrahydropyrimidine-2,4-dione | **MPRO** | **-7.617443** | Cc1c(Cl)cccc1-n1c(SCC(=O)c2c(N)n(C)c(=O)[nH]c2=O)nc2ccccc2c1=O |
| 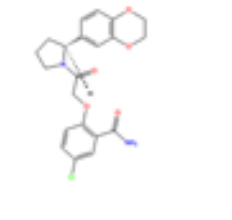 | 5-chloro-2-{2-[2-(2,3-dihydro-1,4-benzodioxin-6-yl)pyrrolidin-1-yl]-2-oxoethoxy}benzamide | **MPRO** | **-7.637594** | NC(=O)c1cc(Cl)ccc1OCC(=O)N1CCCC1c1ccc2OCCOc2c1 |
| 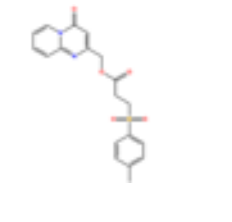 | {4-oxo-4H-pyrido[1,2-a]pyrimidin-2-yl}methyl 3-(4-methylbenzenesulfonyl)propanoate | **MPRO** | **-7.770157** | Cc1ccc(cc1)S(=O)(=O)CCC(=O)OCc1cc(=O)n2ccccc2n1 |
| 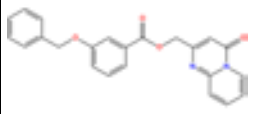 | {4-oxo-4H-pyrido[1,2-a]pyrimidin-2-yl}methyl 3-(benzyloxy)benzoate | **MPRO** | **-7.60401** | O=C(OCc1cc(=O)n2ccccc2n1)c1cccc(OCc2ccccc2)c1 |
| 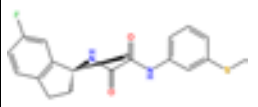 | N-(6-fluoro-2,3-dihydro-1H-inden-1-yl)-N'-[3-(methylsulfanyl)phenyl]ethanediamide | **TMPRSS2** | **-7.129755** | CSc1cccc(NC(=O)C(=O)NC2CCc3ccc(F)cc23)c1 |
| 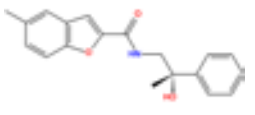 | N-(2-hydroxy-2-phenylpropyl)-5-methyl-1-benzofuran-2-carboxamide | **TMPRSS2** | **-7.090923** | Cc1ccc2oc(cc2c1)C(=O)NCC(C)(O)c1ccccc1 |
| 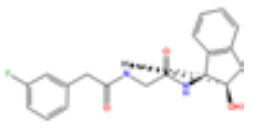 | 2-(3-fluorophenyl)-N-{[(2-hydroxy-2,3-dihydro-1H-inden-1-yl)carbamoyl]methyl}acetamide | **TMPRSS2** | **-7.083628** | OC1Cc2ccccc2C1NC(=O)CNC(=O)Cc1cccc(F)c1 |
| 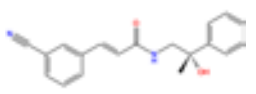 | 3-(3-cyanophenyl)-N-(2-hydroxy-2-phenylpropyl)prop-2-enamide | **TMPRSS2** | **-7.100698** | CC(O)(CNC(=O)C=Cc1cccc(c1)C#N)c1ccccc1 |
| 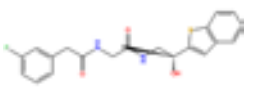 | N-[2-(1-benzothiophen-2-yl)-2-hydroxyethyl]-2-[2-(3-fluorophenyl)acetamido]acetamide | **TMPRSS2** | **-7.173833** | OC(CNC(=O)CNC(=O)Cc1cccc(F)c1)c1cc2ccccc2s1 |
| 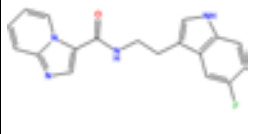 | N-[2-(5-fluoro-1H-indol-3-yl)ethyl]imidazo[1,2-a]pyridine-3-carboxamide | **TMPRSS2** | **-7.082199** | Fc1ccc2[nH]cc(CCNC(=O)c3cnc4ccccn34)c2c1 |
| 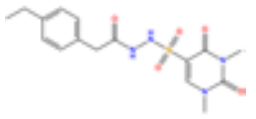 | N'-[(1,3-dimethyl-2,4-dioxo-1,2,3,4-tetrahydropyrimidin-5-yl)sulfonyl]-2-(4-ethylphenyl)acetohydrazide | **TMPRSS2** | **-7.112836** | CCc1ccc(CC(=O)NNS(=O)(=O)c2cn(C)c(=O)n(C)c2=O)cc1 |
| 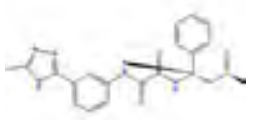 | N'-(2-methanesulfinyl-1-phenylethyl)-N-[3-(5-methyl-1H-1,2,4-triazol-3-yl)phenyl]ethanediamide | **TMPRSS2** | **-7.126673** | Cc1nc(n[nH]1)-c1cccc(NC(=O)C(=O)NC(CS(C)=O)c2ccccc2)c1 |
| 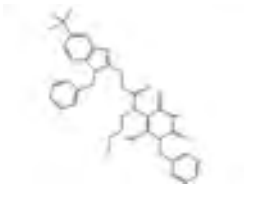 | N-(6-amino-1-benzyl-2,4-dioxo-1,2,3,4-tetrahydropyrimidin-5-yl)-2-{[1-benzyl-5-(trifluoromethyl)-1H-1,3-benzodiazol-2-yl]sulfanyl}-N-butylacetamide | **TMPRSS2** | **-7.099033** | CCCCN(C(=O)CSc1nc2cc(ccc2n1Cc1ccccc1)C(F)(F)F)c1c(N)n(Cc2ccccc2)c(=O)[nH]c1=O |
| 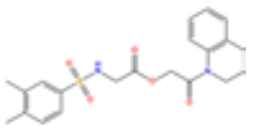 | 2-oxo-2-(1,2,3,4-tetrahydroquinolin-1-yl)ethyl 2-(3,4-dimethylbenzenesulfonamido)acetate | **TMPRSS2** | **-7.148417** | Cc1ccc(cc1C)S(=O)(=O)NCC(=O)OCC(=O)N1CCCc2ccccc12 |
| 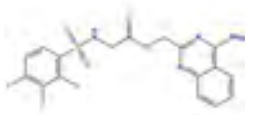 | (4-aminoquinazolin-2-yl)methyl 2-(2,3,4-trifluorobenzenesulfonamido)acetate | **TMPRSS2** | **-7.160395** | Nc1nc(COC(=O)CNS(=O)(=O)c2ccc(F)c(F)c2F)nc2ccccc12 |
| 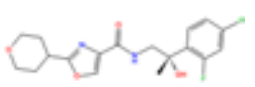 | N-[2-(2,4-difluorophenyl)-2-hydroxypropyl]-2-(oxan-4-yl)-1,3-oxazole-4-carboxamide | **TMPRSS2** | **-7.159434** | CC(O)(CNC(=O)c1coc(n1)C1CCOCC1)c1ccc(F)cc1F |
| 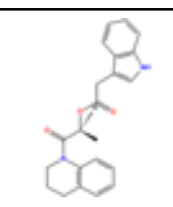 | 1-oxo-1-(1,2,3,4-tetrahydroquinolin-1-yl)propan-2-yl 2-(1H-indol-3-yl)acetate | **TMPRSS2** | **-7.120618** | CC(OC(=O)Cc1c[nH]c2ccccc12)C(=O)N1CCCc2ccccc12 |
| 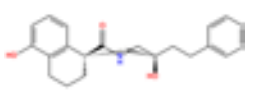 | 5-hydroxy-N-(2-hydroxy-4-phenylbutyl)-1,2,3,4-tetrahydronaphthalene-1-carboxamide | **TMPRSS2** | **-7.139395** | OC(CCc1ccccc1)CNC(=O)C1CCCc2c(O)cccc12 |
| 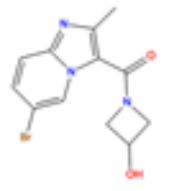 | 1-{6-bromo-2-methylimidazo[1,2-a]pyridine-3-carbonyl}azetidin-3-ol | **TMPRSS2** | **-7.004282** | Cc1nc2ccc(Br)cn2c1C(=O)N1CC(O)C1 |
| 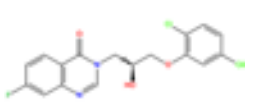 | 3-[3-(2,5-dichlorophenoxy)-2-hydroxypropyl]-7-fluoro-3,4-dihydroquinazolin-4-one | **TMPRSS2** | **-7.190593** | OC(COc1cc(Cl)ccc1Cl)Cn1cnc2cc(F)ccc2c1=O |
| 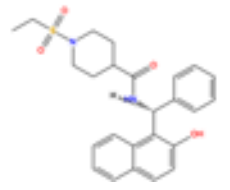 | 1-(ethanesulfonyl)-N-[(2-hydroxynaphthalen-1-yl)(phenyl)methyl]piperidine-4-carboxamide | **TMPRSS2** | **-7.089504** | CCS(=O)(=O)N1CCC(CC1)C(=O)NC(c1ccccc1)c1c(O)ccc2ccccc12 |
| 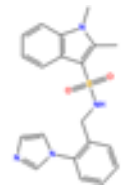 | N-{[2-(1H-imidazol-1-yl)phenyl]methyl}-1,2-dimethyl-1H-indole-3-sulfonamide | **TMPRSS2** | **-7.124544** | Cc1c(c2ccccc2n1C)S(=O)(=O)NCc1ccccc1-n1ccnc1 |
| 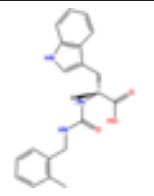 | 3-(1H-indol-3-yl)-2-({[(2-methylphenyl)methyl]carbamoyl}amino)propanoic acid | **TMPRSS2** | **-7.196331** | Cc1ccccc1CNC(=O)NC(Cc1c[nH]c2ccccc12)C(O)=O |
| 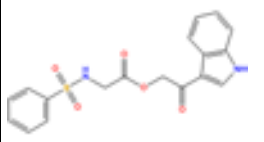 | 2-(1H-indol-3-yl)-2-oxoethyl 2-benzenesulfonamidoacetate | **TMPRSS2** | **-7.106572** | O=C(CNS(=O)(=O)c1ccccc1)OCC(=O)c1c[nH]c2ccccc12 |
| 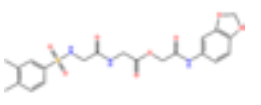 | [(2H-1,3-benzodioxol-5-yl)carbamoyl]methyl 2-[2-(3,4-dimethylbenzenesulfonamido)acetamido]acetate | **TMPRSS2** | **-7.118744** | Cc1ccc(cc1C)S(=O)(=O)NCC(=O)NCC(=O)OCC(=O)Nc1ccc2OCOc2c1 |
| 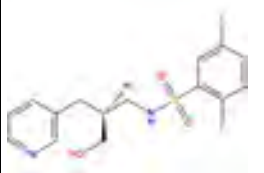 | 2,5-difluoro-N-{3-hydroxy-2-[(pyridin-3-yl)methyl]propyl}benzene-1-sulfonamide | **TMPRSS2** | **-7.123749** | OCC(CNS(=O)(=O)c1cc(F)ccc1F)Cc1cccnc1 |
| 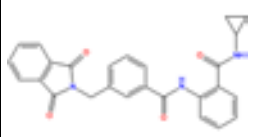 | N-cyclopropyl-2-{3-[(1,3-dioxo-2,3-dihydro-1H-isoindol-2-yl)methyl]benzamido}benzamide | **TMPRSS2** | **-7.180751** | O=C(Nc1ccccc1C(=O)NC1CC1)c1cccc(CN2C(=O)c3ccccc3C2=O)c1 |
| 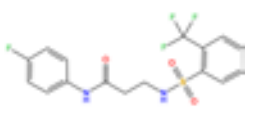 | N-(4-fluorophenyl)-3-[2-(trifluoromethyl)benzenesulfonamido]propanamide | **TMPRSS2** | **-7.088645** | Fc1ccc(NC(=O)CCNS(=O)(=O)c2ccccc2C(F)(F)F)cc1 |
| 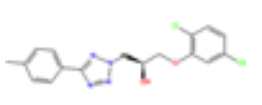 | 1-(2,5-dichlorophenoxy)-3-[5-(4-methylphenyl)-2H-1,2,3,4-tetrazol-2-yl]propan-2-ol | **TMPRSS2** | **-7.115764** | Cc1ccc(cc1)-c1nnn(CC(O)COc2cc(Cl)ccc2Cl)n1 |
| 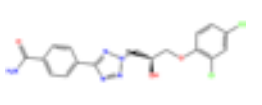 | 4-{2-[3-(2,4-dichlorophenoxy)-2-hydroxypropyl]-2H-1,2,3,4-tetrazol-5-yl}benzamide | **TMPRSS2** | **-7.113681** | NC(=O)c1ccc(cc1)-c1nnn(CC(O)COc2ccc(Cl)cc2Cl)n1 |
| 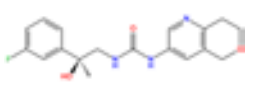 | 1-[2-(3-fluorophenyl)-2-hydroxypropyl]-3-{5H,7H,8H-pyrano[4,3-b]pyridin-3-yl}urea | **TMPRSS2** | **-7.095425** | CC(O)(CNC(=O)Nc1cnc2CCOCc2c1)c1cccc(F)c1 |
| 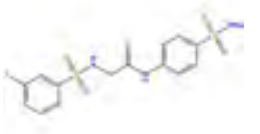 | 2-(3-fluorobenzenesulfonamido)-N-(4-sulfamoylphenyl)acetamide | **TMPRSS2** | **-7.139361** | NS(=O)(=O)c1ccc(NC(=O)CNS(=O)(=O)c2cccc(F)c2)cc1 |
| 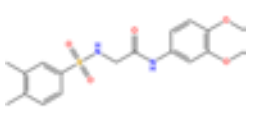 | N-(3,4-dimethoxyphenyl)-2-(3,4-dimethylbenzenesulfonamido)acetamide | **TMPRSS2** | **-7.113242** | COc1ccc(NC(=O)CNS(=O)(=O)c2ccc(C)c(C)c2)cc1OC |
| 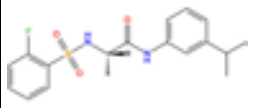 | 2-(2-fluorobenzenesulfonamido)-N-[3-(propan-2-yl)phenyl]propanamide | **TMPRSS2** | **-7.193332** | CC(C)c1cccc(NC(=O)C(C)NS(=O)(=O)c2ccccc2F)c1 |
| 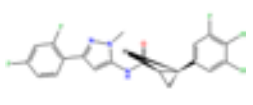 | N-[3-(2,4-difluorophenyl)-1-methyl-1H-pyrazol-5-yl]-2-(3,4,5-trifluorophenyl)cyclopropane-1-carboxamide | **TMPRSS2** | **-7.139957** | Cn1nc(cc1NC(=O)C1CC1c1cc(F)c(F)c(F)c1)-c1ccc(F)cc1F |
| 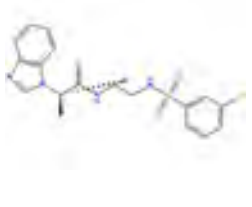 | 2-(1H-1,3-benzodiazol-1-yl)-N-[2-(3-chlorobenzenesulfonamido)ethyl]propanamide | **TMPRSS2** | **-7.125183** | CC(C(=O)NCCNS(=O)(=O)c1cccc(Cl)c1)n1cnc2ccccc12 |
| 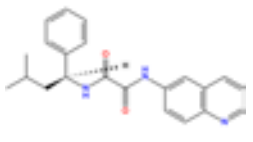 | N'-(3-methyl-1-phenylbutyl)-N-(quinolin-6-yl)ethanediamide | **TMPRSS2** | **-7.149125** | CC(C)CC(NC(=O)C(=O)Nc1ccc2ncccc2c1)c1ccccc1 |
| 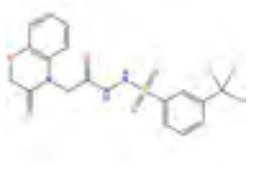 | 2-(3-oxo-3,4-dihydro-2H-1,4-benzoxazin-4-yl)-N'-[3-(trifluoromethyl)benzenesulfonyl]acetohydrazide | **TMPRSS2** | **-7.143432** | FC(F)(F)c1cccc(c1)S(=O)(=O)NNC(=O)CN1C(=O)COc2ccccc12 |
| 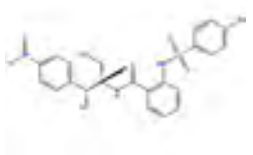 | 2-(4-bromobenzenesulfonamido)-N-[1,3-dihydroxy-1-(4-nitrophenyl)propan-2-yl]benzamide | **TMPRSS2** | **-7.119263** | OCC(NC(=O)c1ccccc1NS(=O)(=O)c1ccc(Br)cc1)C(O)c1ccc(cc1)[N+]([O-])=O |
| 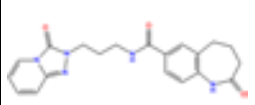 | 2-oxo-N-(3-{3-oxo-2H,3H-[1,2,4]triazolo[4,3-a]pyridin-2-yl}propyl)-2,3,4,5-tetrahydro-1H-1-benzazepine-7-carboxamide | **TMPRSS2** | **-7.104517** | O=C(NCCCn1nc2ccccn2c1=O)c1ccc2NC(=O)CCCc2c1 |
| 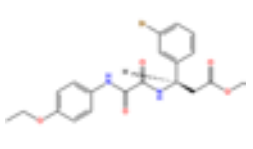 | methyl 3-(3-bromophenyl)-3-{[(4-ethoxyphenyl)carbamoyl]formamido}propanoate | **TMPRSS2** | **-7.082909** | CCOc1ccc(NC(=O)C(=O)NC(CC(=O)OC)c2cccc(Br)c2)cc1 |
| 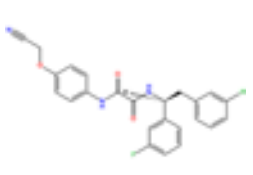 | N'-[1,2-bis(3-fluorophenyl)ethyl]-N-[4-(cyanomethoxy)phenyl]ethanediamide | **TMPRSS2** | **-7.198013** | Fc1cccc(CC(NC(=O)C(=O)Nc2ccc(OCC#N)cc2)c2cccc(F)c2)c1 |
| 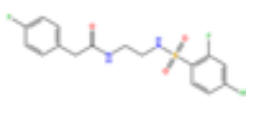 | N-[2-(2,4-difluorobenzenesulfonamido)ethyl]-2-(4-fluorophenyl)acetamide | **TMPRSS2** | **-7.121373** | Fc1ccc(CC(=O)NCCNS(=O)(=O)c2ccc(F)cc2F)cc1 |
| 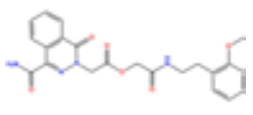 | {[2-(2-methoxyphenyl)ethyl]carbamoyl}methyl 2-(4-carbamoyl-1-oxo-1,2-dihydrophthalazin-2-yl)acetate | **TMPRSS2** | **-7.108118** | COc1ccccc1CCNC(=O)COC(=O)Cn1nc(C(N)=O)c2ccccc2c1=O |

| | | |
|---|---|---|
| 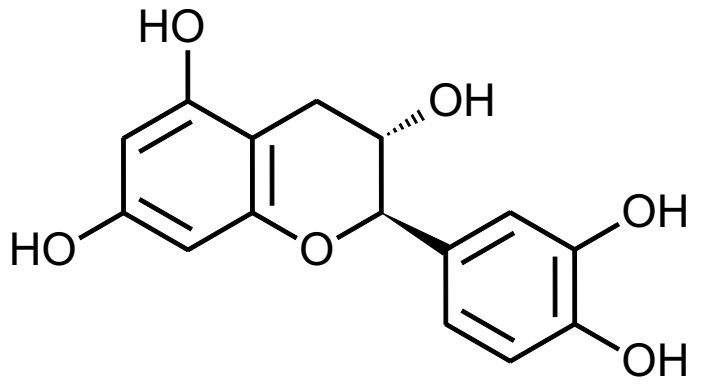<br>title: S4722 (+)-Catechin<br>ENZYME: MPro<br>docking score: -6.73 | 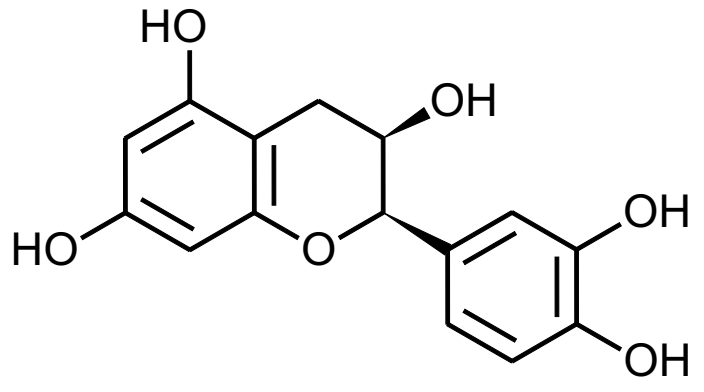<br>title: S4723 (-)Epicatechin<br>ENZYME: MPro<br>docking score: -6.322 | 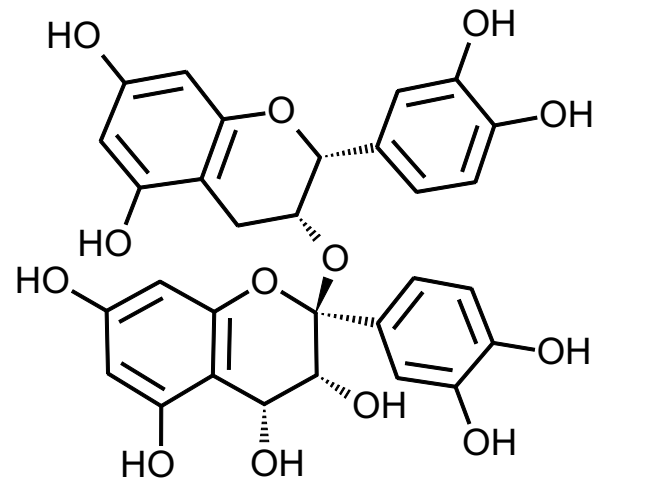<br>title: S5105<br>ENZYME: MPro<br>docking score: -6.19 |
| 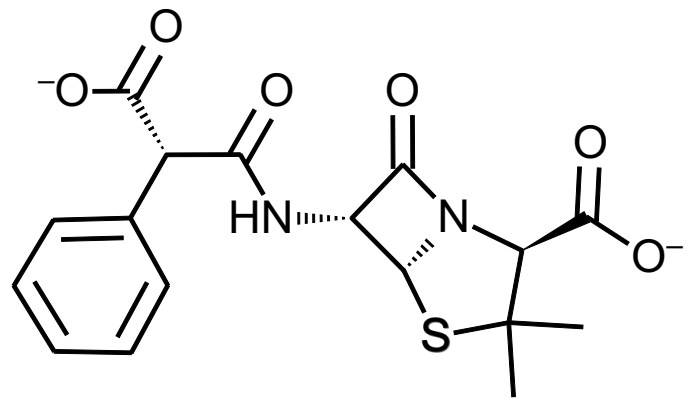<br>title: Carbenicillin disodium<br>ENZYME: MPro<br>docking score: -5.779 | 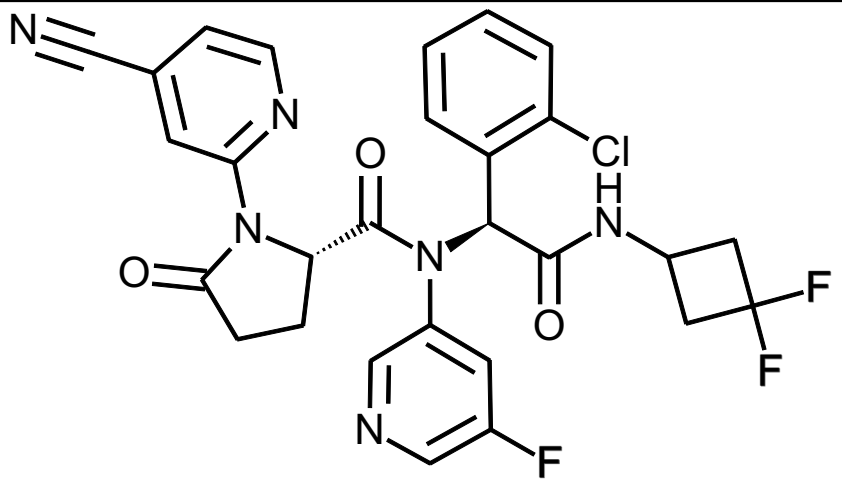<br>title: AG-120 (Ivosidenib)<br>ENZYME: MPro<br>docking score: -5.522 | 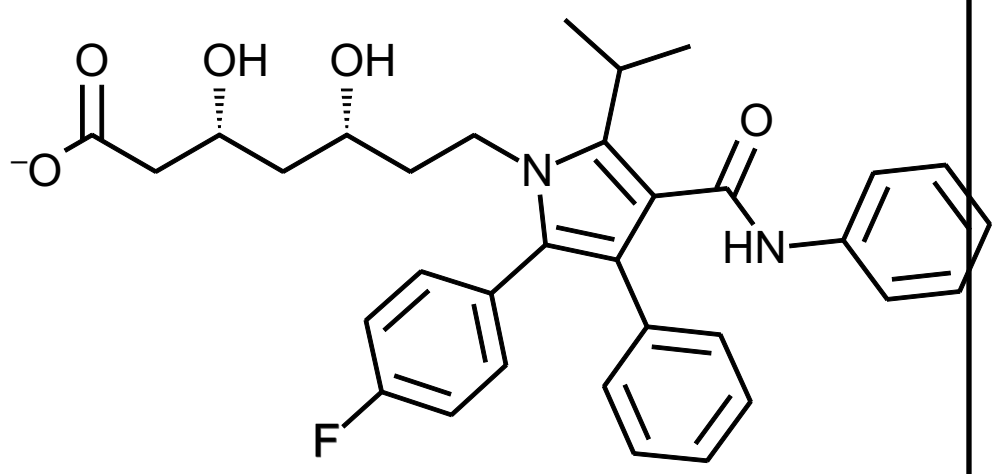<br>title: Atorvastatin calcium<br>ENZYME: MPro<br>docking score: -5.39 |
| 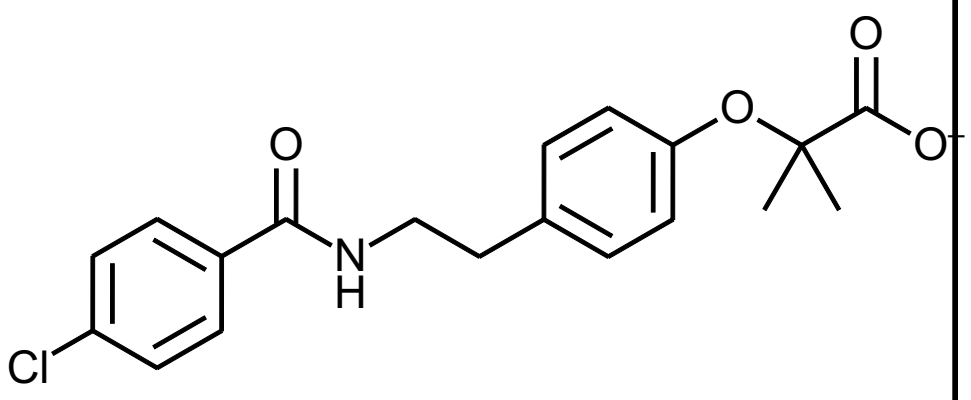<br>title: Bezafibrate<br>ENZYME: MPro<br>docking score: -4.933 | 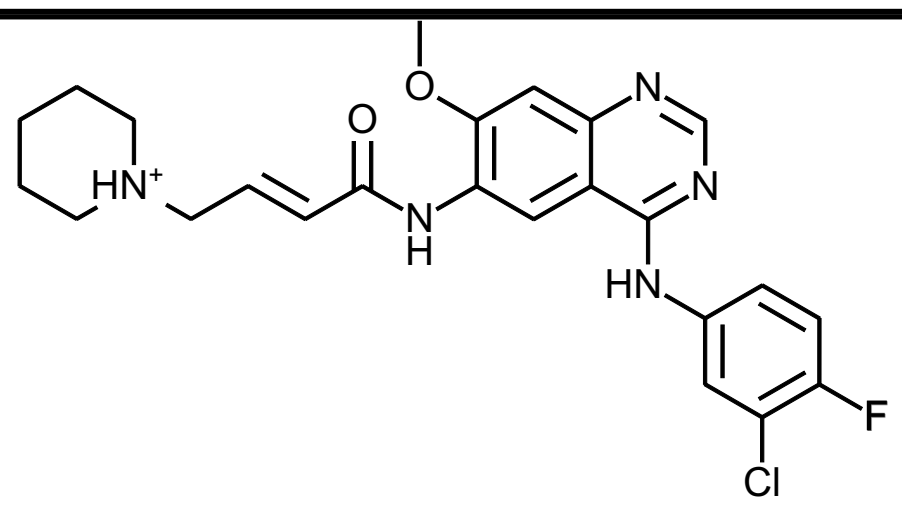<br>title: PF299804<br>ENZYME: MPro<br>docking score: -4.339 | 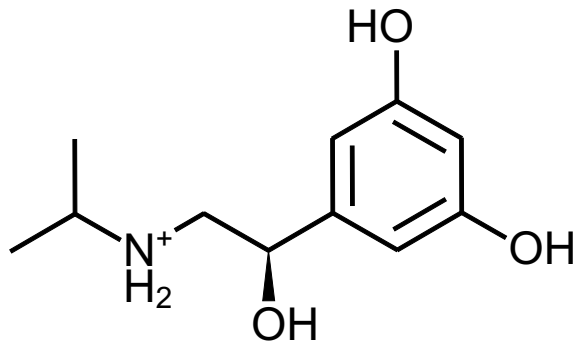<br>title: metaproterenol sulfate(orciprenaline sulfate)<br>ENZYME: Ace2<br>docking score: -8.048 |
| 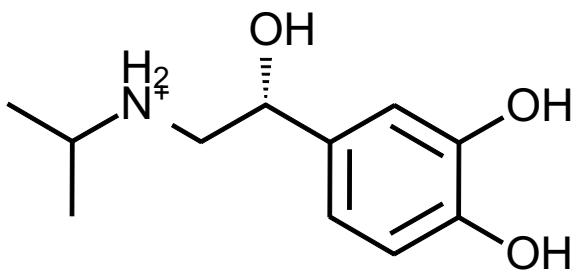<br>title: Isoprenaline hydrochloride<br>ENZYME: Ace2<br>docking score: -7.442 | 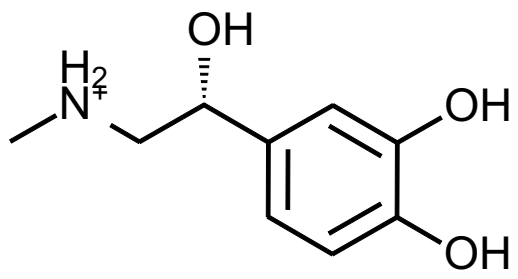<br>title: Epinephrine HCl<br>ENZYME: Ace2<br>docking score: -7.115 | 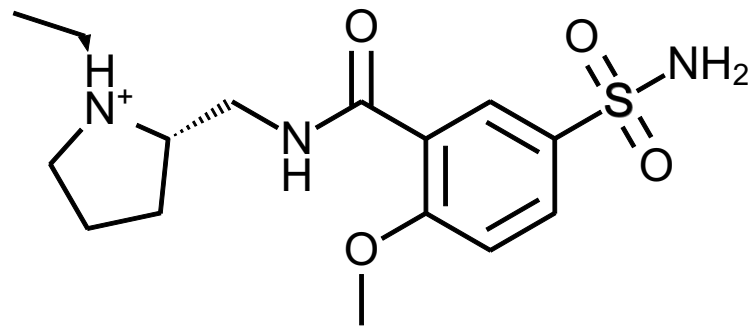<br>title: Levosulpiride<br>ENZYME: Ace2<br>docking score: -6.874 |

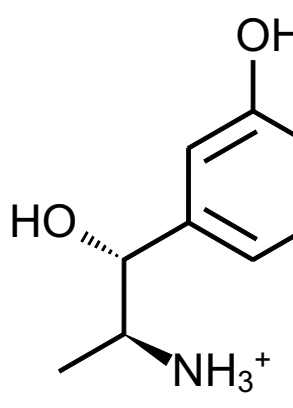

title: metaraminol bitartrate
ENZYME: Ace2
docking score: -6.838

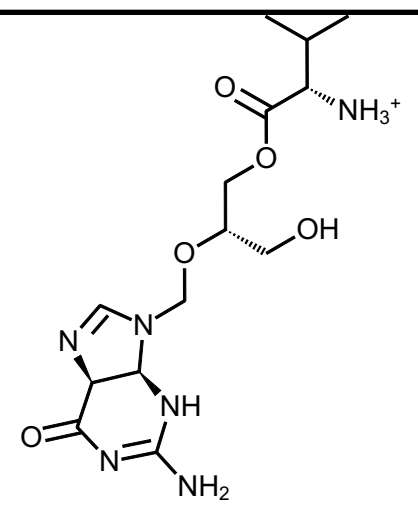

title: valganciclovir hydrochloride
ENZYME: Ace2
docking score: -6.584

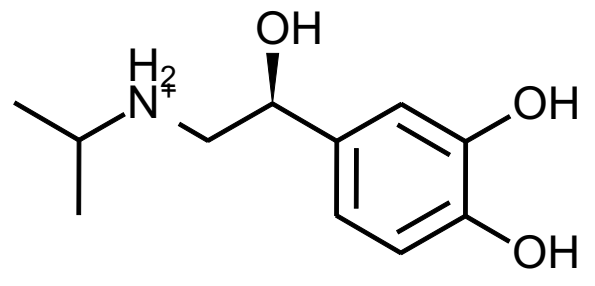

title: Isoprenaline hydrochloride
ENZYME: Ace2
docking score: -6.454

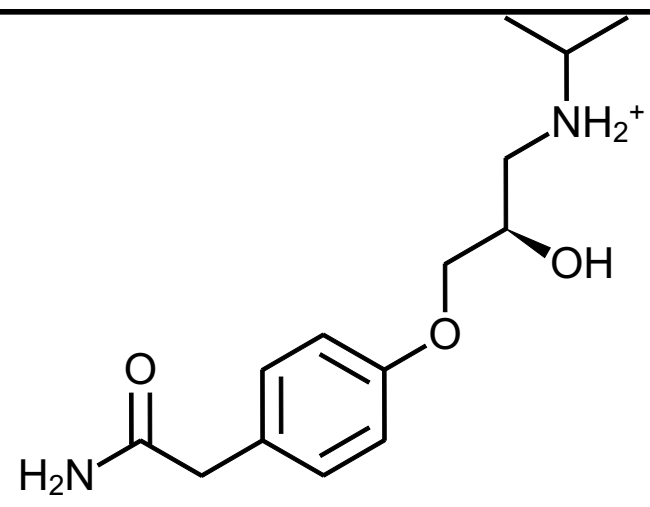

title: S4817 Atenolol
ENZYME: Ace2
docking score: -6.346

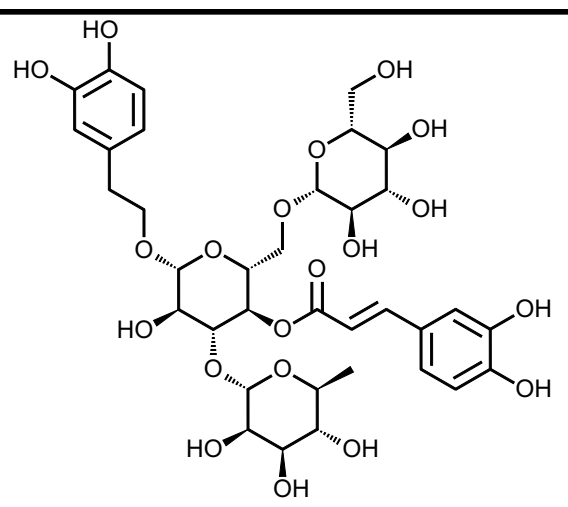

title: S3783 Echinacoside
ENZYME: Ace2
docking score: -6.086

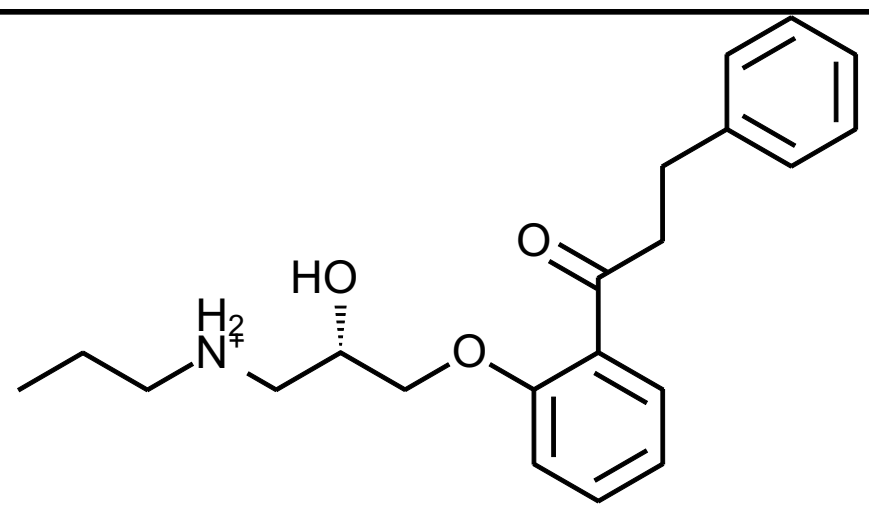

title: Propafenone
ENZYME: Ace2
docking score: -6.04

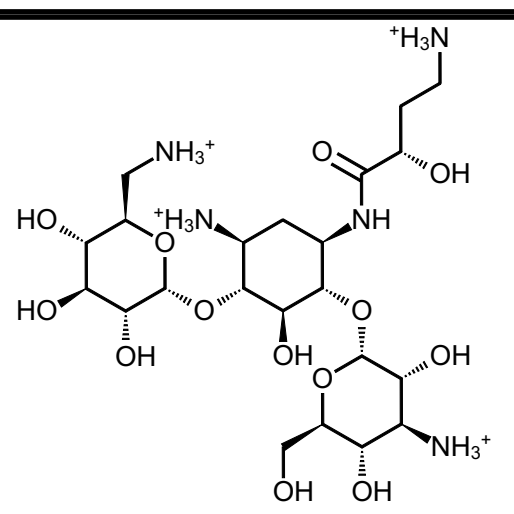

title: Amikacin sulfate
ENZYME: Ace2
docking score: -5.976

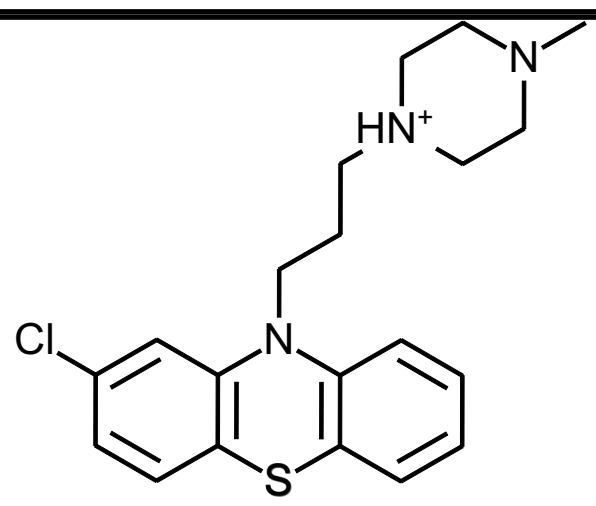

title: Prochlorperazine dimaleate salt
ENZYME: Ace2
docking score: -5.793

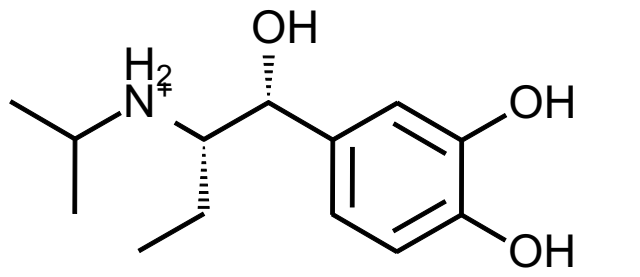

title: isoetharine mesylate
ENZYME: Ace2
docking score: -5.467

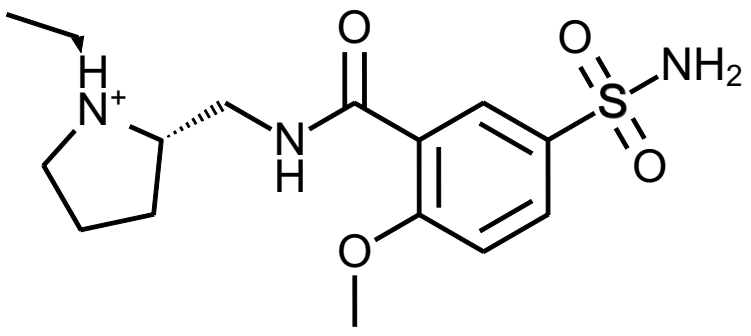

title: Levosulpiride
ENZYME: Ace2
docking score: -6.874

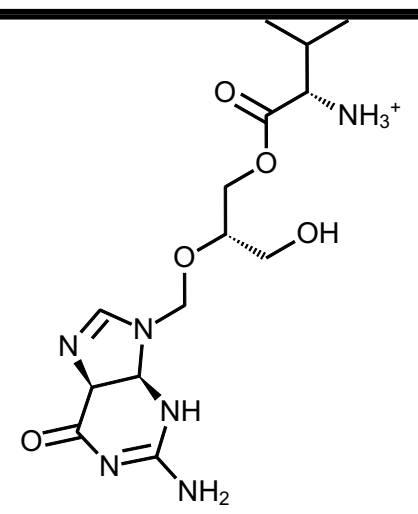

title: valganciclovir hydrochloride
ENZYME: Ace2
docking score: -6.366

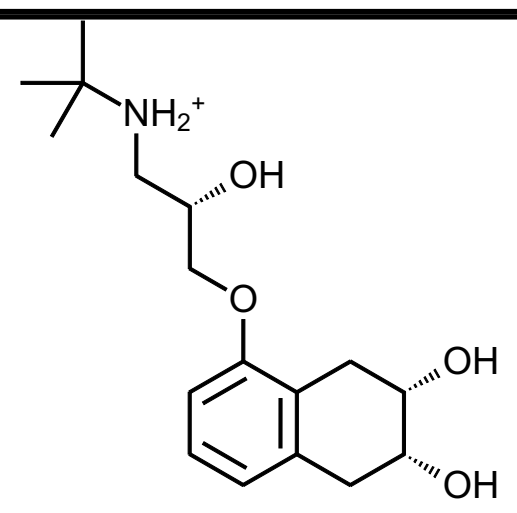

title: S5023 Nadolol
ENZYME: Ace2
docking score: -5.158

| | | |
|---|---|---|
| 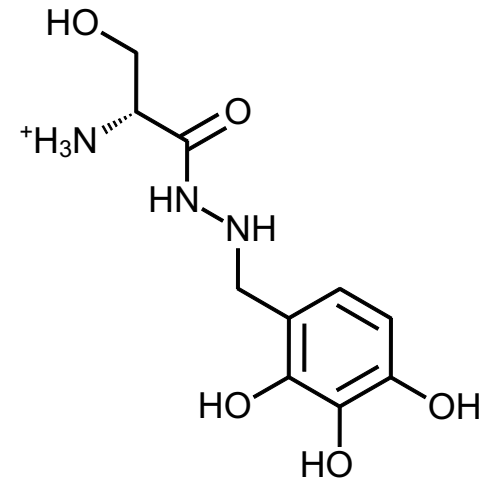<br>title: Benserazide hydrochloride<br>ENZYME: Ace2<br>docking score: -5.928 | 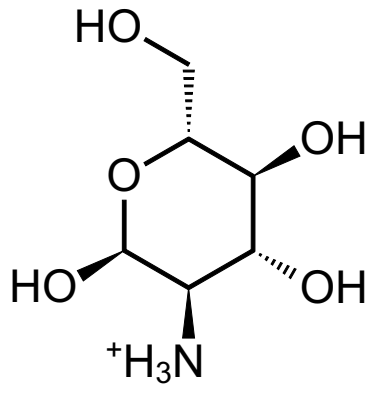<br>title: S3694 Glucosamine (hydrochloride)<br>ENZYME: Ace2<br>docking score: -5.568 | 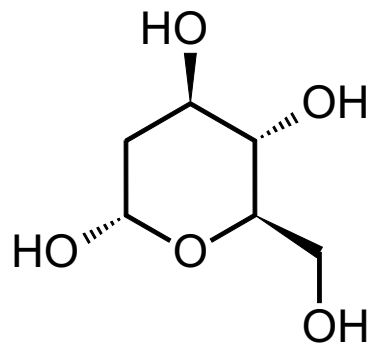<br>title: S4701 2-Deoxy-D-glucose<br>ENZYME: Ace2<br>docking score: -5.181 |
| 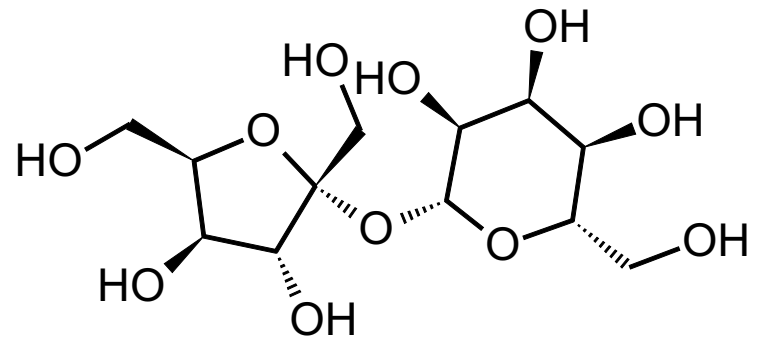<br>title: Inulin<br>ENZYME: Ace2<br>docking score: -5.175 | 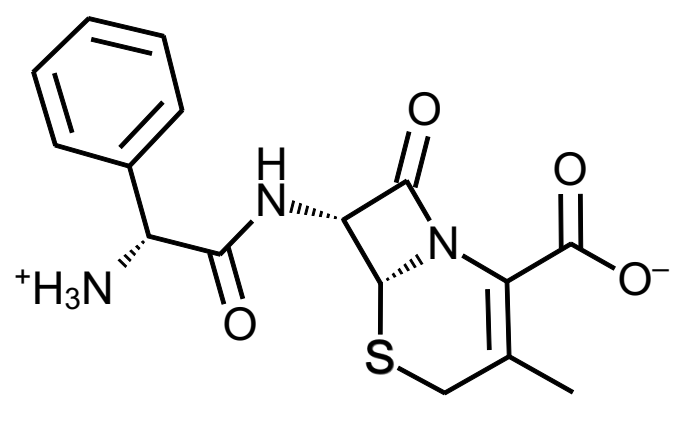<br>title: CEPHALEXIN (cephalexin)<br>ENZYME: Ace2<br>docking score: -5.108 | 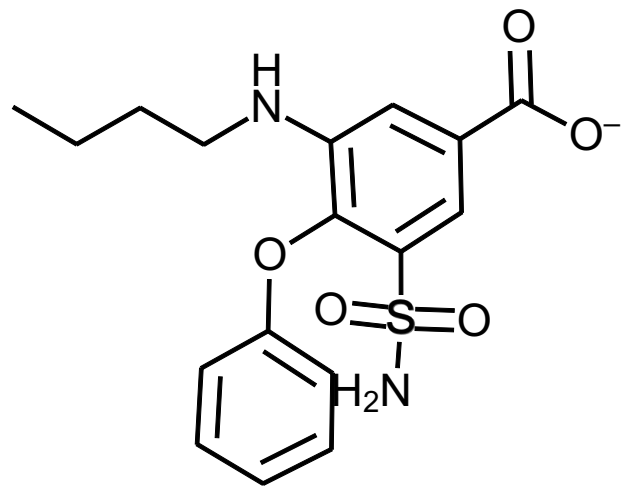<br>title: Bumetanide<br>ENZYME: TMPRSS2<br>docking score: -6.495 |
| 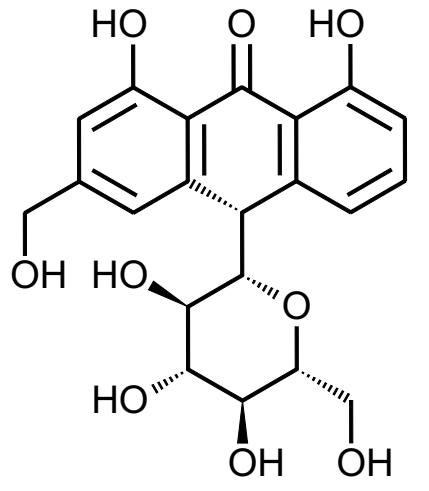<br>title: Aloin<br>ENZYME: TMPRSS2<br>docking score: -6.451 | 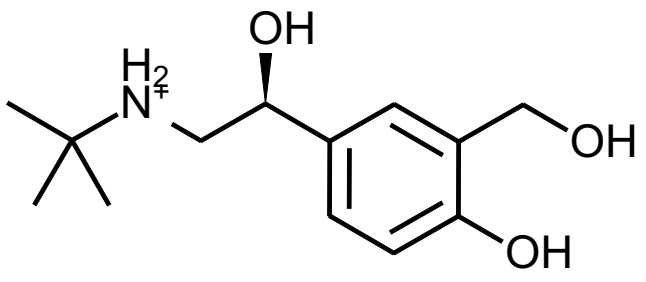<br>title: Salbutamol sulfate<br>ENZYME: TMPRSS2<br>docking score: -6.1 | 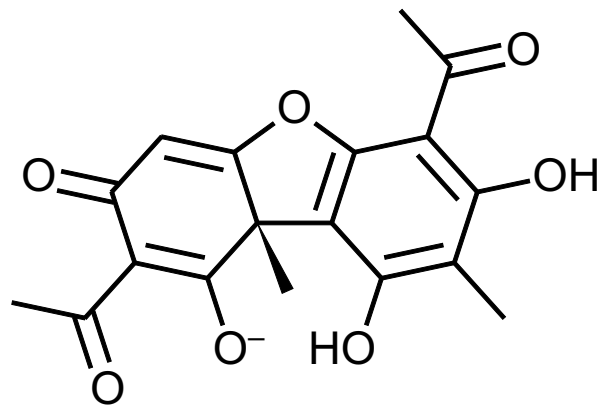<br>title: S4953 Usnic acid<br>ENZYME: TMPRSS2<br>docking score: -5.8 |
| 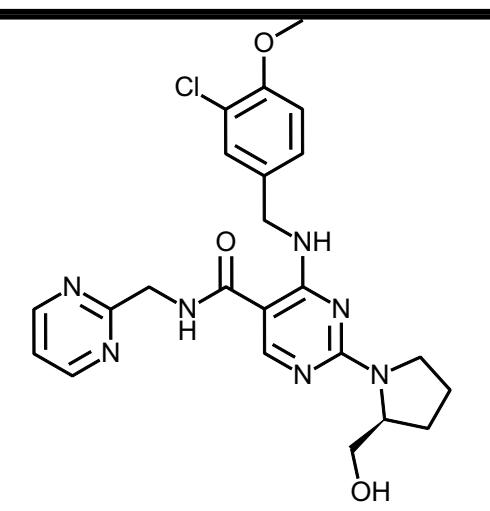<br>title: Avanafil<br>ENZYME: TMPRSS2<br>docking score: -5.616 | 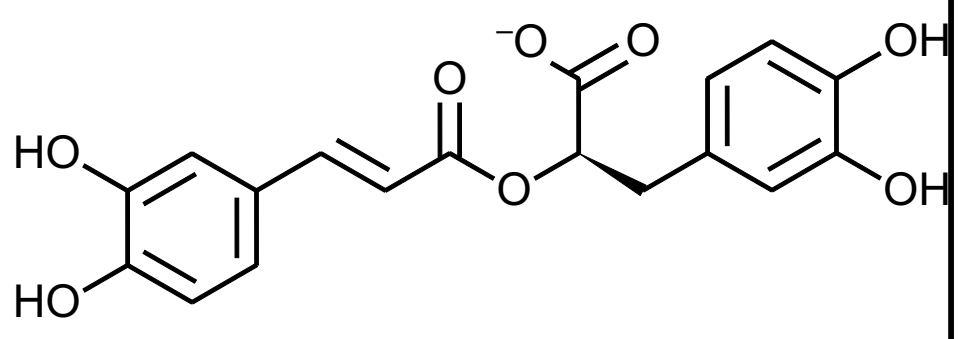<br>title: S3612 Rosmarinic acid<br>ENZYME: TMPRSS2<br>docking score: -5.604 | 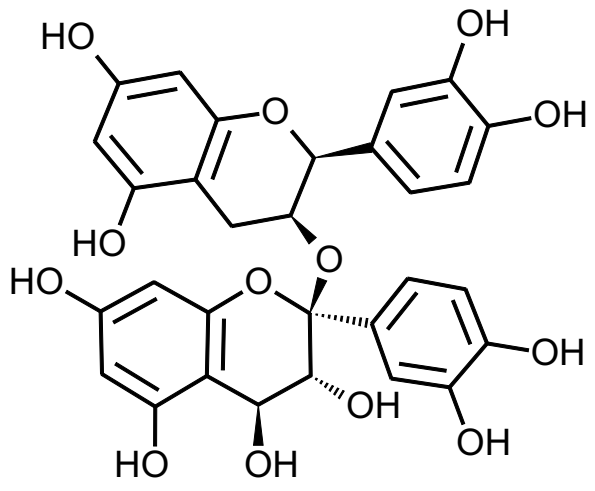<br>title: S5105<br>ENZYME: TMPRSS2<br>docking score: -5.505 |

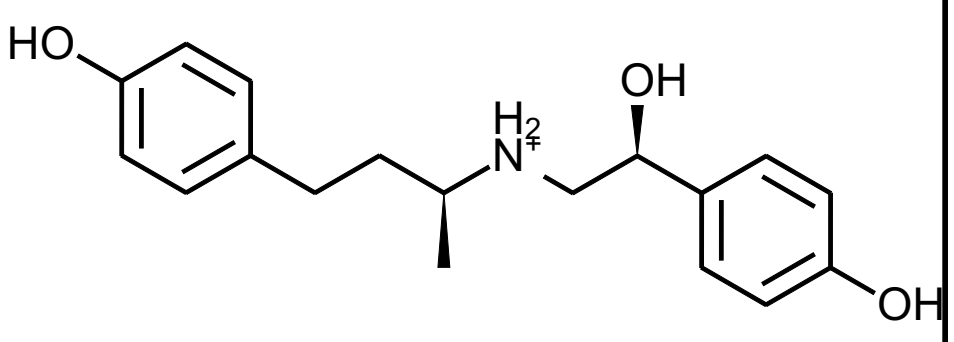

title: ractopamine hydrochloride
ENZYME: TMPRSS2
docking score: -5.218

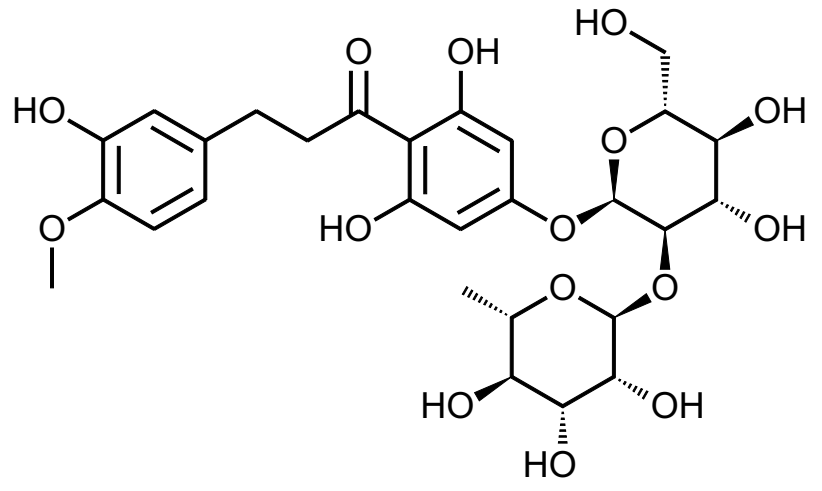

title: Neohesperidin dihydrochalcone
ENZYME: TMPRSS2
docking score: -5.202

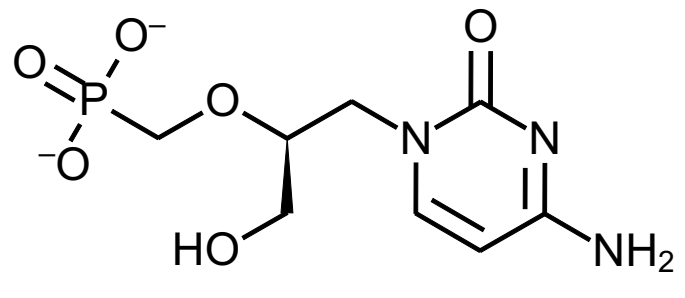

title: Cidofovir
ENZYME: TMPRSS2
docking score: -5.178

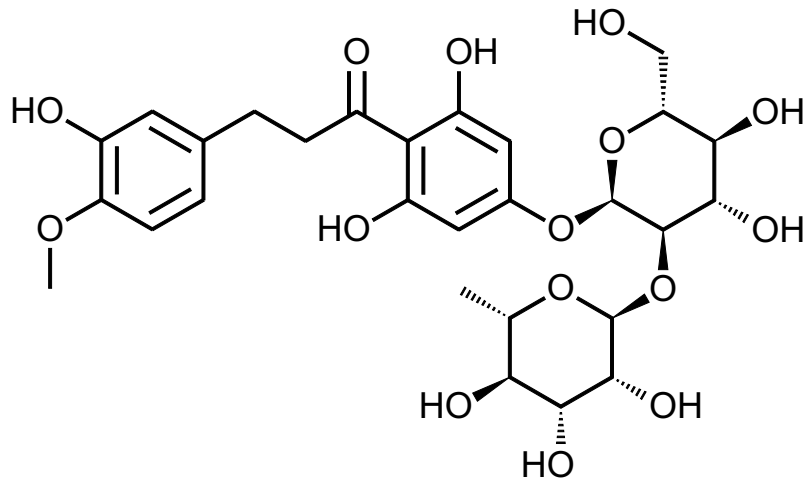

title: Neohesperidin dihydrochalcone
ENZYME: TMPRSS2
docking score: -5.029

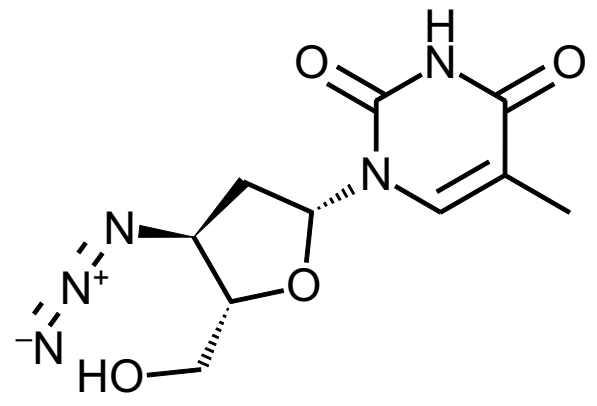

title: Zidovudine
ENZYME: TMPRSS2
docking score: -5.016